\documentclass[]{aa}   

\usepackage{graphicx}
\usepackage{multirow}
\usepackage{arydshln}
\usepackage{amsmath}
\usepackage{natbib}
\usepackage{siunitx}
\usepackage{lipsum}
\usepackage{multirow}
\usepackage{placeins}
\usepackage{hyperref}
\usepackage{threeparttable}
\usepackage{tablefootnote}
\usepackage{mathtools}
\usepackage[thinc]{esdiff}

\DeclareRobustCommand{\rchi}{{\mathpalette\irchi\relax}}
\newcommand{\irchi}[2]{\raisebox{\depth}{$#1\chi$}} % inner command, used by \rchi

\usepackage[varg]{txfonts}
\def \bp{$\beta$\,Pic}
\def \betapictoris{$\beta$\,Pictoris}

\def\feii{Fe\,{\sc ii}}
\def\fei{Fe\,{\sc i}}

\def\mnii{Mn\,{\sc ii}}

\def\niii{Ni\,{\sc ii}}
\def\nii{Ni\,{\sc i}}
\def\caii{Ca\,{\sc ii}}
\def\crii{Cr\,{\sc ii}}

\def\siii{Si\,{\sc ii}}

\def\aliii{Al\,{\sc iii}}

\def \A{$\si{\angstrom}$}
\def\feplus{Fe$^+$}

\def\niplus{Ni$^+$}
\def\siplus{Si$^+$}

\def\mnplus{Mn$^+$}

\def\crplus{Cr$^+$}

\begin{document}
\title{Population of excited levels of \feplus, \niplus\ and \crplus\ \\ in exocomets gaseous tails}
\titlerunning{Excitation of \feplus, \niplus\ and \crplus\ in exocomets}

\author{
T.\ Vrignaud\inst{1}
\and
A. Lecavelier des Etangs\inst{1}
}

\institute{
Institut d'Astrophysique de Paris, CNRS, UMR 7095, Sorbonne Université, 98$^{\rm bis}$ boulevard Arago, 75014 Paris, France 
}

\date{Received June 19, 2024 ; accepted September 20, 2024}

% - SNR => S/N      DONE

% - Sect. 2.1: Hubble Space Telescope (HST) => HST      DONE

% - remove "Plot of the" from the captions that start like that      DONE

% - run a spell check specific to your language preference (US or UK)     DONE

% - remove any italics used to denote emphasis or a specific meaning      DONE

% - check that all abbreviations and acronyms are introduced when first used (both in the abstract and in the main text), and that all instrument acronyms are introduced in the main text     DONE 

% - turn some of you lists into plain paragraphs     DONE

% - reduce the size of Figs 4 and 6 to 1-column width    DONE

% - Upload your appendices A-E and G to Zenodo and add the url in a Data availability section displayed before acknowledgments. Remove them from your article

% - keep only your appendix F as an appendix appended to your article.

% - The appendices are now published as camera-ready material. Please refer to https://www.aanda.org/for-authors/latex-issues/appendices to prepare the appendices for publication

\abstract{The star \betapictoris\ is widely known for harbouring a large population of exocomets, which create variation absorption signatures in the stellar spectrum as they transit the star. While the physical and chemical properties of these objects have long remained elusive, \cite{Vrignaud24} recently introduced the exocomet curve of growth approach, enabling, for the first time, the estimate of exocometary column densities and excitation temperature using absorption measurements in many spectral lines.
Using this new tool, we present a refined study of a \bp\ exocomet observed on December 6, 1997 with the \emph{Hubble} Space Telescope. We first show that the comet's signature in \feii\ lines is well explained by the transit of two gaseous components, with different covering factors and opacities. Then, we show that the studied comet is detected in the lines of other species, such as \niii\ and \crii. These species are shown to experience similar physical conditions than \feii\ (same radial velocity profile, same excitation temperature), hinting that they are well-mixed. 
Finally, using almost 100 \feii\ lines rising from energy levels between 0 and $33\,000 \ \si{cm^{-1}}$, we derive the complete excitation diagram of \feplus\ in the comet. The transiting gas is found to be populated at an excitation temperature of $8190 \pm 160 \ \si{K}$, very close to the stellar effective temperature ($8052 \ \si{K}$). Using a model of radiative and collisional excitation, we show that the observed excitation diagram is compatible with a radiative regime, associated with a close transit distance ($\leq 60$\,R$_\star\,\sim0.43\,$au) and a low electronic density ($\leq 10^7\,\si{cm^{-3}}$). In this regime, the excitation of \feplus\ is controlled by the stellar flux, and do not depend on the local electronic temperature or density. These results allow us to derive the \niplus/\feplus\ and \crplus/\feplus\ ratios in the December 6, 1997 comet, at $8.5 \pm 0.8 \cdot 10^{-2}$ and $1.04 \pm 0.15 \cdot 10^{-2}$ respectively, close to solar abundances.

}

\keywords{ Techniques: spectroscopic - Stars: individual: $\beta$\,Pic - Comets: general - Exocomets - Transit spectroscopy }

\maketitle

\section{Introduction}

The star \bp\ is a young \citep[20 Myr,][]{Miret-Roig_2020} nearby star, embedded in a large debris disk seen edge-on and detected through spectroscopy \citep{Vidal-Majar_1986,Roberge_2006,Brandeker_2016}, and imagery \citep{Smith_1984,Apai_2015}. Two massive planets were also discovered, \bp\ b \citep{Lagrange_2010,Snellen_2018} and \bp\ c \citep{Lagrange_2019,Nowak2020}. Thanks to these features, the \bp\ system offers a unique opportunity to investigate % very first stages of planetary formation. 
a young planetary system in the last stages of planetary formation.

In addition to the disk and the two planets, \bp\ is also well known for harbouring many exocomets, frequently transiting the star \citep{Vidal-Madjar_1998}. These objects are the analogues of comets in our own solar system, that is, icy bodies on highly eccentric orbits, whose surface sublimate when approaching periastron, producing extended gaseous and dusty tails. When the cometary tails happen to transit their host star, they become detectable, using either spectroscopy \citep{Ferlet_1987,Beust_1990,Vidal-Madjar_1994,Kiefer_2014, Kennedy_2018} or photometry \citep{Zieba_2019, Pavlenko_2022, Lecavelier_2022}. Although such objects have been detected around many other stars, such as 49~Cet \citep{Roberge_2014, Miles_2016}, HD~172555 \citep{Kiefer_2014b, Kiefer_2023}, KIC~8462852 \citep{Kiefer_2017}, and KIC~3542116 \citep{Rappaport_2018}, \bp\ is, by far, the system showing the most intense exocometary activity, easily observable thanks to the favourable orientation of the disc and the close proximity of the star. As a result, this system allows an in-depth study of the role played by small bodies in the evolution of planetary systems as a whole: for instance, the presence of cold CO within the circumstellar disk has been linked to the continual evaporation of icy comets \citep{Vidal-Madjar_1994, Jolly1998, Roberge2000}.

However, despite decades of observations, the physical properties of \bp\ exocomets remain widely unknown, as most previous studies were focussed on confirming that the observed spectroscopic events are indeed due to transiting exocomets \citep{Lagrange_1989, Beust_1993,Vidal-Madjar_1994} or developing of numerical simulations \citep{Beust_1990}. Almost no direct measurement of exocometary properties were ever published, with the notable exceptions of the works by \cite{Mouillet_1995}, which provided lower limits for the electronic temperature and density within a \bp\ transiting exocomet, and by \cite{Kiefer_2014}, which revealed the presence of two exocomet families around \bp\ with different evaporation efficiencies and orbital properties. More recently, \cite{Lecavelier_2022} also showed that the size distribution of cometary nuclei in the \bp\ system is consistent with observations in the solar system.

With the aim of studying the physical properties of exocomets with greater details, \cite{Vrignaud24} recently introduced the exocomet curve of growth technique. The gist of this technique is to combine multiple observations of a transiting comet in many ($\sim 10 - 100$) spectral lines, in order to link the comet's absorption depth in a given line to the intrinsic properties of the line (oscillator strength, lower level energy...). This relationship (or curve of growth) encapsulates the physical properties of the comet (e.g. covering factor, optical thickness), which can thus be inferred through model fitting. Applying this approach to the \feii\ absorption lines of a \bp\ exocomet observed on December 6, 1997 with the \emph{Hubble} Space Telescope (HST), \cite{Vrignaud24} provided estimates of the comet's covering factor ($\sim$36$\%$), \feplus\ column density ($\sim$$10^{15}\,\si{cm^{-2}}$) and \feplus\ excitation temperature ($\sim$10\,000\,$\si{K}$). The objective of the present study is to refine the analysis of this comet and to provide a more accurate description of its physical properties. In particular, we aim to extend the analysis to species other than \feplus\ (e.g.\ \niplus, \crplus), and to use the observed excitation diagrams to constrain the comet density and distance to the star. 

The present paper is divided as follows: the HST spectra used in our study are presented in Sect. \ref{Sect. Studied observations}; the complete \feii\ curve of growth of the December 6, 1997 comet is analysed in Sect. \ref{Sect. A refined curve of growth model}; other metallic species are discussed in Sect. \ref{Sect. Study of other species}, and the excitation diagram of \feplus\ is compared with an excitation model in Sect. \ref{Sect. Excitation model}. We then discuss our results and conclude in Sect. \ref{Sect. Discussion} and \ref{Sect. Conclusion}.

\section{Studied observations}
\label{Sect. Studied observations}

\subsection{Selected data}
\label{Sect. Raw data}

Our study is based on archival UV spectra of \betapictoris\ obtained with the Space Telescope Imaging Spectrograph (STIS) onboard the HST. 

The STIS data were collected within four different guest-observer programs, at 11 separate epochs, using the high resolution echelle gratings \emph{E140H} and \emph{E230H}  (R$\sim$114\,000$\sim$2.6\,$\si{km/s}$). Basic information on these observations is gathered in table \ref{recap_observations_stis}. The different data sets cover various wavelength domains, although the two 1997 visits (December 6 and December 19) are unique in all the \bp\ observations ever made with HST, as they cover the entire range from 1500 to 2900 \A. The data were processed using the version 3.4.2 of the {\tt calstis} pipeline released on 2018/01/19, which provides 1-D flux calibrated spectra publicly available in the MAST archive. The wavelengths were calibrated in the heliocentric reference frame.

\subsection{Data reduction}
\label{Sect. Data reduction}

We reduced the HST/STIS spectra using a method very similar to that of \cite{Vrignaud24}. First, all spectral orders from each exposure were resampled on a common wavelength table with a resolution of 18 m\A, and averaged into a single spectrum (taking advantage of the overlap between orders). The spectra were also corrected from the radial motion of \bp\ ($20 \ \si{km/s}$, \cite{Gontcharov2007}. Then, to correct for variation of the flux calibration between one exposure and another, we used cubic splines to renormalise our data to a common flux level. The method is described in \cite{Vrignaud24} ; the gist here is to calculate a cubic spline for each spectrum based on flux measurements on stable spectral regions, and then to divide each spectrum by its corresponding spline. Fig. 1 from \cite{Vrignaud24} provides a complete description of this flux correction method. Finally, each set of spectra collected during the visits of December 6, 1997, December 19, 1997, and August 26, 2001 was averaged into a single spectrum, since no significant temporal variation was detected between the different sub-exposures of each visit.

As a result, the HST/STIS data studied in the following analysis consist of 11~individual \bp\ spectra obtained at different epochs, renormalised to a common flux continuum, and covering various wavelength domain between 1500 and 3000\,\A. This region includes a large set of spectral from various species, such as \feplus, \crplus\ and \niplus .

\subsection{The exocomet-free spectrum}
\label{Sect. The exocomet-free spectrum}

Because of the high rate of exocometary transits in front of \bp\ \citep{Kiefer_2014}, all STIS spectra show strong cometary absorption in many spectral lines, in particular in the \feii\ lines series at 2400, 2600 and 2750\,\A. However, to make quantitative measurements on these absorptions, one must first recover the exocomet-free spectrum (EFS) of \bp. 

\begin{figure}[h]
    \centering    
    \includegraphics[clip,  trim = 0 0 0 50, scale = 0.44]{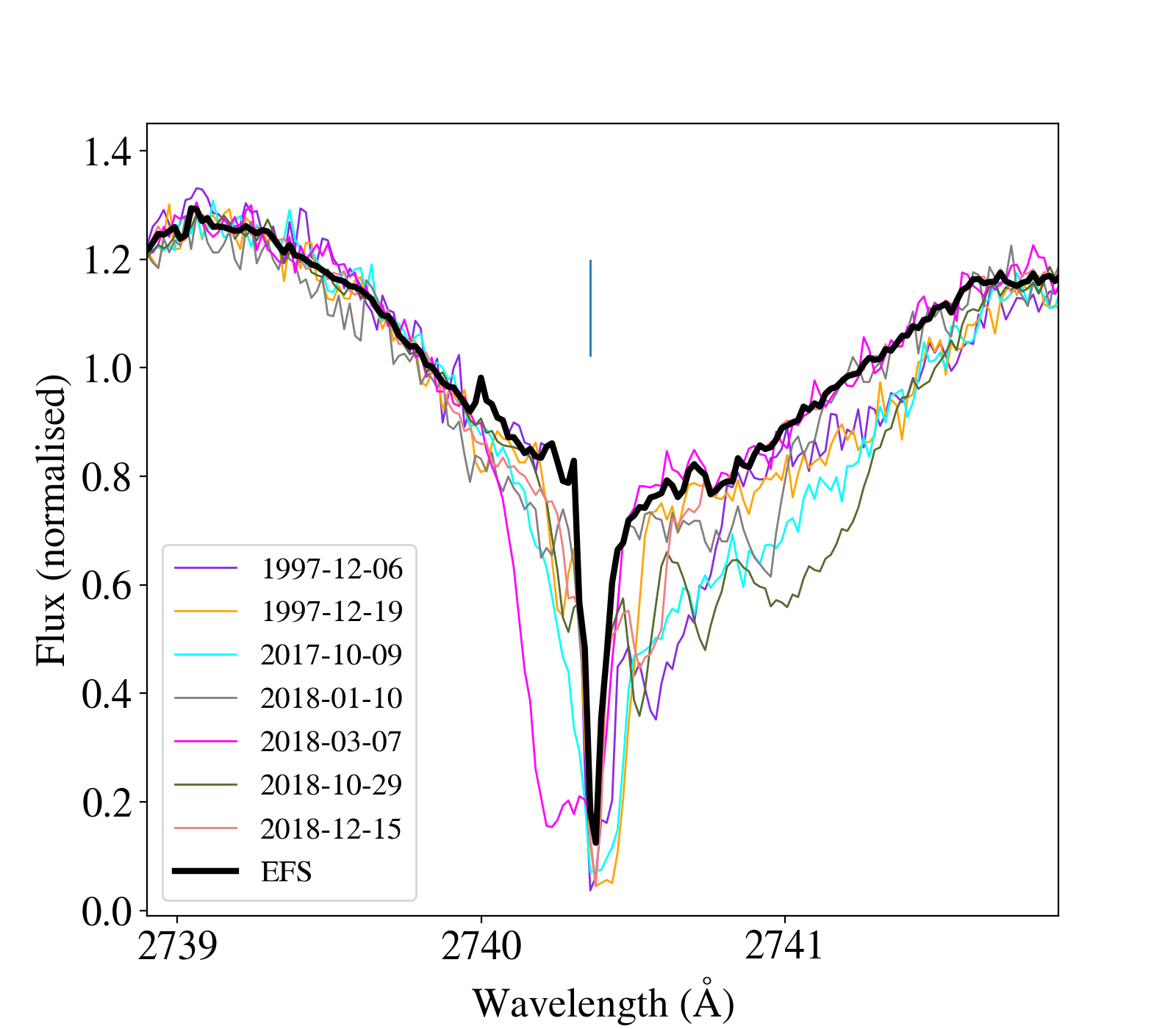}

    \caption[]{Available \bp\ STIS observations of the 2740 \A\ \feii\ line. The black line shows the recovered exocomet-free spectrum. Broad, deep exocometary absorptions are visible at all epochs. The central absorption is due to the circumstellar gas; since this absorption is stable in time, it can be included in the reference spectrum.}
    \label{Fig. 2740 A line all}
\end{figure}

The EFS was calculated using the same method as described in \cite{Vrignaud24}. First, we identified, for each visit, the lines affected by cometary absorptions, and the radial velocity domains at which these absorptions occur. This identification was done by studying strong \feii\ lines with many available observations (e.g. 2740 \A, Fig.~\ref{Fig. 2740 A line all}). Thanks to this identification, we could then predict, for each wavelength pixel, which spectra are affected by cometary absorption and which ones are not. The unnoculted flux of \bp\ was thus recovered by averaging, for each wavelength, all the spectra for with no comet absorption is expected. 

The radial velocities (RV) ranges in which each visit shows cometary absorption are provided in table \ref{recap_observations_stis}. These ranges apply to the strongest \feii\ lines (e.g. 2740 \A); for weaker lines, slightly smaller RV ranges were often used, in order to improve the signal-to-noise ratio (S/N) of the exocomet-free spectrum. For the lines where very few observations are available (such as the 2400 and 2600 \A\ \feii\ lines series, observed only three times with STIS), it is often not possible to recover the exocomet-free spectrum at all radial velocities. This lack remains a challenging obstacle to the systematic analysis of all the cometary features detected in the HST data.

\subsection{The 1997-12-06 comet}
\label{Sect. The 1997-12-06 comet}

The 1997-12-06 exocomet already studied by \cite{Vrignaud24} is particularly rich in information, for several reasons. First, its absorption is rather strong, and noticeably redshifted (between 10 and 50 $\si{km/s}$). It is thus not blended with the strong, narrow circumstellar absorption near 0 km/s (Fig. \ref{Fig. 2740 A line all}), which makes the study easier. Second, the December 6, 1997 spectrum covers an extensive wavelength range, between 1500 and 3000\,\A, providing access to hundreds of spectral lines of interest, such as the \feii\ lines series at 2400, 2600 and 2750\,\A. Finally, the other spectrum taken two weeks later in 1997 (December 19, covering the same wavelength range) is not polluted by exocomets absorption between 20 and 50 $\si{km/s}$. This guarantees that, for any line where the December 6, 1997 comet is observed, the recovery of the stellar comet-free spectrum is possible, at least in the $[+20, +50] \ \si{km/s}$ radial velocity range.
    
\begin{figure}[h]
    \centering    
    \includegraphics[clip,  trim = 0 0 0 50, scale = 0.44]{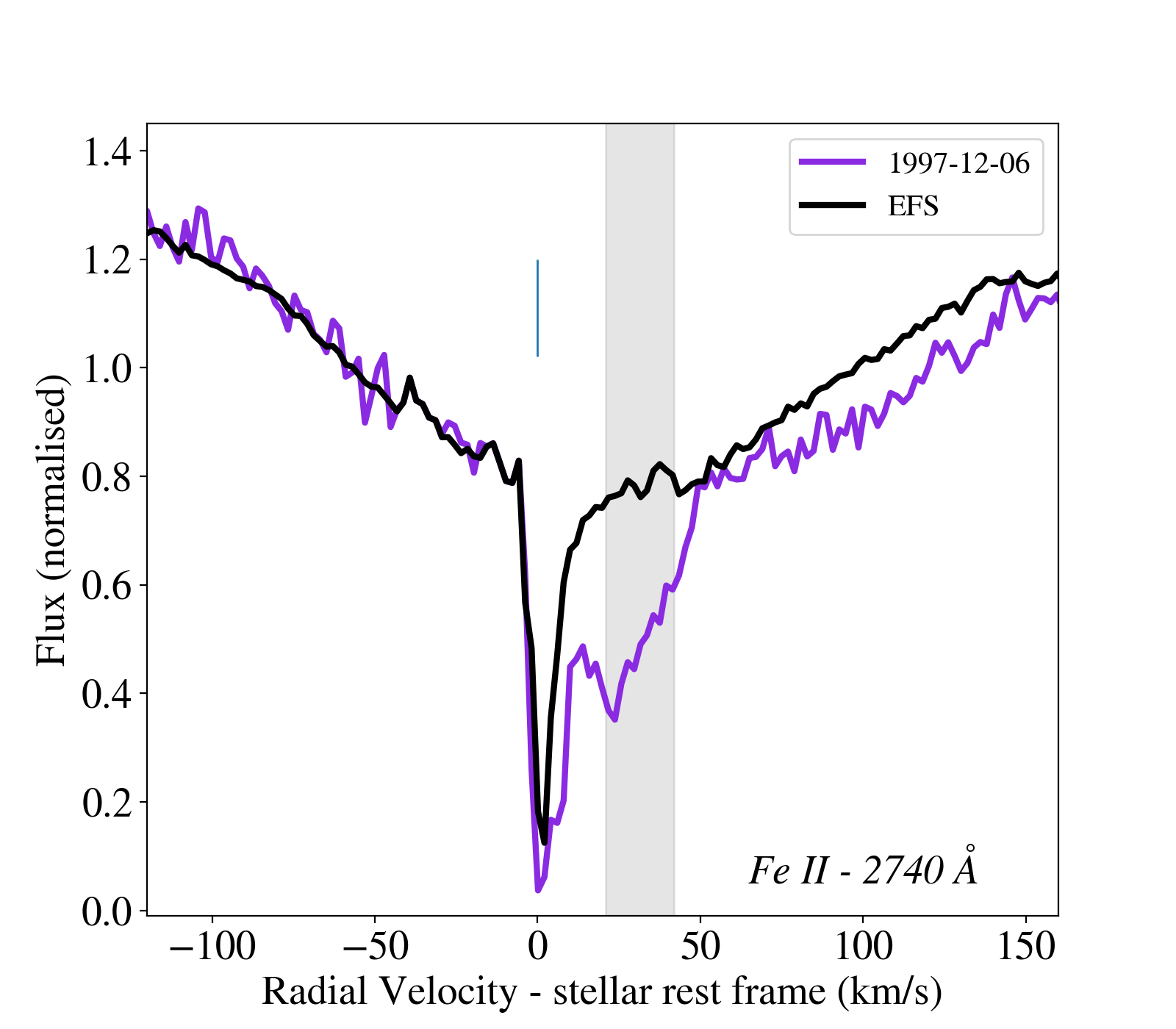}
    \caption[]{Spectrum of \bp\ obtained on December 6, 1997, along with the EFS, around the 2740 \A\ \feii\ line. The shaded area indicates the radial velocity domain ($ 21 - 42 \ \si{km s^{-1}}$) used to measure the December 6, 1997 comet's absorption depth in Sect. \ref{Sect. A refined curve of growth model} to \ref{Sect. Discussion}.}
    \label{Fig. 2740 A line 1997}
\end{figure}

\begin{figure}[h]
    \centering    
    \includegraphics[clip,  trim = 0 0 0 50, scale = 0.44]{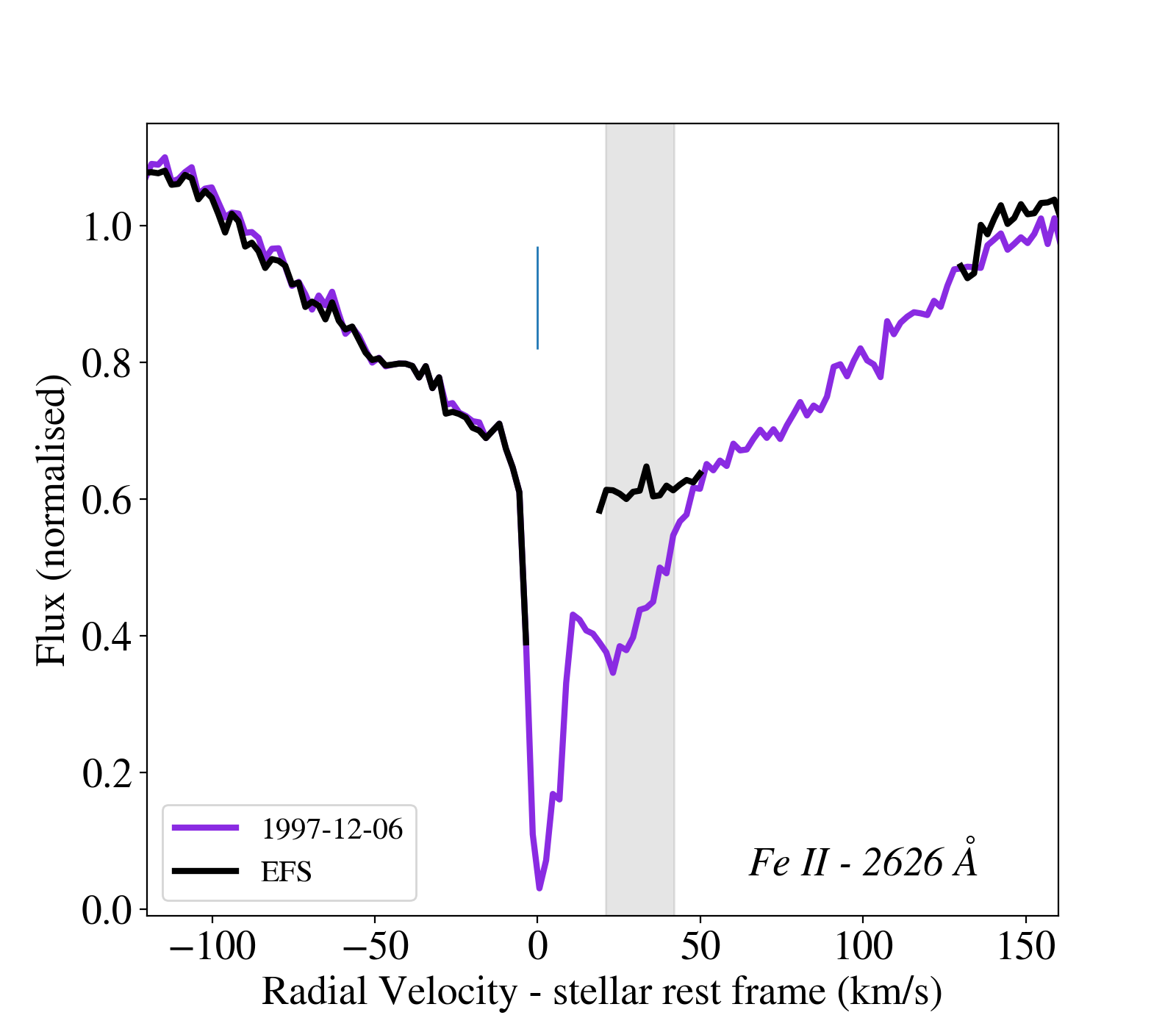}
    \caption[]{Spectrum of \bp\ obtained on December 6, 1997, along with the EFS, around the 2626 \A\ \feii\ line. The grey area emphasises the same RV domain as in Fig. \ref{Fig. 2740 A line 1997}. Here, the estimate of the EFS is incomplete, due to the low number (2) of available observations in this line.}
    \label{Fig. 2626 A line 1997}
\end{figure}

Two examples of \feii\ lines in which the December 6, 1997 comet shows strong absorption are shown on Fig. \ref{Fig. 2740 A line 1997} and \ref{Fig. 2626 A line 1997}, along with the recovered exocomet-free spectrum.

\section{A refined curve of growth model}
\label{Sect. A refined curve of growth model}

\subsection{The original model}
\label{Sect. The original model}

Presented in \cite{Vrignaud24}, the curve of growth technique is an efficient tool to retrieve the physical properties of a transiting comet from absorption measurements in a large number of lines from a given species (such as \feplus) with various oscillator strengths and excitation energies. The model introduced in \cite{Vrignaud24} is a one-zone model, which approximates a given comet by an homogeneous gaseous cloud, covering a fraction $\alpha$ of the stellar disk, and characterised by a uniform excitation temperature $T$. Under this hypothesis, the measured absorption depth ${\rm abs}_{lu}$ in any spectral line of the studied species is given by (see \cite{Vrignaud24} for details):
\begin{equation}
\hspace{1.1 cm} \overline{{\rm abs}}_{lu} = \overline{\boldsymbol{\alpha}} \cdot \left(    1 - \exp\bigg(- \ \boldsymbol{\gamma} \cdot \frac{\lambda_{lu}}{\lambda_0} g_l f_{lu} \cdot e^{-E_l/k_B \boldsymbol{T}}\bigg)
\right),
\label{Eq. Model 1}
\end{equation}
where the $l$ and $u$ indexes denote the lower and upper energy levels of the line, $\lambda_{lu}$, $f_{lu}$,  $g_l$ and $E_l$ its wavelength, oscillator strength, lower level multiplicity and lower level energy, $k_B$ the Boltzmann constant, and  where the model parameters $\overline{\alpha}$, $\gamma$ and $T$ (in bold) represent respectively the average covering factor, typical optical thickness and excitation temperature of the transiting gas. These parameters are specific to the studied chemical species (e.g. \feplus) and to the radial velocity range in which the comet's average absorption depths are measured. The wavelength $\lambda_0$ can be chosen arbitrarily; in the following we will fix its value to the wavelength of the 2756 \A\ \feii\ line, as in \cite{Vrignaud24}.

The key of Eq. \ref{Eq. Model 1} is that it holds for any line of the studied species (e.g. \feplus). Fitting this model to the comet absorption depths ${\rm abs}_{lu}$ measured in a great number of lines can thus provide accurate estimates of the gas physical properties, including its average covering factor ($\overline{\alpha}$) in the studied RV range, the excitation temperature of the studied species ($T$), and its column density (through the estimate of $\gamma$). For instance, the study of \cite{Vrignaud24} yielded $\overline{\alpha} = 36 \pm 1 \%$, $T = 10\,500 \pm 500 \ \si{K}$ and $\gamma = 4.72 \pm 0.25$ for the December 6, 1997 comet, allowing to constrain its \feii\ column density in the [+25,+40] km/s range to $6.7 \pm 0.4 \cdot 10^{14} \ \si{cm^{-2}}$. 

However, these estimates were obtained by fitting only weak \feii\ lines ($g_l f_{lu}$ < 0.7): a discrepancy was noticed for the very strongest lines ($g_l f_{lu} \geq 0.7$), in which the absorption of the December 6, 1997 comet is much deeper than predicted by the extrapolation of the curve of growth.

To illustrate this discrepancy, we fitted again the original curve of growth model (Eq. \ref{Eq. Model 1}) to the December 6, 1997 comet absorption depths in 63 \feii\ lines, with wavelengths between 2249 and 2769 \A. This set includes all lines from the three main line series at 2400, 2600 and 2750 \A, except those mixed in intricate multiplets (e.g. 2750 \A\ triplet) or close to very strong lines (e.g. 2599 \A\ line, blended with the strong 2600 \A\ line). They arise from rather low excitation levels ($E_l \leq 12\,000 \ \si{K}$), but have very diverse $gf$ values, ranging from 0.01 for the weakest lines to 3.20 for the strongest 2382\,\A\ line. 

Other \feii\ lines at shorter wavelength (e.g. 1608 or 1702 \A, in which the December 6 comet is also detected) were not included in the fit, because the stellar flux is rather weak in these lines (resulting in a much poorer S/N on absorption measurements), and because the oscillator strengths are poorly determined \citep[see the NIST database,][] {NIST_ASD}. The full list of studied lines is provided in Table \ref{Tab. list lines Fe II}; they can also be visualised from the full stellar spectra, Fig. \ref{Fig. Full spectrum}. 

The comet's absorption depths were measured in the $[+21, +42]\,\si{km/s}$ RV range, where the strongest absorption occurs. This range is slightly increased compared to \cite{Vrignaud24}, in order to improve the S/N in the weakest lines (\feii, \crii). Uncertainties on the absorption depths were calculated taking into account photon noise (generally $\sim 1 \%$) and systematic errors introduced by our reduction algorithm ($\sim 1 \%$). 

\begin{figure}[h]
    \centering    
    \includegraphics[clip,  trim = 63 20 0 50,  scale = 0.415]{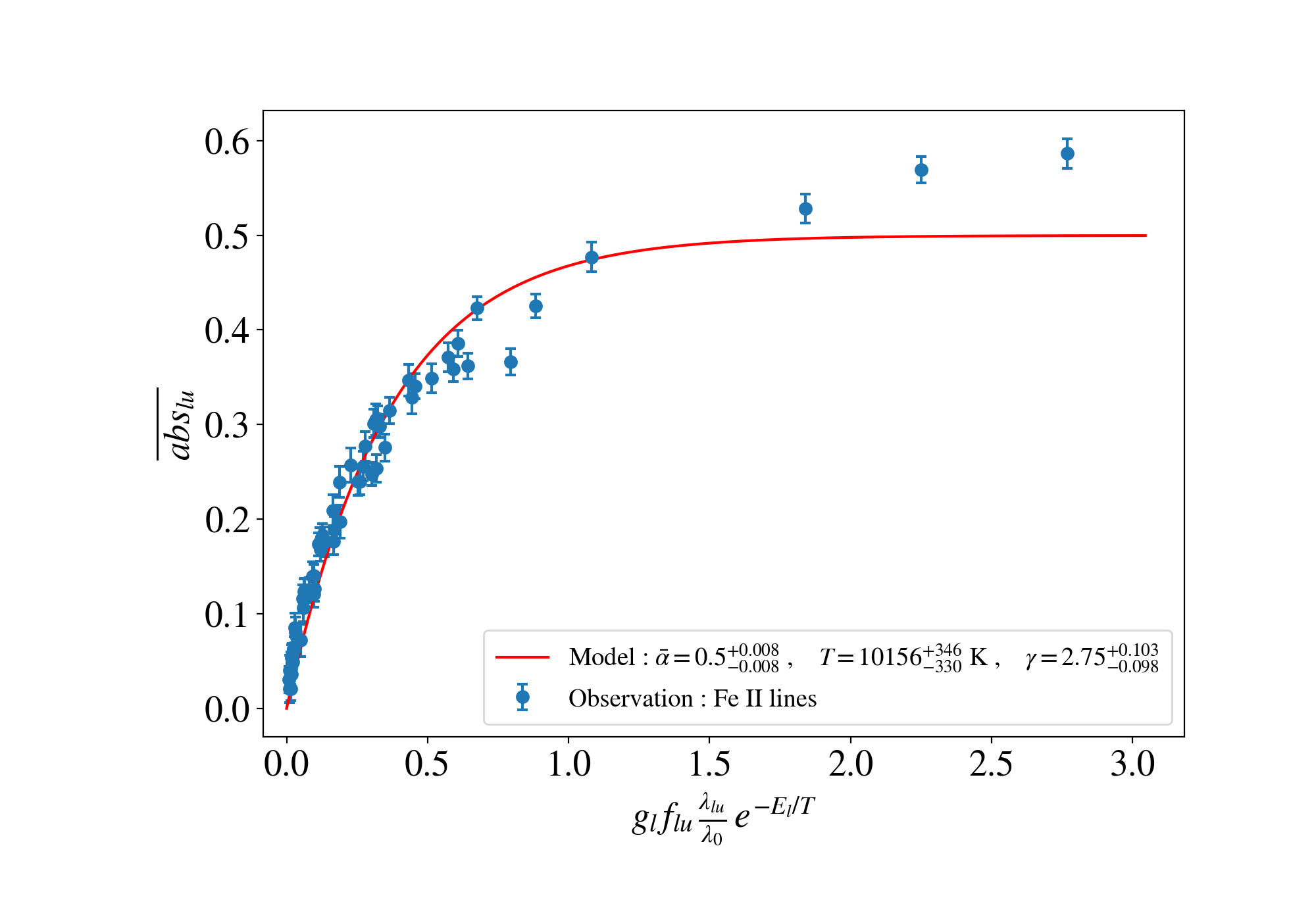}
    \includegraphics[clip,  trim = 63 0 0 20,  scale = 0.415]{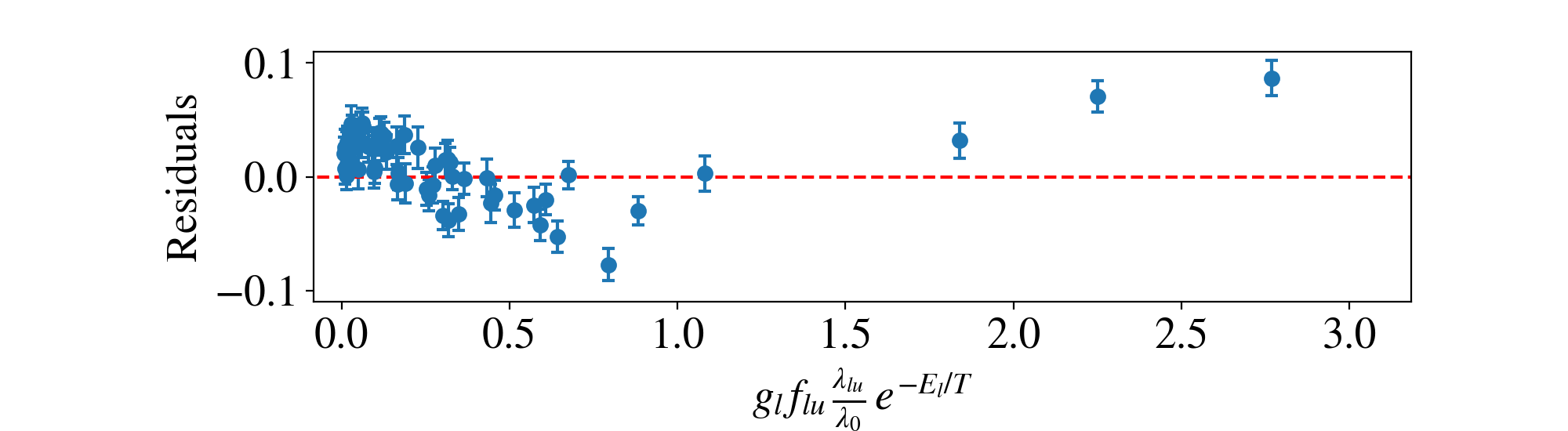}
    \caption[]{\textbf{Top}: Curve of growth of \feii\ in the December 6, 1997 comet. The blue dots indicate the measured absorption depth in 63 \feii\ lines, while the red line shows the best fitted 1-component model. \textbf{Bottom}: Residuals of the previous fit. A clear remaining trend is visible in the data, hinting that the fit could be improved in order to retrieve more information from the comet's absorption spectrum.}
    \label{Fig. Cog_1_component}
\end{figure}

As shown on Fig.~\ref{Fig. Cog_1_component}, the agreement between the curve of growth model given by \cite{Vrignaud24} and the December 6, 1997 comet absorption depths is rather poor (reduced $\rchi^2$ of 5.01). The comet is found to have very strong absorptions in both the weakest lines ($g_l f_{ul} e^{-E_/k_BT} \sim 0.1$) and the strongest ($g_l f_{ul} e^{-E_/k_BT} \geq 2$), well above the best-fit model. Here, the comet's covering factor is constrained to be about $50\%$ of the stellar disk, but, as already noted by \cite{Vrignaud24}, the observed absorption depths in the very strongest lines hint that the comet actually covers a greater stellar area. We also note that the fit constrains the \feplus\ excitation temperature to be $10\,150 \pm 350 \ \si{K}$, similar to the value found in \cite{Vrignaud24}.

\subsection{Two components model}
\label{Sect. Two components model}

\begin{figure}[h]
    \centering    
    \includegraphics[clip,  trim = 63 20 0 50,  scale = 0.415]{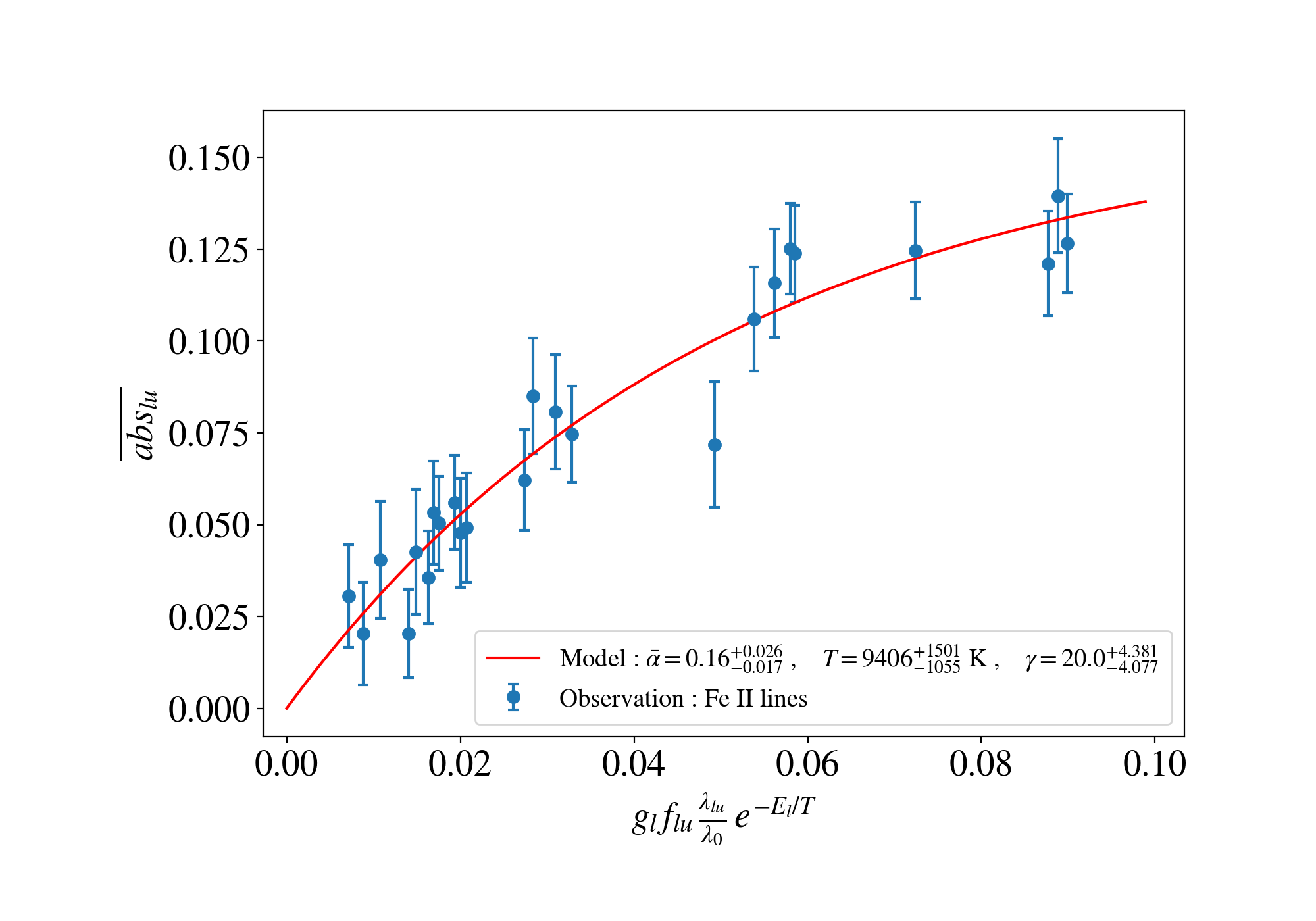}
    \includegraphics[clip,  trim = 63 20 0 50,  scale = 0.415]{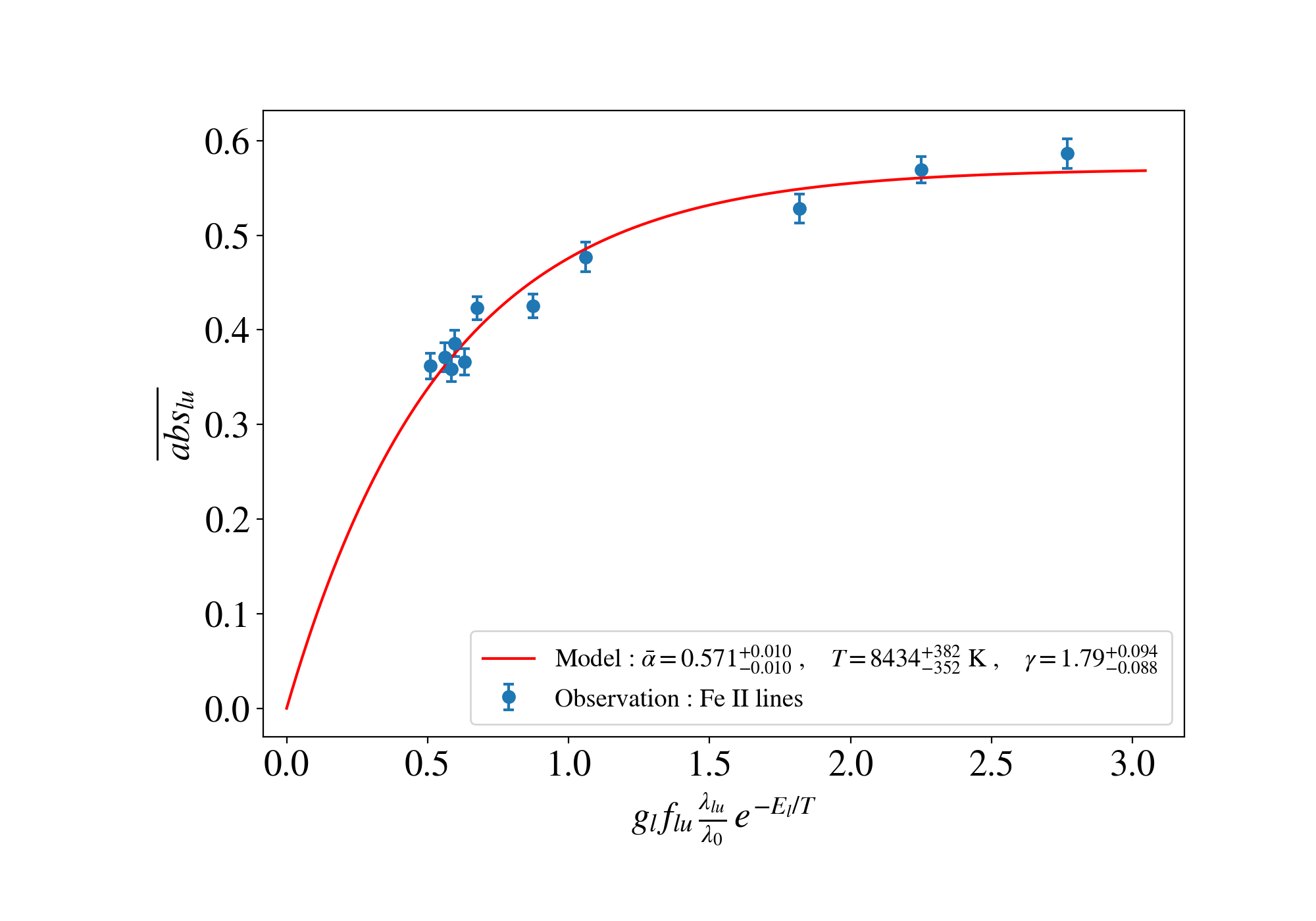}
    \caption[]{Curve of growth of \feii\ lines in the December 6, 1997 comet, focusing only the weakest lines (top, absorption below 14 \%) and the strongest lines (bottom, absorption above 34 \%). The curve were fitted with the one-component curve of growth model (Eq. \ref{Eq. Model 1}).}
    \label{Fig. cog_1_component_only_weak_cog}
\end{figure}

As shown by Fig.~\ref{Fig. Cog_1_component}, the whole set of absorption measurements in the December 6, 1997 comet is poorly described by a one component model. To better understand the geometry of the transiting comet, it is useful to performed again the previous fit, but this time focusing on limited subsets of lines. For instance, Fig.~\ref{Fig. cog_1_component_only_weak_cog} provides the comet's curve of growth in the 24 weakest \feii\ lines (absorption depth below 14\%, top panel) and in the 11 strongest (absorption depth above 34\%, bottom panel). Both curves of growth were fitted using the one-component model (Eq. \ref{Eq. Model 1}). The comparison between the two fits shows that, while the measured excitation temperatures are similar ($\sim$9000\,$\si{K}$), the covering factors ($\alpha$=16$\pm$2\% in weak lines and $\alpha=$57$\pm$1\% in strong lines) and optical thicknesses ($\gamma$=20 and 1.8) are very different. This hints that the column density distribution within the comet is not homogeneous: a small part of the comet appears to be very dense ($\gamma \gg 1$), resulting in a rapid increase of the absorption depth in the weakest lines (top panel), while the remaining part of the comet seems to be larger but much thinner, saturating at a slower pace ($\gamma \sim 1$, bottom panel).

To better fit the December 6, 1997 comet's absorption depths, we thus introduce a two component model for the curve of growth. This new model describes a given transiting comet as such: 
\begin{itemize}
    \item The gaseous cloud is characterised by a total covering factor $\boldsymbol{\overline{\alpha}_{t}}$, averaged over the radial velocity range where the cometary absorption depths are measured;
    \item The cloud is described by two components : a dense core and a more extended thinner envelope. The relative sizes of these two components is parameterised by the core component covering factor, $\boldsymbol{\overline{\alpha}_{\rm c}}$;
    \item The excitation temperature of the studied species ($\boldsymbol T$) is assumed to be uniform within the whole cloud;
    \item The optical thickness ratio between the external and core components is denoted 
    % $\boldsymbol{\frac{\gamma_{\rm e}}{\gamma_{\rm c}}}$;
    $\boldsymbol{\gamma_{\rm e}/\gamma_{\rm c}}$;
    \item Finally, the total column density of the studied species (e.g. \feplus) is parameterised by its optical thickness in the core component, $\boldsymbol{\gamma_{\rm c}}$.    
\end{itemize}

This new model has thus five parameters (highlighted in bold font). It should be noted that only the last ($\gamma_{\rm c}$) is completely specific to the considered species; the same value for the first four parameters can be used simultaneously for several species if their excitation temperatures and spatial distributions are the same. This parameterisation leads to the following expression for the absorption depth in a given spectral line :
\begin{equation}
    \hspace{1.4 cm} \overline{\rm abs}_{lu}  = \overline{\alpha}_{\rm c}  \cdot \left(  1 - e^{- \tau_{\textrm{c, } lu}} \right) + \ \overline{\alpha}_{\rm e} \cdot \left(  1 - e^{- \tau_{\textrm{e, } lu}} \right),
\label{Eq. Model 2}
\end{equation}
with $\tau_{\textrm{c, } lu}$ and $\tau_{\textrm{c, } lu}$ the line optical thicknesses in the core and external components: 
$$
\tau_{\textrm{c, } lu} = \gamma_{\rm c} \cdot \frac{\lambda_{lu}}{\lambda_0} g_l f_{lu} e^{-E_l/k_B T},
$$
$$\tau_{\textrm{e, } lu} = \gamma_{\rm e} \cdot \frac{\lambda_{lu}}{\lambda_0} g_l f_{lu} e^{-E_l/k_B T},
$$
and with $\overline{\alpha}_{\rm e}$ the covering factor and $\gamma_{\rm e}$ the characteristic optical thicknesses of the external component, easily expressed as a function of the model parameters:

$$
\begin{array}{ll}
     \overline{\alpha}_{\rm e} = \overline{\alpha}_{\rm t} - \overline{\alpha}_{\rm c},  \\
     \gamma_{\rm e} = \gamma_{\rm c} \cdot \frac{\gamma_{\rm e}}{\gamma_{\rm c}}. 
\end{array}
$$

\subsection{Fit}
\label{Sect. Fit 2 components}

\begin{figure}[h]
    \centering    
    \includegraphics[clip,  trim = 63 20 0 50,  scale = 0.415]{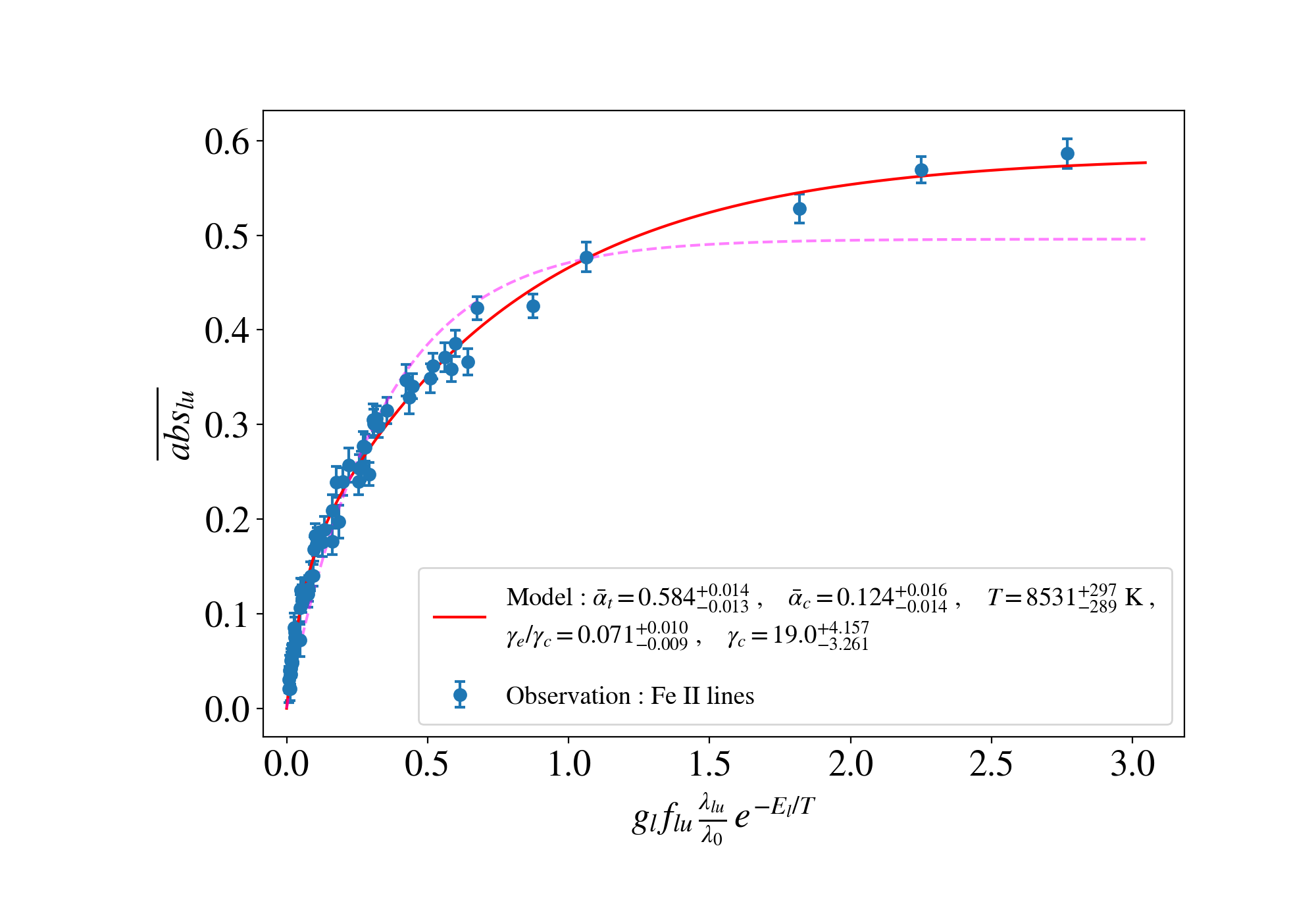}
    \includegraphics[clip,  trim = 63 0 0 20,  scale = 0.415]{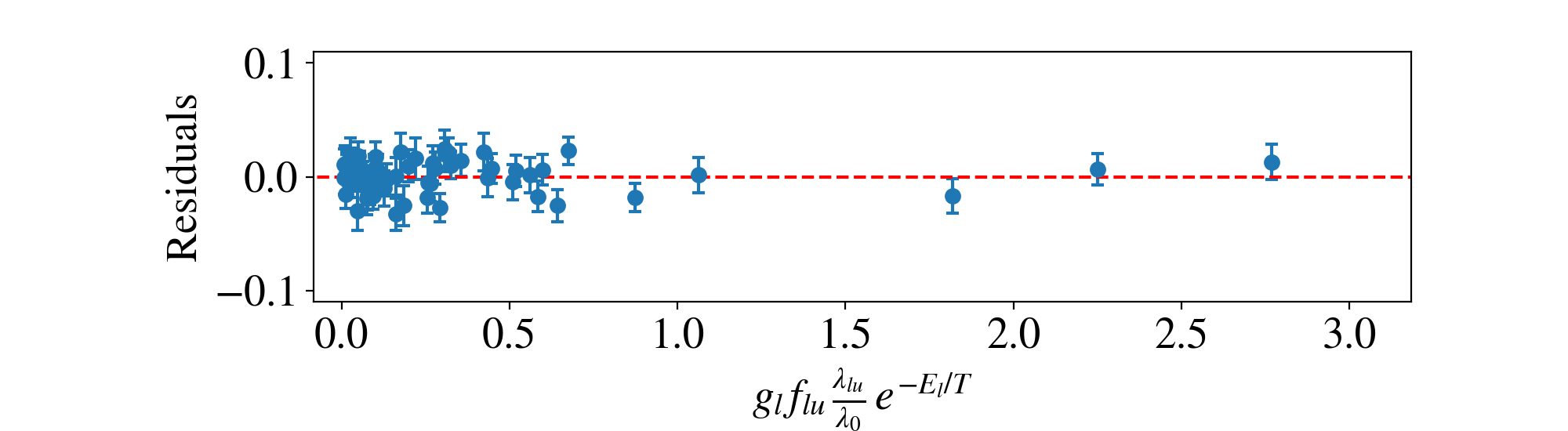}
    \caption[]{Same figure as \ref{Fig. Cog_1_component}, but this time the absorption depths of the December 6, 1997 comet were fitted with a 2-component curve of growth model (red). The purple, dotted line indicates the single component model fitted in Fig. \ref{Fig. Cog_1_component}. The agreement between data and model is much better with the two-component model.}
    \label{Fig. Cog_2_component}
\end{figure}

The two-component model was tested on the December 6, 1997 comet, using the same set of \feii\ lines as in Sect.~\ref{Sect. The original model} (i.e., 62 lines rising from the 2400, 2600 and 2750 \A\ series) and the same radial velocity range ($[+21, +42]\,\si{km/s}$). The result of the fit is presented in Fig.~\ref{Fig. Cog_2_component}; the agreement between the measured average absorption depths and the two-component model is a lot better than with a single component (reduced $\rchi^2$ of 1.08, compared to 5.01 for Fig.~\ref{Fig. Cog_1_component}). Here, the total covering factor of the comet is constrained to 58.4$\pm$1.4\,\%, significantly higher than the value obtained with the one-component model (50 \%). This total size is then distributed into a dense component, covering about 12 \% of the stellar disk surface and getting rapidly saturated ($\gamma_c$$\sim$19), and a thinner one, covering 46\% of the stellar disk and with a much lower optical thickness ($\gamma_e = \gamma_c \cdot \left( \gamma_e/\gamma_c \right) \sim 1.35$, about 14 times fainter than the core component). These values are consistent with the rough estimates from Fig.~\ref{Fig. cog_1_component_only_weak_cog}, obtained by fitting the one-component model to restricted samples of lines.

We note that it is difficult to know whether the two gaseous components identified in the December 6, 1997 spectrum actually belong to the same inhomogeneous cometary tail or correspond to two different comets with different covering factors and optical thicknesses, but similar radial velocities. In the following, we will adopt the first hypothesis, keeping referring to the 'December 6, 1997 comet'; however, one should keep in mind that the studied object is probably fairly complex, and could actually be better described as a string of cometary nuclei.

Using the fitted parameters from the two-component curve of growth model, it is possible to estimate the \feplus\ column density within the December 6, 1997 comet, averaged above the whole stellar surface, through \citep[see][for details]{Vrignaud24}: 
\begin{equation}
\hspace{0.8 cm} {\rm N}_{\rm tot} = \frac{4 \varepsilon_0 m_e c}{e^2 \lambda_0} (v_2 - v_1) \ Z\big(T\big)  \cdot \bigg ( \gamma_{\rm c} \overline{\alpha}_{\rm c} + \gamma_{\rm e} \overline{\alpha}_{\rm e}\bigg).
\label{Eq. N_tot}
\end{equation}
This quantity depends on the studied radial velocity range $[v_1, v_2]$. For the RV range $[+21,+42]\,\si{km/s}$, we find ${\rm N}_{\rm tot}=5.0\pm0.4\cdot10^{14} \ \si{cm^{-2}}$. For comparison, \cite{Vrignaud24} estimated a \feplus\ column density of $\sim 6.7 \cdot 10^{14} \, \si{cm^{-2}}$, in a similar radial velocity range ([+25,+40] km/s). However, this last value corresponds to the \feplus\ column density within the cometary tail, found to cover $36 \%$ of the stellar surface. Converting this column density into the average \feplus\ column density above the whole stellar disc, we get a value of $\sim 2.4 \cdot 10^{14} \ \si{cm^{-3}}$, much below our current estimate. The omission of a dense component in the study of \cite{Vrignaud24} thus led to a significant underestimation of the \feplus\ column density in the transiting comet.

\subsection{Excitation diagram}
\label{Sect. Excitation diagram}

The fit of the curve of growth with a two-component model constrains the \feplus\ excitation temperature in the December 6, 1997 comet to be 8530$\pm$300\,$\si{K}$ (Fig.~\ref{Fig. Cog_2_component}). To better visualise this excitation temperature, it is convenient to build the excitation diagram of \feplus . This diagram is obtained using the fact that, once the comet geometry is constrained, there is, for each line, a direct relationship between the Boltzmann factor $e^{-E_l/k_BT}$ and the comet absorption depth ${\rm abs_{lu}}$ (Eq.~\ref{Eq. Model 2}). This relation can be written as:
$$
{\rm abs_{lu}} = \varphi_{lu} \big( e^{-E_l/k_BT} \big),
$$
where the function $\varphi_{lu}$ depends on the comet geometry ($\overline{\alpha}_{\rm c}, \, \overline{\alpha}_{\rm e}, \, \gamma_{\rm c}, \, \gamma_{\rm e}$) and on the line properties ($g_l f_{lu}, \, \lambda_{lu}$). If the gas follows a Boltzmann distribution, we have $e^{-E_l/k_BT} = Z(T) \frac{{\rm N}_l}{g_l {\rm N}_{\rm tot}}$. Inverting - numerically - the above equation, we see that each absorption measurement in a given line provides an estimate of the abundance of its lower level $l$, through:

\begin{equation}
    \hspace{3.1 cm}  \frac{{\rm N}_l}{g_l {\rm N}_{\rm tot}} = \frac{1}{Z(T)} \varphi^{-1}_{lu} \big( {\rm abs_{lu}} \big).
    \label{Eq. Excitation diagram}
\end{equation}
It can then be shown that this equation still provides a good estimate of $\frac{{\rm N}_l}{g_l {\rm N}_{\rm tot}}$ even if the transiting gas is not perfectly following a Boltzmann distribution, as long as Eq. \ref{Eq. N_tot} holds.

\begin{figure}[h]
    \centering    
    \includegraphics[clip,  trim = 55 55 0 50, scale = 0.4]{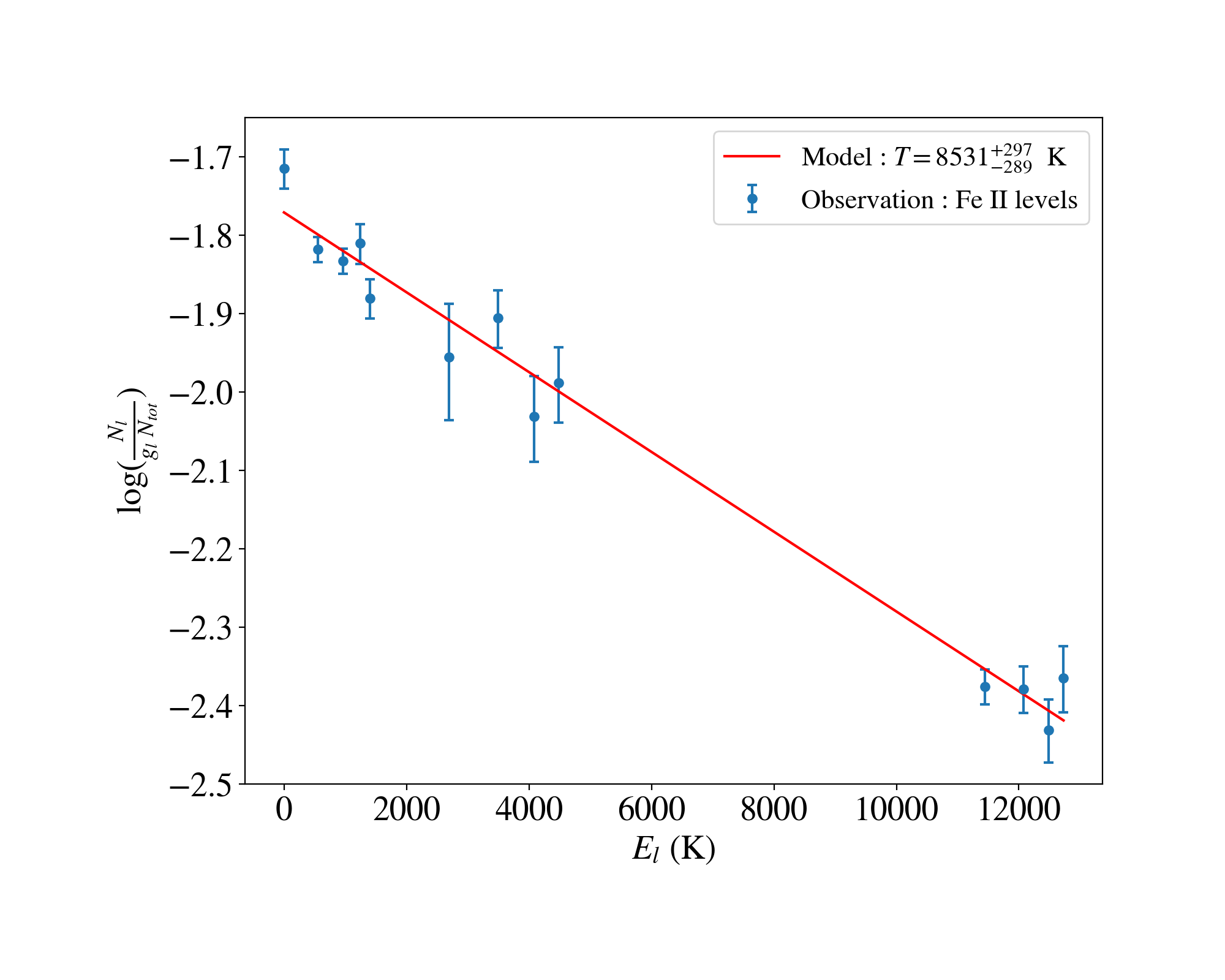}

    \caption[]{Excitation diagram of \feplus\ in the December 6, 1997 comet, derived after fitting the two-components curve of growth model. Each blue dot shows the relative abundance of one excitation level, normalised by its multiplicity $g_l$. The solid red line shows the Boltzmann distribution at $T = 8530 \ \si{K}$  }
    \label{Fig. Cog_2_component_ex}
\end{figure}

The excitation diagram of \feplus\ in the December 6, 1997 comet is shown on Fig.~\ref{Fig. Cog_2_component_ex}. Since a given energy level can be probed by several lines independently (Table \ref{Tab. list lines Fe II}), all abundance measurements from lines rising from the same energy level were averaged altogether. Despite a few outliers (most notably the ground level), the data points are fairly well aligned with the model. The excitation temperature (inversely proportional to the slope) is here constrained to be $T$=8530$\pm$300\,$\si{K}$, significantly lower than the value obtained with the one-component model ($T$\,$\sim$\,$10\,150$\,$\si{K}$, Sect.~\ref{Sect. The original model}). Given the much better fit quality for the two-component model, this new value is probably a better estimation of the excitation temperature of low-lying ($E_l \leq 12\,000 \ \si{K}$) \feplus\ levels in the studied comet.

\section{Study of other species}
\label{Sect. Study of other species}

The analysis of the December 6, 1997 comet's absorption signatures in \feii\ lines now seems rather satisfactory : the comet is well fitted by a two-component curve of growth model, yielding a total covering factor of $\sim$58\% and an excitation temperature of $\sim$8500\,$\si{K}$. 
However, the December 6, 1997 comet is also detected in \siii, \niii, \mnii\ and \crii\ lines (see Fig.~\ref{Fig. comparaison espèces I}, \ref{Fig. comparaison espèces II} and \ref{Fig. exemples Cr II}). As of now, we still have no insight whether all these species are well-mixed (i.e., their spatial distributions are all the same) or have completely different behaviours - due, for instance, to different radiation pressures or production mechanisms. The goal of this section is to address this question.

\subsection{Similar absorption profiles in Ni, Mn, Si and Cr lines}
\label{Sect. Similar absorption profiles in Ni, Mn, Si and Cr lines}

Figs.~\ref{Fig. comparaison espèces I} and \ref{Fig. comparaison espèces II} show \bp\ lines from various ions (\feii, \niii, \mnii, \siii), as observed on December 6 and December 19, 1997. Since a slight change in shape for the December 6, 1997 comet's absorption profile is observed between the two orbits of the December 6, 1997 HST visit, we restrained the comparison between lines observed within the same orbit. 

The studied comet is clearly detected in all lines, with a very similar absorption profile : no matter the species, the absorption always spans from $\sim$10 to $\sim$50\,km/s, and exhibits a sharp maximum near 22\,km/s (particularly visible in \feii\ and \niii, thanks to a stronger absorption). The main difference between species is the depth of the absorption, due to weaker line strength (\siii) or elemental abundances (\mnii). Despite much fainter absorption depth, \crii\ is also detected at similar radial velocities (Fig.~\ref{Fig. exemples Cr II}).

Overall, this comparison hints that exocometary species tend to stay 'mixed' in gaseous tails, keeping the same spatial and velocity distributions. This may appear puzzling, as the observed species have very different radiation pressure to gravity ratios \citep[0.15 for \siii, 1.39 for \niii, 4.87 for \feii\ and 11.0 for \mnii,][]{Lagrange1998}, and should therefore have different dynamics. This mixing is probably due to the high ionisation of the gas, which leads the detected species to efficiently exchange momentum via Coulomb scattering \citep[as suggested in][]{Beust1989} and couples their dynamics. In this scenario, the transiting gas would act as a single fluid, subject to a radiation pressure with contributions from all different ions.

\subsection{The excitation temperature of \niplus\ and \crplus}
\label{Sect. The excitation temperature of niplus and crplus}

The \niii\ and \crii\ lines are particularly interesting, as they are fairly numerous (more than 15 lines each) and rise from a wide range of excitation levels, with energies from 0 to 15\,000\,$\si{cm^{-1}}$ for \niii\ and from 0 to 12\,000\,$\si{cm^{-1}}$ for \crii. Fitting the curves of growth of the two species, it is therefore possible to estimate their excitation temperatures in the December 6, 1997 comet. 

To perform these two fits, we assumed that the comet geometry constrained in Sect.~\ref{Sect. Fit 2 components} for \feplus\ also applies to \niplus\ and \crplus, as suggested by the similarity of the absorption profiles (Sect.~\ref{Sect. Similar absorption profiles in Ni, Mn, Si and Cr lines}). We thus fitted the two-components curve of growth model to the \crii\ and \niii\ absorption depths of our studied comet, imposing gaussian priors on $\overline{\alpha}_{\rm t}$, $\overline{\alpha}_{\rm c}$ and $\gamma_{\rm e}/\gamma_{\rm c}$ accordingly to the results of Sect. \ref{Sect. Fit 2 components}. For each of the two species, the remaining free parameters were thus $\gamma_{\rm c}$ - linked to the typical opacity of the studied species within the comet - and $T$ - its excitation temperature. The studied lines are listed in Table \ref{Tab. list lines Cr II} for \crii\ and \ref{Tab. list lines Ni II} for \niii. They can also be visualised through the full \bp\ spectrum provided in Fig. \ref{Fig. Full spectrum}. 

\begin{figure}[h]
    \centering    
    \includegraphics[clip,  trim = 55 55 0 50, scale = 0.4]{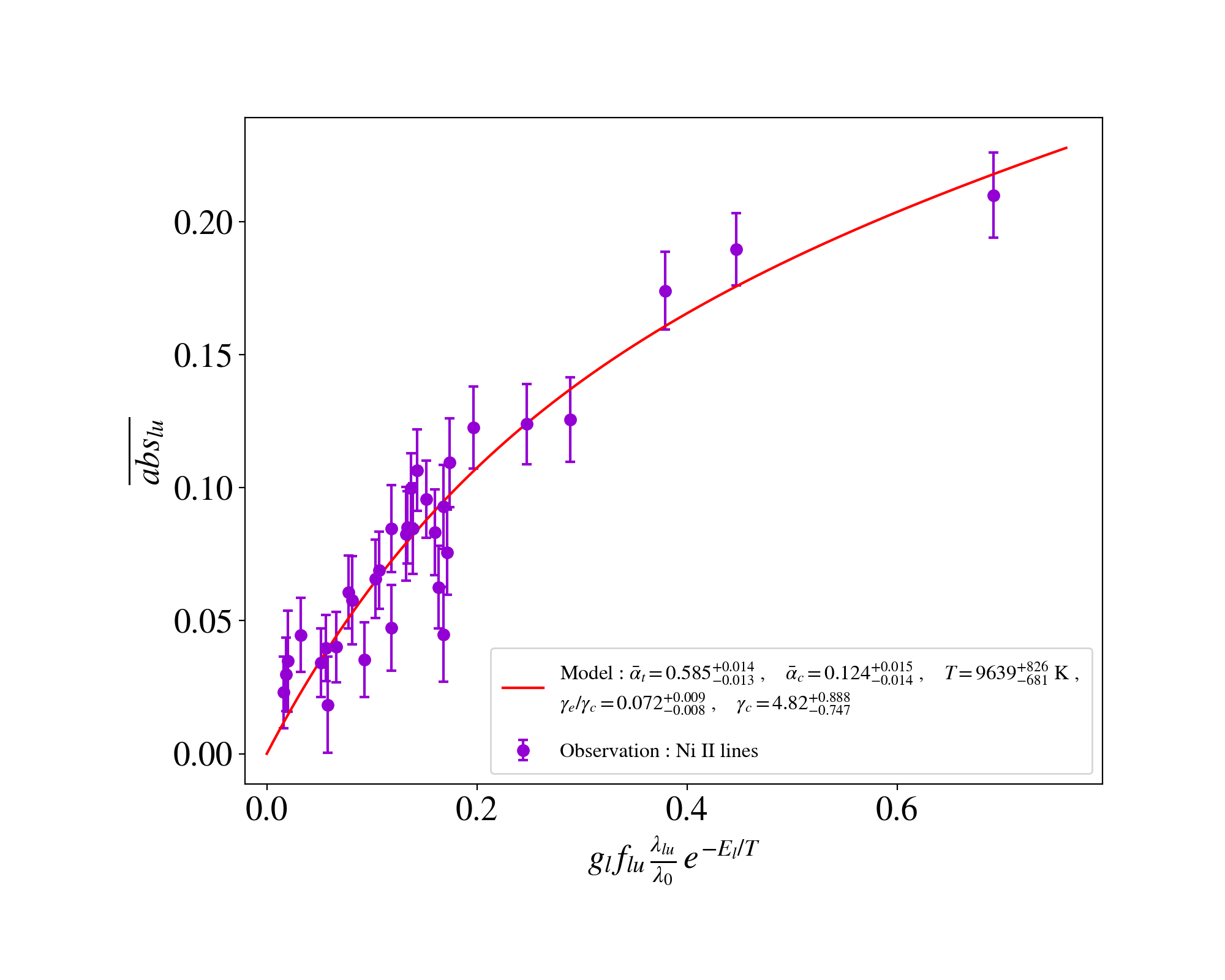}
    \includegraphics[clip,  trim = 55 55 0 50, scale = 0.4]{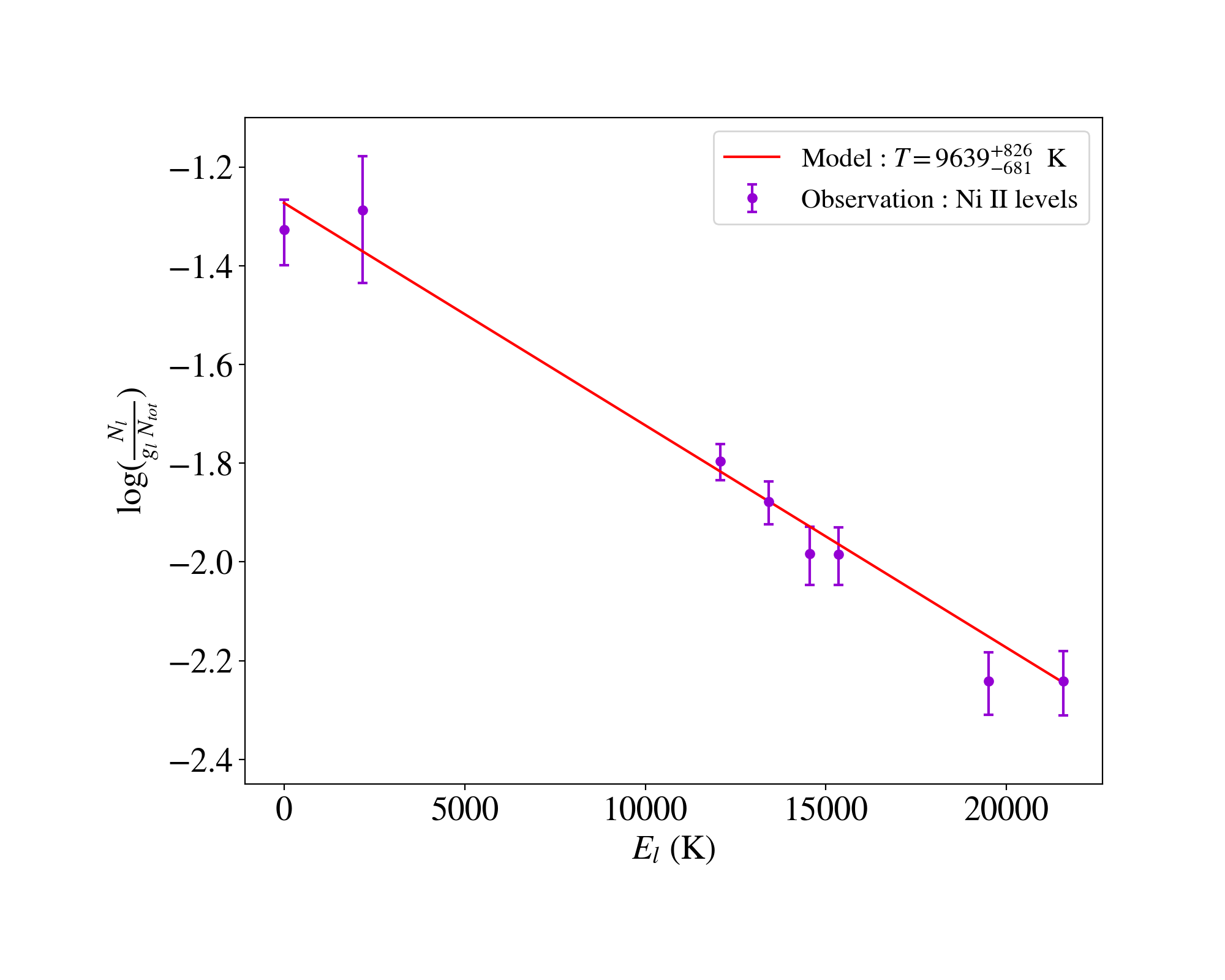}

    \caption[]{\textbf{Top}: Curve of growth of \niii, in the December 6, 1997 comet. Purple dots indicate the measured absorptions for 33 \niii\ lines in the [+21,\,+42] km/s range, and the red line shows the best fit with the two-component model. 
    %Gaussian priors were assumed for $\overline{\alpha}_{\rm t}$, $\overline{\alpha}_{\rm c}$ and $\frac{\gamma_{\rm e}}{\gamma_{\rm c}}$, according to the result of Fig. 6 for \feii. 
    \\
    \textbf{Bottom}: Excitation diagram of \niplus\ (same legend as in Fig.~\ref{Fig. Cog_2_component_ex}.)}
    \label{Fig. COG Ni 2}
\end{figure}

\begin{figure}[h]
    \centering    
    \includegraphics[clip,  trim = 55 55 0 50, scale = 0.4]{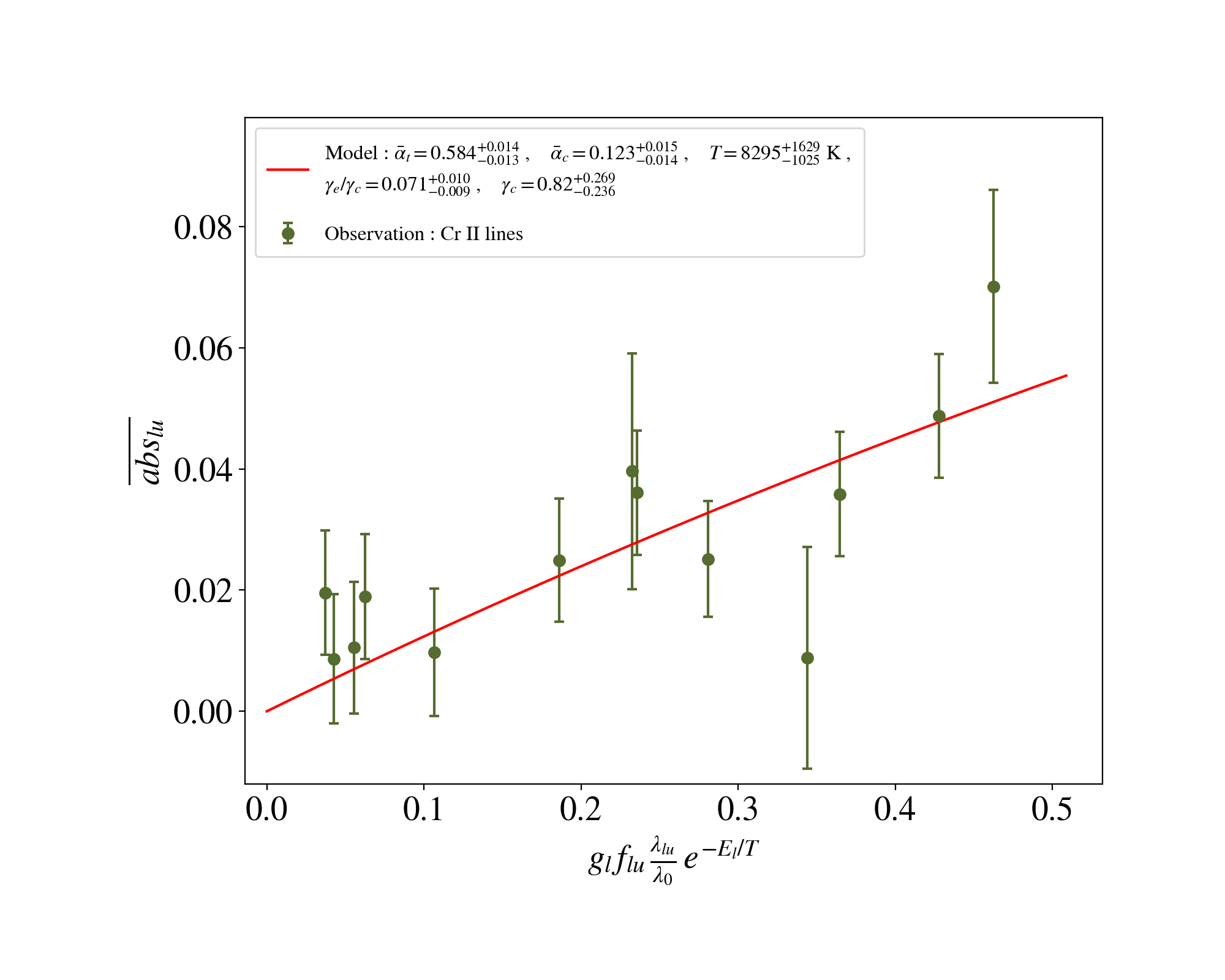}
    \includegraphics[clip,  trim = 55 55 0 50, scale = 0.4]{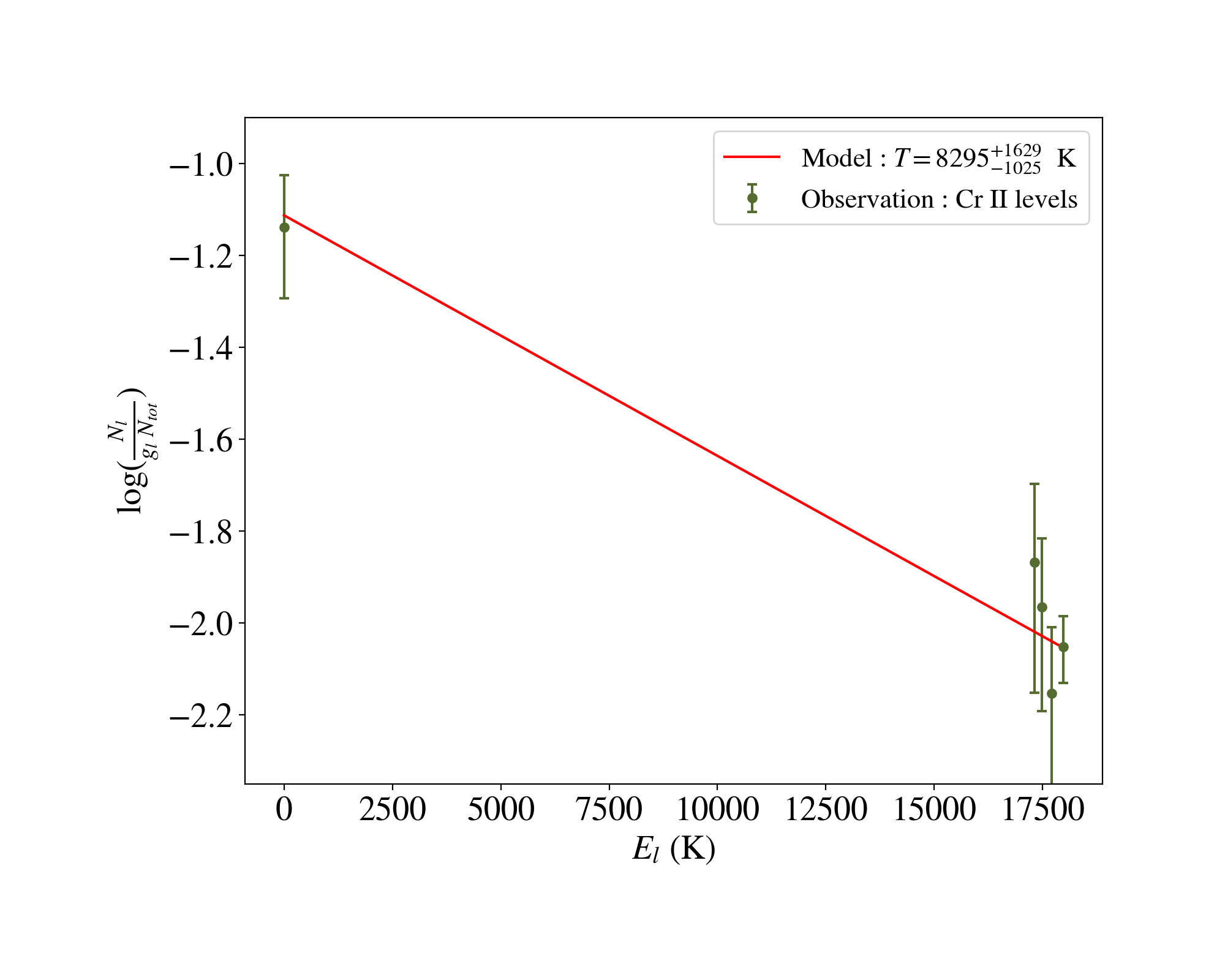}

    \caption[]{\textbf{Top}: Curve of growth of \crii\ in the December 6, 1997 comet. Green dots indicate the measured absorptions for the 17 \crii\ lines in the [+21,\,+42]\,$\si{km/s}$ range, while the red line shows the best fit with the two-component model. 
    %As for Fig. \ref{Fig. COG Ni 2}, gaussian priors were assumed for $\overline{\alpha}_{\rm t}$, $\overline{\alpha}_{\rm c}$ and $\frac{\gamma_{\rm e}}{\gamma_{\rm c}}$. 
    \\ 
    \textbf{Bottom}: Excitation diagram of \crplus .}
    \label{Fig. COG Cr 2}
\end{figure}

The result of the two fits are presented Fig. \ref{Fig. COG Ni 2} and \ref{Fig. COG Cr 2}, respectively. The curve of growth of \niplus\ in the December 6, 1997 comet appears to be consistent with the geometrical parameters - $\overline{\alpha}_{\rm t}$, $\overline{\alpha}_{\rm c}$ and $\gamma_{\rm e}/\gamma_{\rm c}$ - found for \feplus. This validates our hypothesis that the many chemical species detected in this comet are well-mixed, and have similar spatial distributions above the stellar disk. For \crii, the measured absorption depths are very low (a few \%), and very likely optically thin. As a result, they do not allow to compare the shape of the December 6, 1997 comet's curve of growth in \crii\ and \feii\ lines. 
    
In addition, the excitation temperature measured for \niplus\ ($9600_{-700}^{+800} \ \si{K}$) and \crplus\ ($8300_{-1000}^{+1600}\,\si{K}$) are found to be similar to that of \feplus\ (8530$\pm$300\,$\si{K}$). Again, this suggests that the three species are well-mixed in the transiting tail, and experience similar physical conditions. This shared dynamics among all detected ions is a key feature of \bp\ exocomets, as it could allow us to derive exocometary abundance ratios based on direct comparison between lines from different species. This point will be discussed in greater detail below, in Sect. \ref{Sect. Other metallic species}.

\section{Excitation model}
\label{Sect. Excitation model}

In Sect. \ref{Sect. A refined curve of growth model} and \ref{Sect. Study of other species}, we showed that the excitation temperatures of \feii, \niii\ and \crii\ in the gaseous tail of the December 6, 1997 comet are all similar, close to $\approx 9000\,\si{K}$. To interpret this result, \cite{Vrignaud24} proposed that the transiting gas follows a collisional regime, associated with a high enough electronic density ($n_e\,\sim\,10^7$\,$\si{cm^{-3}}$). Here we explore alternative scenarios, investigating the role played by radiative processes in the excitation of \feplus.

\subsection{Complete excitation diagram of \feplus}
\label{Sect. Complete excitation diagram of feplus}

Until now, we focussed on lines rising from rather weakly excited energy levels of \feii, \niii\ and \crii\ ($E_l \leq 20\,000 \ \si{K}$). However, as already noted in \cite{Vrignaud24}, the December 6, 1997 comet is also detected in large number of \feii\ lines rising from much more energetic levels (between $23\,000$ and $47\,000$\,$\si{K}$). The comet's absorption depth in these lines is often very low (typically between 0 and 10 \%, much less than in the three main \feii\ series at 2400, 2600 and 2750\,\A), but, when combined, they can provide valuable information on the population of highly excited levels of \feii. 

\begin{figure}[h]
    \centering    
    \includegraphics[clip,  trim = 55 55 0 50, scale = 0.4]{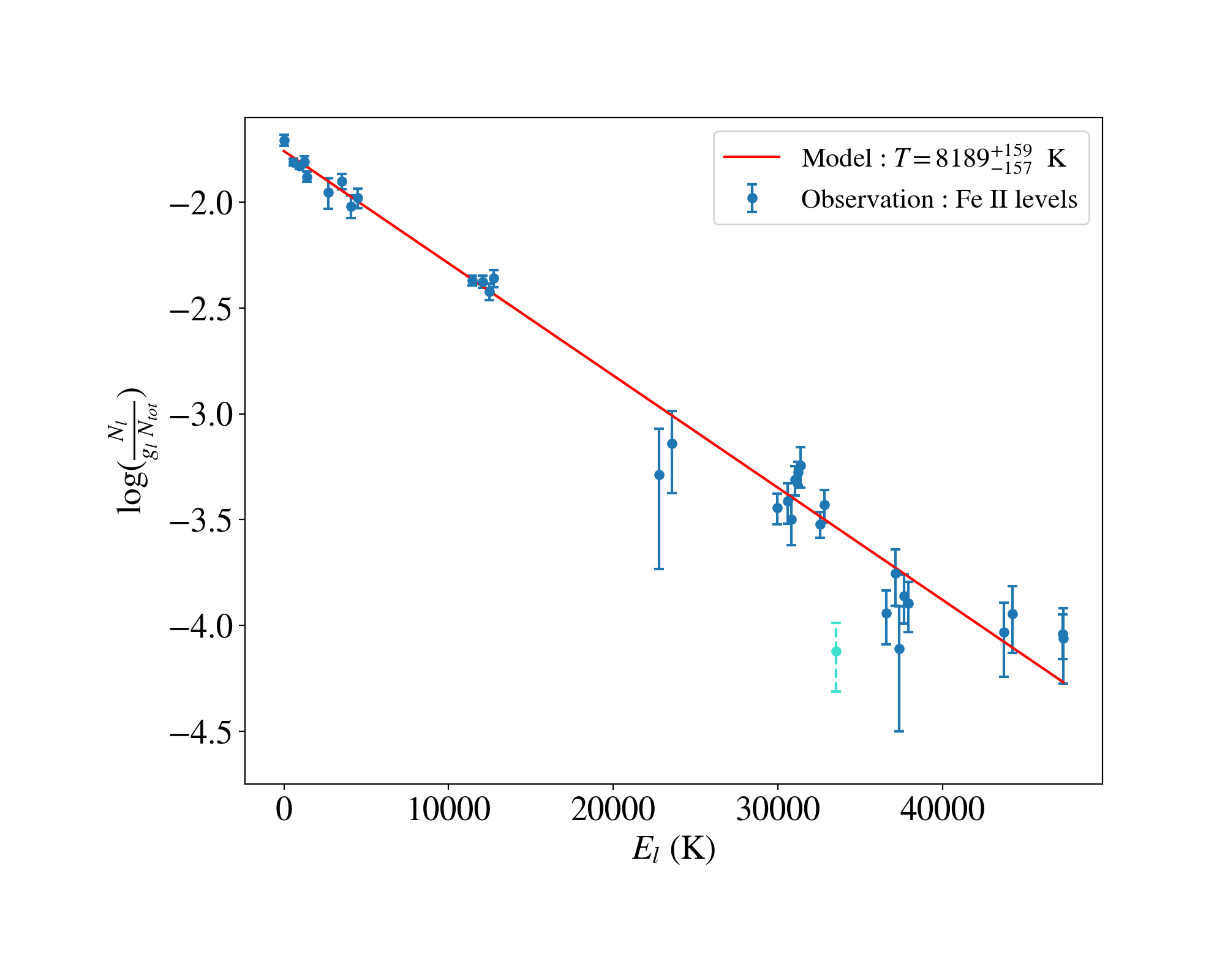}

    \caption[]{Excitation diagram of \feplus\ in the December 6, 1997 comet, built from the analysis of 96 \feii\ lines rising from a wide range of energy levels ($E_l \sim 0 - 33\,000 \ \si{cm^{-1}} \sim 0 - 47\,000 \ \si{K}$). The turquoise point indicate the peculiar a$^6$S$_{5/2}$ level, which is severely depleted compared to the fitted model.}
    \label{Fig. Excitation Fe 2 complete}
\end{figure}

Using these lines and those studied in Sect. \ref{Sect. A refined curve of growth model}, we performed a joint fit of the December 6, 1997 comet's absorption depths in 96 \feii\ lines, with wavelength between 1700 and 2800 \A\ and lower level energies ranging from 0 to 47\,000 K (see Table \ref{Tab. list lines Fe II}). This set of lines includes 63~lines rising from low-lying \feii\ states ($E_l \leq 12\,000 \ \si{K}$), already studied in Sect.~\ref{Sect. Fit 2 components}, and 33~additional lines rising from much higher excitation levels ($E_l = 23\,000 - 47\,000 \ \si{K}$). To select these last lines, we focussed on strong transitions ($g_l \cdot f_{lu} \geq 1.5$) rising from high multiplicity \feii\ levels ($g_l \gtrsim 10$). The excitation diagram derived from this analysis is shown on Fig.~\ref{Fig. Excitation Fe 2 complete}. The global excitation temperature is constrained to be 8190$\pm$160\,$\si{K}$, slightly lower than the value obtained by fitting only low-level lines (8530$\pm$300\,$\si{K}$, Sect.~\ref{Sect. Fit 2 components}). This difference is due to the slight over-abundance of the a$^4$D term ($E_l\sim12\,000 \ \si{K}$, see Fig. \ref{Fig. Excitation Fe 2 complete}) compared to the fitted Boltzmann distribution, which increases the apparent excitation temperature of \feplus\ when considering only low-lying levels.

Figure 10 shows that the population of \feii\ excited levels in the studied comet is overall close to a Boltzmann distribution. We note however one clear outlier: the a$^6$S$_{5/2}$ level ($E_l = 33\,550 \ \si{K}$) appears to be severely depleted, by a factor of $\approx 4$, compared to levels with similar energies. This outlier is shown with a dotted, turquoise error bar on Fig.~\ref{Fig. Excitation Fe 2 complete}; it was not included in the fit of the excitation temperature. The peculiarity of this \feii\ level is discussed below.

Using Eq.~\ref{Eq. Excitation diagram} and the model parameters obtained from the curve of growth fit, we compute a new total \feplus\ column density in the [+21,+42]\,km/s RV range of 4.8$\pm$0.3$\cdot 10^{14}$\,$\si{cm^{-2}}$, slightly less than in Sect.~\ref{Sect. Excitation diagram}. This value is probably a better estimate of the total \feii\ column density, since it is based on the study of a larger number of energy levels with a wider energy range.

\subsection{Model}
\label{Sect. Excitation model description}

To interpret the observed excitation diagram in the December 6, 1997 comet (Fig. \ref{Fig. Excitation Fe 2 complete}), we refined the excitation model introduced in \cite{Vrignaud24} by including stellar radiation. We calculated the statistical equilibrium of \feplus\ under various physical conditions, using a model of radiative and collisional excitation characterised by three parameters: $d$, the comet's distance to the star, $T$, its electronic temperature, and $n_e$, its electronic density.

\begin{figure}[h]
    \centering    
    \includegraphics[clip,  trim = 0 0 0 30, scale = 0.37]{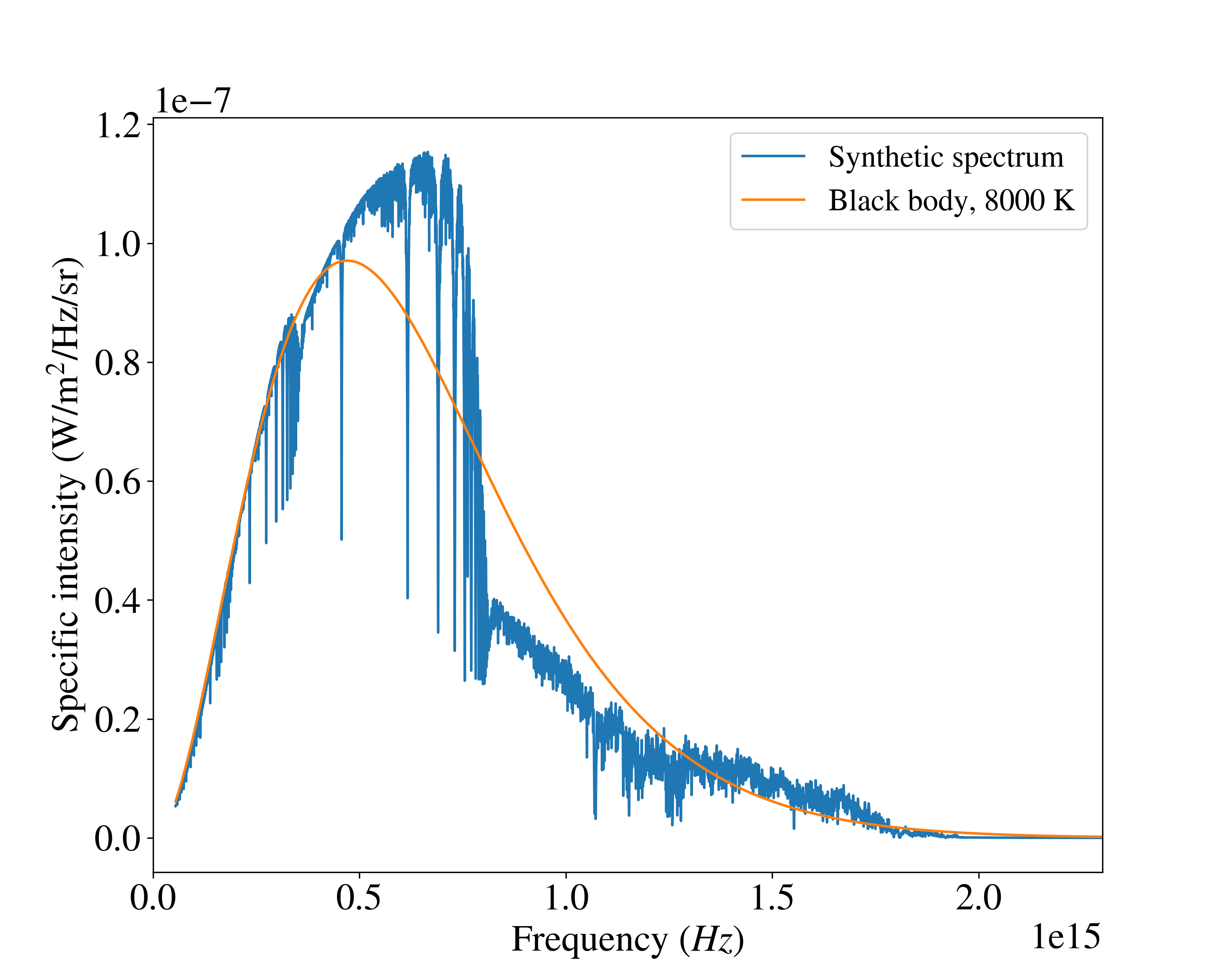}

    \caption[]{Synthetic \bp\ spectrum used in our model, along with the corresponding black body ($T_{\rm eff} = 8000 \ \si{K}$).}
    \label{Fig. spectrum model}
\end{figure}

The detailed equations leading to the energy level distribution of \feplus\ are provided in App. \ref{Anx. Statistical equilibrium}. In summary, the problem reduces to solving the linear system: 
$$
K \vec{X} = \vec{0},
$$
with $\vec{X}$ the vector containing the population ratios of all energy levels, and K a matrix depending on the model parameters ($d$, $T$, $n_e$), on the electron collision strengths and Einstein coefficients of \feii, and on the input stellar spectrum. The \feii\ data was taken from \cite{Tayal2018}, which provides extensive calculations for transition probabilities and collision strengths of \feii\ on a wide range of excitation levels ($0 - 140\,000 \ \si{cm^{-1}}$) and electronic temperatures ($10^2 - 10^5 \ \si{K}$). The stellar spectrum was obtained from the PHOENIX library \citep{PHOENIX}, using stellar parameters close to that of \bp\ ($T_{\rm eff} = 8000 \ \si{K}$, $\log(g) = 4.5$, [Fe/H]~=~0, [$\alpha$/M] = 0). The entire spectrum is shown on Fig. \ref{Fig. spectrum model}; it was renormalised so that the wavelength integrated flux matches that of a black-body at $T_{\rm eff} = 8000 \ \si{K}$. The chromospheric emission of \bp\ \citep{bouret2002} was not taken into account, as it is only detected in a few lines below 1300 \A\ (O\,{\sc vi}, C\,{\sc iii}), very far from the main \feii\ lines (2200-2800 \A). 

\subsection{Modelled energy distribution}
\label{Sect.Modelled energy distribution}

Depending on the chosen model parameters ($n_e$, $T$, $d$), the energy distribution of \feii\ can take very diverse shapes.

\subsubsection{Radiative regime}
\label{Sect. Radiative regime}

At low density ($n_e \leq 10^7 \ \si{cm^{-3}}$) and close distances ($d \leq 60 \ R_\star$), the \feplus\ gas follows a radiative regime (Fig. \ref{Fig. Result excitation Radiative}). Under this regime, the excitation temperature of low-energy ($E_l \leq 55\,000 \ \si{K}$) levels is set to the stellar effective temperature, $T_{\rm ex} = T_{\rm eff} = 8000 \ \si{K}$, while more excited levels ($E_l \geq 55\,000 \ \si{K}$) are severely depleted.

The presence of these two branches is linked to the radiative properties of \feplus: transitions between any pair of levels within each of the two branches are forbidden ($A_{ji} \sim 1 \ \si{s^{-1}}$), while transitions from one branch to another are often optically allowed ($A_{ji} \sim 10^8 \ \si{s^{-1}}$). At low density ($n_e \leq 10^6 \ \si{cm^{-3}}$), the role played by electronic collisions is negligible; as a result, spontaneous de-excitation lead the upper branch to be severely depleted (Fig. \ref{Fig. Result excitation Radiative}). 

The very close match between the \feplus\ excitation temperature and the stellar effective temperature is due to the fact that all \feplus\ allowed transitions are in the ultraviolet, where the stellar specific intensity is rather close to a black body, and where Wien's approximation ($h\nu \gg k_BT$) applies: 

$$
I_\nu \, \sim \, B_\nu(T) \, \sim \, \frac{2h\nu^3}{c^2} e^{-hv/k_BT},
$$
with $B_\nu(T) = \frac{2h\nu^3}{c^2} \frac{1}{e^{h\nu/k_BT} - 1}$ (Planck function). This approximation introduces Boltzmann factors in the statistical equilibrium equation (Sect. \ref{Sect. Statistical equilibrium}), which end up in the calculated energy distribution. More detailed calculations on this phenomenon can be found in \cite{Manfroid2021} (see the \emph{Three level atom} section).

\subsubsection{Semi-collisional regime}
\label{Sect. Semi-collisional regime}

For close distances  ($d \leq 60 \ R_\star$) but higher electronic densities ($n_e \geq 10^7 \ \si{cm^{-3}}$), the gas enters in a semi-collisional regime: collisions start to play a significant role in the \feplus\ energy distribution, particularly for the least excited levels. This regime is illustrated on Figs. \ref{Fig. Result excitation Semi-col 1}, \ref{Fig. Result excitation Semi-col 2} and \ref{Fig. Result excitation Semi-col 3} for an electronic density of $10^8 \ \si{cm^{-3}}$ and various kinetic temperatures. 

In the semi-collisional regime, the energy distribution of the least excited branch ($E_l \leq 55\,000 \ \si{K}$) is found to be a mix between a radiative regime (which would lead to $T_{\rm ex} = T_{\rm eff} = 8000 \ \si{K}$) and a fully collisional regime (leading to $T_{\rm ex} = T_e$). However, a peculiar case arises when the electronic temperature happens to match the stellar effective temperature ($8000 \ \si{K}$): in this case, both collisions and radiation tend to populate \feplus\ at the same excitation temperature, making their respective effects hard to disentangle (see Fig. \ref{Fig. Result excitation Semi-col 3}. 

Increasing further the electronic density, a fully collisional regime (imposing $T_{\rm ex} = T_e$) is eventually reached. Depending on the comet's distance to the star and electronic temperature, we found that such regime is expected for electronic densities above $10^{12} - 10^{16} \ \si{cm^{-3}}$.

\subsubsection{Large distance regime}
\label{Sect. Large distance regime}

Finally, for large comet-to-star distance ($d \geq 60 \ R_\star$), the gas gradually enters a composite regime, with a more complex energy distribution. Here, forbidden transitions start to play a significant role, and lead to a depletion of all energy levels except the ground state (see Fig. \ref{Fig. Result excitation large distance}, for $d = 280\, R_\star$). If the electronic density is sufficient ($n_e \geq 10^4 \ \si{cm^{-3}}$), collisional excitation also play a significant role in the energy distribution.

\subsection{Comparison with data}
\label{Sect. Comparison with data}

The lower panels of Fig.~\ref{Fig. Result excitation Radiative} to \ref{Fig. Result excitation large distance} provide a comparison between the observed \feplus\ excitation diagram in the December 6, 1997 comet, and the outcome of our model. This first comparison shows that our observations are compatible with a radiative regime, associated with a rather low electronic density (Fig. \ref{Fig. Result excitation Radiative}). On the other hand, the calculated energy distributions at high electronic densities and electronic temperatures different from 8000~K (Fig. \ref{Fig. Result excitation Semi-col 1} and \ref{Fig. Result excitation Semi-col 2}) appear to be very different from the observed excitation diagram.

We also note that the severe depletion of the a$^6$S$_{5/2}$ level (shown with a turquoise marker in Appendix E) observed in the comet is well reproduced by our model in the case of a radiative regime (Fig. \ref{Fig. Result excitation Radiative}). The explanation for the peculiar behaviour of this level is the following : while the strongest lines of most \feii\ states lie in the 2300 - 2800 \A\ region (see Table \ref{Tab. list lines Fe II}), the lines of the a$^6$S$_{5/2}$ level are located at 1785 \A\ $\sim 1.7 \cdot 10^{15} \ \si{Hz}$, where the flux of \bp\ is particularly strong, well above a black body (see Fig. \ref{Fig. spectrum model}). This results in a more efficient radiative pumping for the a$^6$S$_{5/2}$ level than other \feii\ levels, leading the first to be depleted. Here, the fact that such behaviour is found in our data hints that radiative processes play a prominent role in the excitation of \feplus\ in the studied comet.  

To deeper explore the agreement between our measurements for the December 6, 1997 comet, and the results of our excitation model, we calculated the $\rchi^2$ values of the model on a wide parameter grid, with $T_e$ ranging from $10^2$ to $10^5 \ \si{K}$, $n_e$ from  $10^4$ to $10^{10} \ \si{cm^{-3}}$, and $d$ from 10 to 140 R$_\star$ \citep[$0.07 - 1$ au, assuming $R_\star = 1.53\,R_\odot$,][]{Wang_2016}. The minimal $\rchi^2$ value was found to be $\rchi^2_{\rm min} = 41$, associated with a radiative regime or a semi-collisional regime at $T_e \sim T_{\rm eff}$. With 33 energy levels included in the study, this minimal value corresponds to a reduced $\rchi^2$ of $\rchi^2_r=1.24$. The set of parameters compatible with our observations (within 95 \%) was thus estimated using the following criterion:  

$$\tilde \rchi^2(d, n_e, T) \leq \tilde \rchi^2_{\rm min} + 4,$$

with $\tilde \rchi^2 = \rchi^2/\rchi^2_r$ \ and \ $\tilde \rchi^2_{\rm min} = \rchi^2_{\rm min}/\rchi^2_r$. The resulting maps are shown on Fig. \ref{Fig. Chi2 maps}, along with the $\tilde \rchi^2_{\rm min} + 4$ contour. We can identify three different regimes which are, a priori, compatible with our observations: 

\begin{itemize}
    \item A radiative regime, associated with a close comet-to-star distance ($d \leq 60$ R$_\star$), a low electronic density ($n_e \leq 10^7 \ \si{cm^{-3}}$) and any temperature. This regime is indicated with letter $a$ in Fig.  \ref{Fig. Chi2 maps}. In this case, the \feii\ excitation temperature corresponds to the stellar effective temperature, as observed in the comet.
    \item A semi-collisional densities (letter $b$, Fig.  \ref{Fig. Chi2 maps}), associated with an electronic temperature very close to the stellar effective temperature ($T_e = 8000 \pm 500 \ \si{K}$) and higher electronic densities (up to $10^9 \ \si{cm^{-3}}$). In this peculiar case, both radiative and collisional processes tend to set the \feii\ excitation temperature to 8000 K; the computed excitation diagram thus remains close to a purely radiative regime (Fig. \ref{Fig. Result excitation Semi-col 3}).
    \item A large distance regime (letter $c$), associated with electronic densities around $10^6 \ \si{cm^{-3}}$ and temperatures above 10\,000 K. This case is also peculiar: here, the depletion of all excited \feii\ levels due to spontaneous emission in forbidden lines happens to be exactly compensated by collisions with energetic collisions, leading the computed model to match again the observed excitation diagram. 
\end{itemize}

On the other hand, our observations allow us to fully reject LTE (corresponding to the high density limit), for all values of $d$ and $T_e$. This is due to the observed depletion of the a$^6$S$_{5/2}$ level, which is incompatible with a Boltzmann energy distribution. 

\subsection{A radiative regime in the December 6, 1997 comet}
\label{A radiative regime in the December 6, 1997 comet}

\begin{figure}[h]
    \centering    
    \includegraphics[clip,  trim = 0 0 0 50, scale = 0.44]{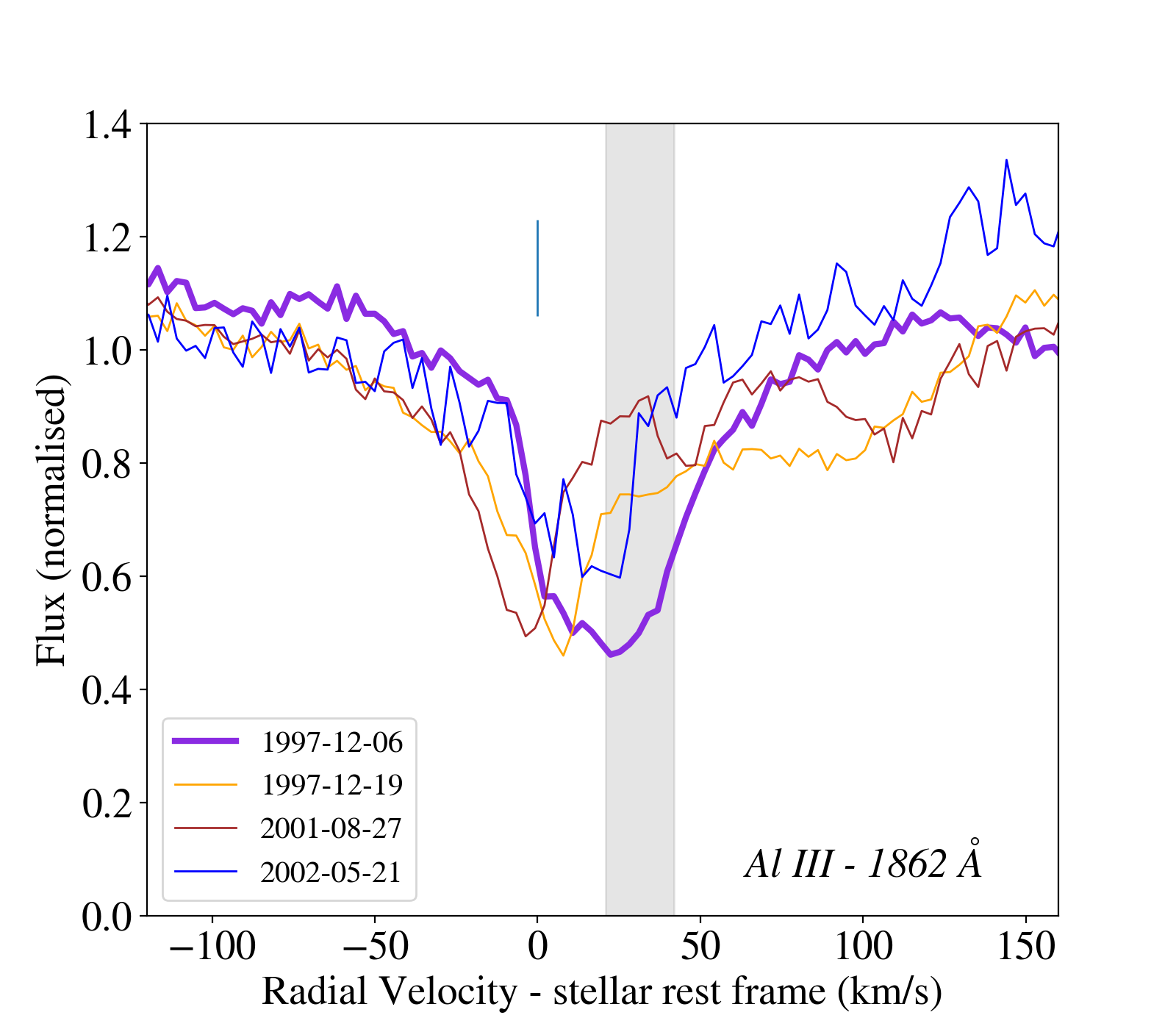}
    \caption[]{Available \bp\ spectra around the 1862 \A\ \aliii\ line, with the December 6, 1997 spectrum shown in purple. Here, the strong comet absorption at all epochs makes the retrieval of an exocomet-free spectrum difficult; however, the December 6, 1997 comet (shaded area) seems detected in this line as well (see Figs. \ref{Fig. 2740 A line 1997}, \ref{Fig. 2626 A line 1997}  and \ref{Fig. comparaison espèces I} for comparison with other species).  }
    \label{Fig. Al III}
\end{figure}

Even though the three regimes mentioned above are all compatible with our data, the last two seem rather unlikely, as they correspond to fortuitous coincidences. Indeed, the semi-collisional regime is possible only if the electronic temperature falls in the $[7500, 8500] \ \si{K}$ range, within most of the comet. This would be rather surprising: it is hard to see why the electronic temperature would be so uniform in such a large comet and coincidental with the stellar effective temperature. In addition, the December 6, 1997 comet seems to be detected in the \aliii\ doublet (1850 \A, see Fig. \ref{Fig. Al III}), which requires very high temperatures to be formed \citep[$\geq 40\,000 \ \si{K}$, see ][]{Beust_1993}. Here, no impact of such high electronic temperature is seen on the excitation diagram of \feii, pointing towards a scenario where collisions play no significant role in the energy distribution in the comet, and that, therefore, the electronic density is rather low ($\leq 10^7 \ \si{cm^{-3}}$). In a similar way, the acceptable parameter space for $d \geq 60$ R$_\star$ is very limited (Fig. \ref{Fig. Chi2 maps}), corresponding to $T_e \geq 10^4 \ \si{K}$ and $n_e \simeq 10^6 \ \si{cm^{-3}}$ (when collisions with hot electrons happen to compensate exactly the depletion of excited levels through forbidden transitions). As for the semi-collisional regime, this coincidence seems unlikely. 

As a result, our analysis points towards a radiative regime in the gaseous tail of the December 6, 1997 comet, associated with a low electronic density ($n_e \leq 10^7 \ \si{cm^{-3}}$) and a close distance to \bp\ ($d \leq 60$ R$_\star$). This conclusion is different from that of \cite{Vrignaud24} where the radiative excitation was not taken into account. Ultimately, it appears that the transiting gas is not at thermodynamical equilibrium, but rather that the excitation of \feplus\ is controlled by the intense stellar radiation received at such close distance to \bp. 

The upper value on the electronic density in the December 6, 1997 comet can also be compared to rough geometrical estimates. Indeed, assuming that the cometary tail has a solar composition (except for Hydrogen, which we can take twice as abundant as Oxygen) and that all elements are singly ionised, we find that the elemental abundance of \feplus\ should be around 1.5 \%. Using this value and the \feii\ column measured in the comet ($\sim 4.8 \cdot 10^{14} \ \si{cm^{-2}}$), we get an electronic column density of $\sim 3 \cdot 10^{16} \ \si{cm^{-2}}$. Dividing this value by the typical size of the comet - about one stellar radius -, we obtain an electronic volume density of $\approx 4 \cdot 10^5 \ \si{cm^{-3}}$, fully compatible with our upper limit for a radiative regime.

\section{Discussion}
\label{Sect. Discussion}

\subsection{Dependency of $T_{\rm ex}$ with radial velocity}

\begin{figure}[h]
    \centering    
    \includegraphics[clip,  trim = 0 0 0 60, scale = 0.36]{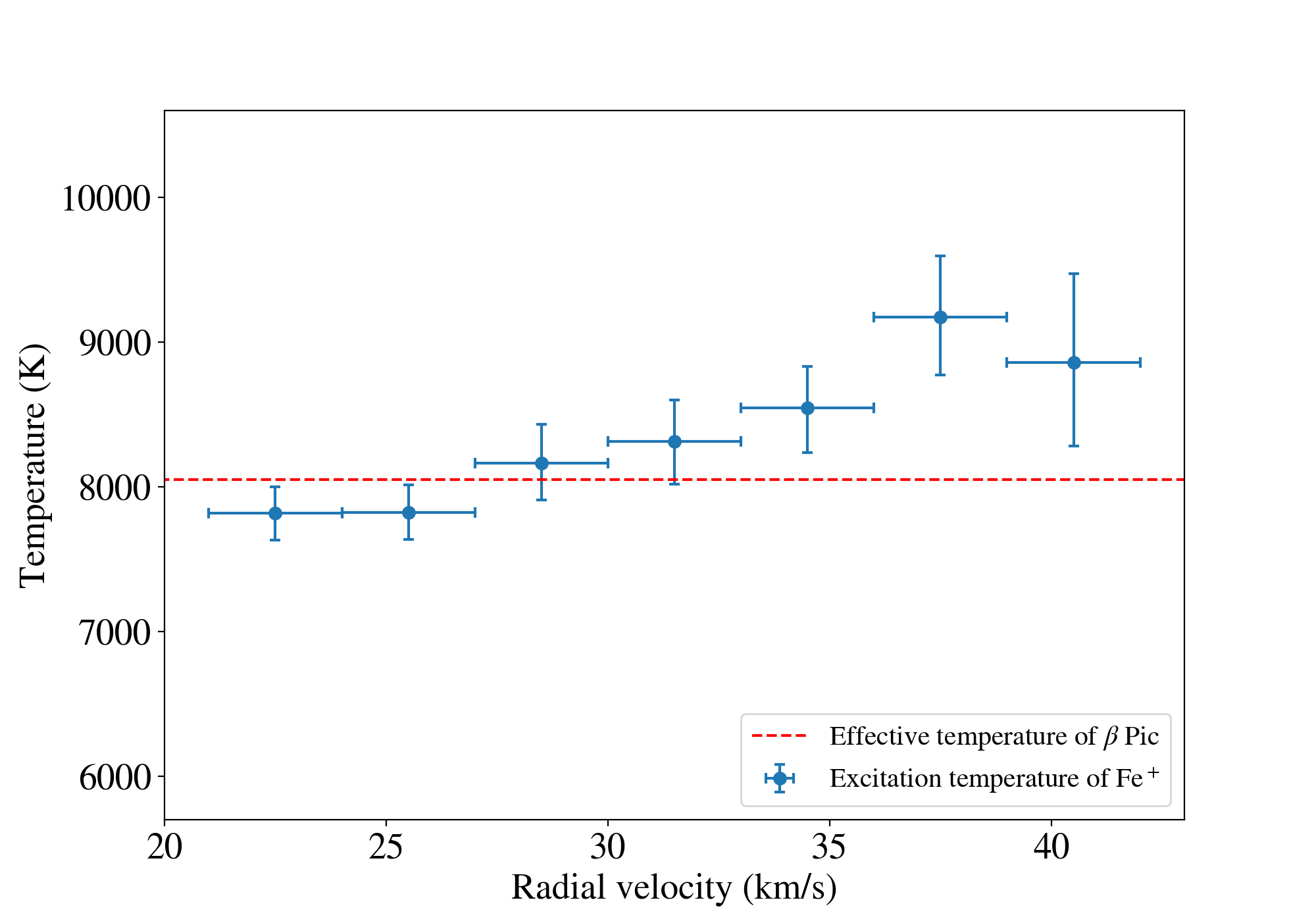}
    \caption[]{Excitation temperature of \feii\ in the December 6, 1997 comet, as a function of the radial velocity. The temperatures were estimated by building excitation diagrams similar to Fig. \ref{Fig. Excitation Fe 2 complete}, but based on absorption measurement in short RV bins (3 km/s width). The horizontal line emphasises the stellar effective temperature, $T_{\rm eff} = 8052 \ \si{K}$ \citep{Gray2006}.}
    \label{Fig. Profil_T_RV}
\end{figure}

\begin{figure}[h]
    \centering    
    \includegraphics[clip,  trim = 30 50 0 60, scale = 0.38]{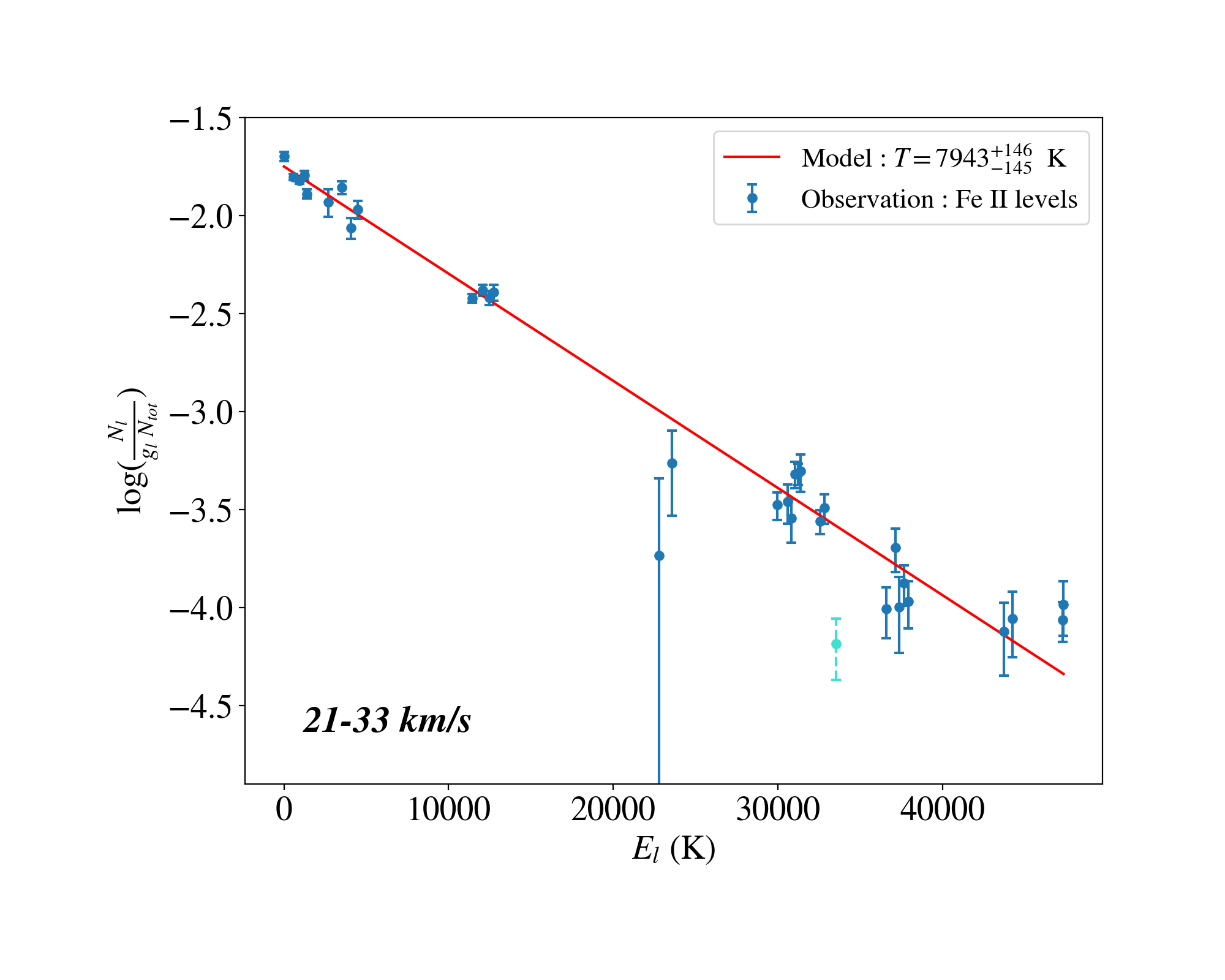}
    \includegraphics[clip,  trim = 30 50 0 60, scale = 0.38]{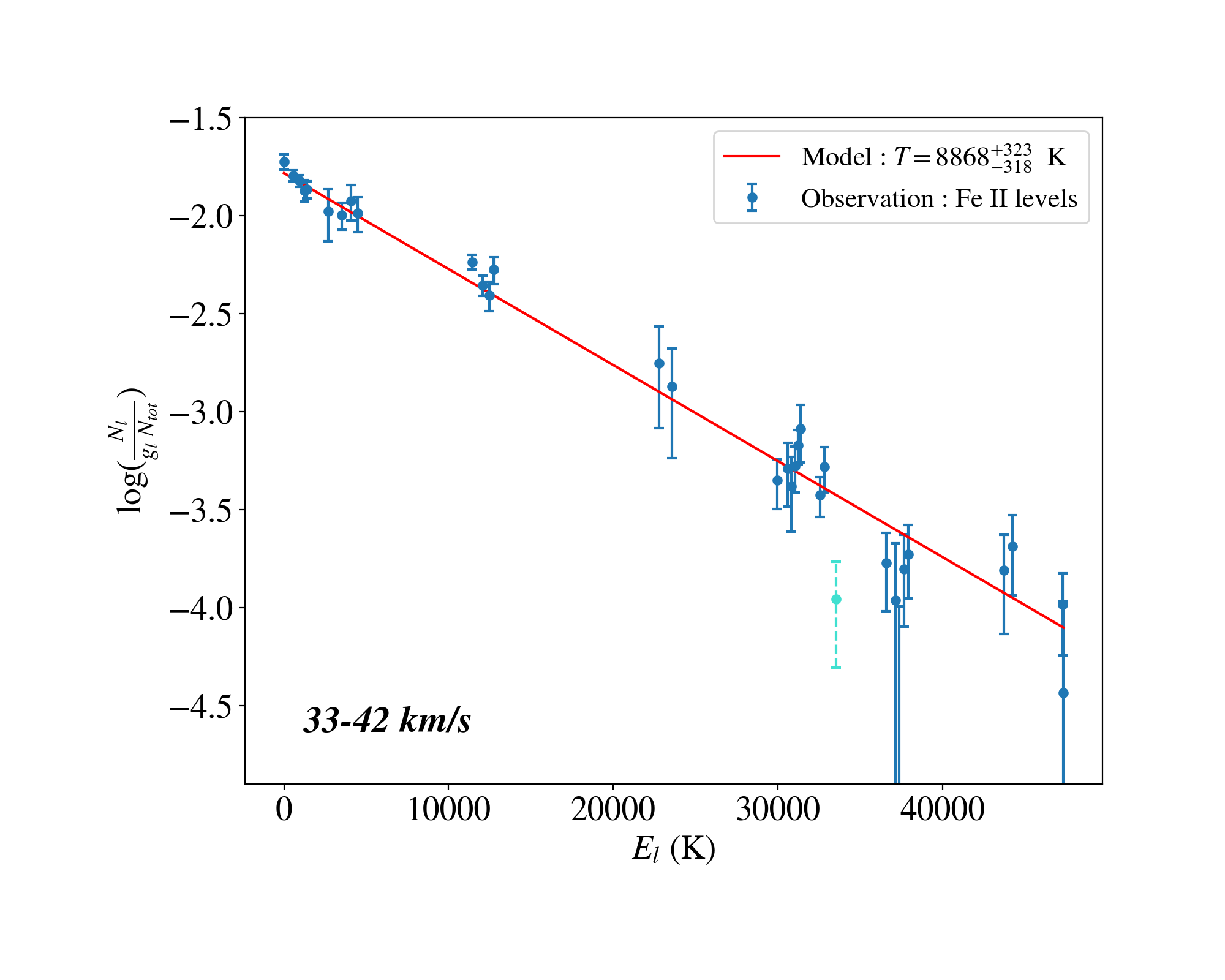}
    \caption[]{Excitation diagram of \feii\ in the December 6, 1997 comet in two different radial velocity range : 21 - 33 km/s (top) and 33 - 42 km/s (bottom). The excitation temperature seems to be slightly higher in the most redshifted part of the comet, due to the increased abundance of some excited levels (e.g. the a$^4$D term at $\sim 12\,000$ K). As for Fig. \ref{Fig. Excitation Fe 2 complete}, the a$^6$S$_{5/2}$ state was not included in the fit, as it is severely depleted compared to the other \feii\ excitation levels.}
    \label{Fig. excitation RV ranges}
\end{figure}

Until now, we assumed that the gas excitation state in the December 6, 1997 comet is homogeneous, and in particular that it does not depend on radial velocity. To check this hypothesis, we estimated the excitation temperature of \feplus\ in the comet at different radial velocities, using the two-component curve of growth model and the same set of 96 \feii\ lines as in Sect. \ref{Sect. Complete excitation diagram of feplus}. The measured excitation temperatures are provided on Fig. \ref{Fig. Profil_T_RV}: we note that $T_{\rm ex}$ remains fairly constant throughout the comet absorption profile, close to the stellar effective temperature ($8052 \ \si{K}$, \cite{Gray2006}). This confirms that the gas is in a radiative regime: the energy distribution of \feplus\ is decoupled from the local properties of the gas (kinetic temperature, electronic density), and is only the result of radiative processes stimulated by the stellar luminosity. We note however a slight trend in the data: the global excitation temperature of \feplus\ slightly increases with the radial velocity, varying from 7800 $\si{K}$ near 25 km/s to 9000 $\si{K}$ at 40 km/s. To illustrate this trend, we provide on Fig. \ref{Fig. excitation RV ranges} the excitation diagrams of \feii\ within the 21-33 km/s RV range (top panel, yielding $T_{\rm ex} = 7940 \pm 150 \ \si{K}$) and the 33-42 km/s range (bottom panel, yielding $T_{\rm ex} \simeq 8870 \pm 330 \ \si{K}$). The difference between the two diagrams is subtle, but it seems that some levels get slightly more populated in the most redshifted part of the comet (particularly the a$^4$D term at $12\,000 \ \si{K}$), leading to an increase of the global excitation temperature. The origin of this deviation is unclear; we found no set of parameters $(d, n_e, T_e)$ able to better reproduce the observation in the 33-42 km/s range than a simple radiative regime.

Here, it should be noted that our excitation model is simple, and needs improvement. In particular, self-opacity and several gaseous components with different electronic densities and temperature could be considered. Nonetheless, it remains remarkable that a model with a uniform excitation state throughout a whole comet allows us to reproduce the observed population of \feii\ levels on such a large energy range.

\subsection{Towards abundance ratios}
\label{Sect. Other metallic species}

In Sect. \ref{Sect. Study of other species}, we showed that the December 6, 1997 comet harbours many species (\feplus\ \niplus, \crplus, \mnplus, \siplus) with similar behaviours, probably due to efficient momentum exchange between particles via Coulomb scattering. In particular, we found that the spatial distribution of the detected species are similar: their absorptions are all seen at the same radial velocity, and show similar shapes. We also showed that \feplus, \niplus\ and \crplus\ have similar energy distributions in their low-lying levels ($E_l \leq 20\,000 \ \si{cm^{-1}}$), with excitation temperatures around $9000 \ \si{K}$. This is very likely the result of the three species following a radiative regime, where ion excitation is dominated by radiative processes.

These results open the way for abundance estimates in exocomets. Indeed, the ion mixing in \bp\ exocomets allows the use of global cometary models to fit the lines of several species simultaneously, adopting, for instance, a unique covering factor $\overline{\alpha}$ for all species. Assuming that all exocometary species follow a radiative regime (as found for \feplus, and most likely \niplus and \crplus), it also becomes possible to predict, for any given species, the relative abundances of its excitation levels. The total column density of any absorbing species can thus be inferred from the study of a small number of lines, rising from a limited set of excitation level (e.g. the 2600 \A\ ground-level triplet for \mnii). As a benchmark, using Eq. \ref{Eq. N_tot} and the parameters constrained in Sect. \ref{Sect. Study of other species}, we estimate the total, stellar disk-averaged \niplus\ and \crplus\ column densities in the December 6, 1997 comet  to be $4.1 \pm 0.3 \cdot 10^{13}\,\si{cm^{-2}}$ and $5.0 \pm 0.6 \cdot 10^{12} \, \si{cm^{-2}}$, respectively. Comparing these values with the one found for \feplus\ in the same RV range ($4.8 \pm 0.3 \cdot 10^{14} \ \si{cm^{-2}}$, Sect. \ref{Sect. Fit 2 components}), we can estimate the elemental \niplus/\feplus\ and \crplus/\feplus\ ratios in the December 6, 1997 comet. We find:

$$
\begin{array}{cc}
     &  [\text{Ni}^+/\text{Fe}^+] = 8.5 \pm 0.8 \cdot 10^{-2},  \\
     &  [\text{Cr}^+/\text{Fe}^+] = 1.04 \pm 0.15 \cdot 10^{-2}. 
\end{array}
$$

These ratios are rather close to the solar ratios of the corresponding elements ($5.2 \cdot 10^{-2}$ and $1.38 \cdot 10^{-2}$ respectively, see \cite{Asplund2009}), although \niplus\ seems over-abundant when compared to \feplus, while \crplus\ may be slightly depleted. In any case, further work is needed to confirm these preliminary results, to extend them to other species routinely detected in exocomets (e.g. \caii, \mnii, \siii, \aliii...), and to get a complete a complete overview of the composition of exocomets. 

\subsection{Comparison with similar studies}
\label{Sect. Comparison with similar studies}

Our study confirms the close transit distance of \bp\ exocomets relatively to their host star: for the December 6, 1997 comet, we derive an upper limit of 60 R$_\star$, equivalent to $\sim 0.43$ au. This upper limit is in agreement with the \caii\ study of \cite{Kiefer_2014}, in which the typical transit distance of \bp\ exocomets was estimated to be $10 \pm 3$ R$_\star$ for the shallow family - associated with wide, highly redshifted absorptions, and $19 \pm 4$ R$_\star$ for the D family - associated with deep, narrow absorption close to $0 \ \si{km/s}$. Note that it is unclear to which family the December 6, 1997 comet belongs, as it exhibits features from both the shallow and deep families: its absorption is rather wide (FWHM of $\sim 30 \ \si{km/s}$) and redshifted, but also fairly deep.

The radiative regime found in the December 6, 1997 comet can also be compared to the results of \cite{Manfroid2021}, where a similar analysis was conducted on many solar system comets, using \fei\ and \nii\ emission lines. For all studied comets, the excitation temperature of Fe and Ni was found to be around $4000 \ \si{K}$, close to the solar effective temperature ($5800 \ \si{K}$). Here, the slight difference between the cometary excitation temperatures and the solar effective temperature was attributed to strong absorption lines in the NUV solar spectrum. In the case of \bp, this effect seems to be less important, thanks to the rapid stellar spin \citep[130 km/s,][]{Royer2007}, which smooths the stellar spectrum and brings it rather close to a perfect black-body. This explains why the measured excitation temperature of \feii\ is so close to \bp\ effective temperature.

However, our results seem in contradiction to the work of \cite{Mouillet_1995}, which analysed the \caii\ absorptions of a \bp\ comet observed in 1992 and concluded to a semi-collisional regime, associated with high electronic temperature ($\geq 15\,000 \ \si{K}$) and density ($\geq 10^6 \ \si{cm^{-3}}$). From our perspective, this different conclusion is due to incorrect abundance estimates for the studied \caii\ levels: saturation in the H and K lines was not taken into account, leading the authors to severely underestimate the column density of the ground level. A rough calculation shows that taking saturation into account, the \caii\ cometary features analysed in \cite{Mouillet_1995} are actually compatible with a radiative regime.

\section{Conclusion}
\label{Sect. Conclusion}

Our study allowed us to deeply refine our knowledge of the December 6, 1997 \bp\ comet, which was already studied in \cite{Vrignaud24}. First, analysing the curve of model with a two-component model, we found that the comet's absorption signatures in \feii\ lines can be satisfactorily explained by the transit of multiple gaseous components, with different optical depth and covering factors. In fact, these different components could very well be associated with several evaporating nuclei, with similar orbits following each other on a transiting string. The excitation temperature of \feplus\ in the transiting comet is close to $8500 \ \si{K}$ for low-lying ($E_l \leq 12\,000 \ \si{cm^{-1}}$) levels, or $8200 \ \si{K}$ if we consider a wider energy range ($E_l \sim 0 - 47\,000 \ \si{K}$). Similar excitation temperatures were found for \niplus\ and \crplus\ $(T_{\rm ex} \sim 9000 \ \si{K}$). This result, along with the comparison between the comet's absorption profiles in different lines, led us to conclude that all exocometary species tend to stay well-mixed within the gaseous tails, keeping the same spatial distribution. This mixing is probably the result of the high ionisation of \bp\ exocomets, which couples the dynamics of the different species via Coulomb interaction.

Finally, the full excitation diagram of \feplus\ in the December 6, 1997 comet was compared with the predictions of an excitation model, taking into account both radiative and collisional processes. This comparison showed that \feplus\ very likely follows a radiative regime, associated with a low electronic density ($n_e\leq 10^7 \ \si{cm^{-3}}$) and a rather close distance to \bp\ ($d\leq60$~R$_\star\sim0.43$~au). In this regime, excitation and de-excitation are dominated by stellar photon absorption and emission, leading the excitation temperature of the absorbing gas to match the stellar effective temperature ($8000 \ \si{K}$). Given the excitation diagrams observed for \crplus\ and \niplus, it is very likely that these two species also follow a radiative regime. 

We also showed that the \niplus/\feplus\ and \crplus/\feplus\ ratio in the December 6, 1997 comet are close to solar abundances. Now that our curve of growth analysis and our model of the excitation level population allow the prediction of the energy distribution of any ion in the gaseous tails of exocomets, similar measurements could be obtained for many other species, such as \mnii, \caii\ and \siii. With such measurements, we have now for the first time the possibility to derive the typical composition of exocomets, that can be use to provide new insight on their past history and the physical processes at play in their vicinity. 

\section*{Data Availability}

Appendices B to G are available at: \newline https://doi.org/10.5281/zenodo.13828957.

\begin{acknowledgements}
T.V. \& A.L. acknowledge funding from the Centre National d’\'Etudes Spatiales (CNES), and thank the anonymous referee for his constructive remarks.
\end{acknowledgements}

\bibliographystyle{aa}
\bibliography{bibliography}

\newpage

\begin{appendix}

\section{Statistical equilibrium of \feplus}
\label{Anx. Statistical equilibrium}

This section details the equations that allowed us to calculate the energy distribution of \feplus\ (Sect. \ref{Sect. Excitation model description}), depending on the physical properties of the comet (electronic density $n_e$; electronic temperature $T$) and on its distance to \bp\ ($d$).

\subsection{Collisional excitation}
\label{Sect. Collisional excitation}

Denoting $(i,j)$ any pair of \feplus\ excitation levels, the time variation of the level $i$ population due to collisional excitation with level $j$ writes: 

$$
\diff{n_i}{t} \Big|_{\text{col, }j} = \ C_{ji}\,n_e\,n_j \ - \ C_{ij}\,n_e\,n_i,
$$

with $n_i$ and $n_j$ denoting the levels $i$ and $j$ volume densities, and with $C_{ij}$ and $C_{ji}$ the collisional excitation rates (in\,\,$\si{s^{-1} (cm^{-3})^{-1}}$) for the $i\rightarrow j$ and $j\rightarrow i$ transitions, respectively. These coefficients, which depend on the electronic temperature, are linked by the following equation (when level $i$ is chosen to be the lower level of the transition): 
$$
C_{ij} = \frac{g_j}{g_i} e^{-h \nu_{ij}/k_B T_e} C_{ji} \ \ \ \ \ \ \ \ (j>i),
$$
with $\nu_{ij}$ the frequency associated with the $i\leftrightarrow j$ transition. Collisional rates can also be related to the effective collision strength $\Upsilon_{ji}$, through:
$$
C_{ji} = \beta \frac{1}{g_j \sqrt{T_e}} \Upsilon_{ji} \ \ \ \ \ \ \ \  (j>i),
$$
where $\beta = \Big( \frac{2 \pi \hbar^4}{k_B m_e^3} \Big)^2$. The effective collision strength is a dimensionless parameters characterising how sensitive to collision the $i\leftrightarrow j$ transition is. For our study, we used the values provided by \cite{Tayal2018}.

\subsection{Radiative excitation}
\label{Sect. Radiative excitation}

Keeping the same notation as before, the time variation of the level $i$ volume density due to radiative interaction with level $j$ writes: 

$$
\diff{n_i}{t} \Big|_{rad, \ j} = \left\{
    \begin{array}{ll}    
        \ B_{ji}\,J_{\nu_{ij}}\,n_j - B_{ij}\,J_{\nu_{ji}}\,n_i + A_{ji}\,n_i \ \ \ \ \  \text{if j>i,}\\

        \vspace{0.3 cm} \\
        
        \ B_{ji}\,J_{\nu_{ij}}\,n_j - B_{ij}\,J_{\nu_{ji}}\,n_i - A_{ij}\,n_i \ \ \ \ \  \text{if i>j.}\\
    \end{array}
\right.
$$

with $A_{ji}, B_{ji}$ and $B_{ij}$ the Einstein coefficients for spontaneous emission, stimulated emission and photon absorption (in the case where $j>i$). They can all be expressed in terms of $A_{ji}$, through: 

$$
B_{ji} = \frac{c^2}{8 \pi h \nu_{ij}^3} A_{ji}, \ \ \ \ \ \ B_{ij} = \frac{g_j}{g_i} \frac{c^2}{8 \pi h \nu_{ij}^3} A_{ji}.
$$
As for the effective collision strengths, the $A_{ji}$ values of \feii\ were obtained from \cite{Tayal2018}, 

The stellar flux received by the comet at the transition frequency ($J_{\nu_{ij}}$, expressed in $\si{W / m^2 / Hz}$) can be linked to the stellar specific intensity $I_{\nu_{ij}}$ (in $\si{W / m^2 / Hz / sr}$) and to the comet-to-star distance $d$, by:  

$$
J_{\nu_{ij}} = 2 \pi \ \Big (1 - \sqrt{1 - (R_\star/d)^2 } \ \Big ) \ I_{\nu_{ij}}.
$$
The spectrum used in our model ($I_{\nu}$) is provided in Fig. \ref{Fig. spectrum model}; it was obtained from the PHOENIX stellar library \citep{PHOENIX}.

\subsection{Statistical equilibrium}
\label{Sect. Statistical equilibrium}

The statistical equilibrium condition writes $\diff{n_i}{t}$ = 0, for all $i$. Taking into account both collisional and radiative excitation, this condition yields, for any level $i$: 

$$
\displaystyle \sum_{j<i} C_{ji} n_e n_j - C_{ij} n_e n_i + B_{ji} J_{\nu_{ij}} n_j - B_{ij} J_{\nu_{ij}} n_i - A_{ij} n_i
$$
$$
+  \sum_{j>i} C_{ji} n_e n_j - C_{ij} n_e n_i + B_{ji} J_{\nu_{ij}} n_j - B_{ij} J_{\nu_{ij}} n_i + A_{ji} n_j
\ = \ 0.
$$

This last equation can be rewritten as a linear combination of $(n_1, n_2, \dots, n_N)$, with $N$ the total number of studied excitation levels, so that: 

$$
\displaystyle \sum_{j} k_{ij} \ n_j = 0 \ \ \ \ \ \text{for i = 1, 2, ... N,}
$$
with : 
$$
k_{ij} = \left\{
    \begin{array}{ll}
    
        C_{ji} n_e + B_{ji} J_{\nu_{ij}} \hspace{3.39 cm} \text{if j < i,}\\

        \vspace{0.3 cm} \\
        
        - \displaystyle \sum_{k<i} \Big( C_{ik} n_e + B_{ik} J_{\nu_{ik}} + A_{ik} \Big) - \displaystyle \sum_{k>i} \Big( C_{ik} n_e + B_{ik} J_{\nu_{ik}} \Big) \\
        
        \hspace{5.355 cm} \text{if j = i,}

        \vspace{0.6 cm}\\
        
        C_{ji} n_e + B_{ji} J_{\nu_{ij}} + A_{ji} \hspace{2.595 cm} \text{if j > i}.\\
    \end{array}
\right.
$$

\vspace{0.2 cm}

Solving the statistical equilibrium is thus equivalent to solving the linear system $K \vec{X} = \vec{0}$, with $K = (a_{ij})_{i,j}$ and $\vec{X} = (n_1, n_2, \dots, n_N)$. Note here that matrix K is singular; only the relative densities of the different excitation levels can be constrained. The coefficients of $K$ depend on the A-values and effective collision strengths of \feii, on the chosen stellar spectrum, and on the model parameters, $d$, $T_e$ and $n_e$.

\newpage
\onecolumn
\section{Observations}

\begin{table}[h!]
	\centering  
	\small \renewcommand{\arraystretch}{1.4} \begin{tabular}{c c  c  c  c  c  c  c}          
		   
     \cline{1-8}
    \noalign{\smallskip}
     \cline{1-8}                 
     
     Date & Program ID & Spectrum  & Wavelength & Start time & Exposure time & Grating & RV ranges showing \\
          &            &           &    (\A)    &    (UT)    &     (s)       &         & comet absorption (km/s)    \\

     \cline{1-8}      

    \multirow{6}{*}{1997-12-06} & \multirow{6}{*}{7512} & o4g001010 & 1463 - 1661 & 07:18' & 900 & E140H & \multirow{6}{*}{[-4;+200]} \\ 
    & & o4g001030 & 1628 - 1902 & 07:48' & 679 & E230H & \\ 
    & & o4g001020 & 1879 - 2150 & 07:42' & 80 & E230H & \\ 
    & & o4g001060 & 2128 - 2396 & 09:16' & 288 & E230H & \\ 
    & & o4g001050 & 2377 - 2650 & 09:02' & 360 & E230H & \\ 
    & & o4g001040 & 2620 - 2888 & 08:48' & 360 & E230H & \\ 
    
    \noalign{\bigskip}
    
    \multirow{6}{*}{1997-12-19} & \multirow{6}{*}{7512} & o4g002010 & 1463 - 1661 & 19:51' & 900 & E140H & \multirow{6}{*}{[-17;+19], [+50;+150]} \\ 
    & & o4g002030 & 1628 - 1902 & 20:22' & 679 & E230H & \\ 
    & & o4g002020 & 1879 - 2150 & 20:14' & 80 & E230H & \\ 
    & & o4g002060 & 2128 - 2396 & 21:53' & 288 & E230H & \\ 
    & & o4g002050 & 2377 - 2650 & 21:40' & 360 & E230H & \\ 
    & & o4g002040 & 2620 - 2888 & 21:26' & 360 & E230H & \\ 
    
    \noalign{\bigskip}
    
    \multirow{2}{*}{2001-08-26} & \multirow{2}{*}{9154} & o6fp01010 & 1499 - 1687 & 05:59' & 400 & E140H & \multirow{2}{*}{[-17;-3]} \\ 
    & & o6fp01040 & 1499 - 1687 & 08:52' & 400 & E140H & \\ 
    
    \noalign{\bigskip}
    
    2001-08-27 & 9154 & o6fp02010 & 1628 - 1897 & 01:15' & 210 & E230H & [-17;-3] \\ 
    
    \noalign{\bigskip}
    
    2002-05-21 & 9154 & o6fp03010 & 1834 - 2103 & 17:03' & 200 & E230H & [] \\ 
    
    \noalign{\bigskip}
    
    2002-05-22 & 9154 & o6fp04010 & 2327 - 2606 & 17:09' & 200 & E230H & [-8,40] \\ 
    
    \noalign{\bigskip}
    
    2017-10-09 & 14735 & odaa01010 & 2665 - 2931 & 23:40' & 475 & E230H & [-35;+120] \\ 
    
    \noalign{\bigskip}
    
    2018-01-10 & 14735 & odaa03010 & 2665 - 2931 & 08:12' & 475 & E230H & [-70;0], [+19;+150] \\ 
    
    \noalign{\bigskip}
    
    2018-03-07 & 14735 & odaa62010 & 2665 - 2931 & 02:33' & 475 & E230H & [-40;+13] \\ 
    
    \noalign{\bigskip}
    
    2018-10-29 & 15479 & odt901010 & 2665 - 2931 & 21:49' & 2101 & E230H & [-30;-4], [+8;+140] \\ 
    
    \noalign{\bigskip}
    
    2018-12-15 & 15479 & odt902010 & 2665 - 2931 & 04:22' & 2101 & E230H & [-40;+50] \\ 
    
    \noalign{\smallskip}
    \cline{1-8}                 
    \noalign{\smallskip}
\end{tabular}   
\caption{\centering Log of HST/STIS observations used in the present study.}
\label{recap_observations_stis}
\end{table}

\FloatBarrier

\newpage

\section{Comparison between \feii, \niii, \mnii\ and \siii\ lines}

\begin{figure*}[h!]
\centering
    \includegraphics[scale = 0.4,     trim = 8 0 0 30,clip]{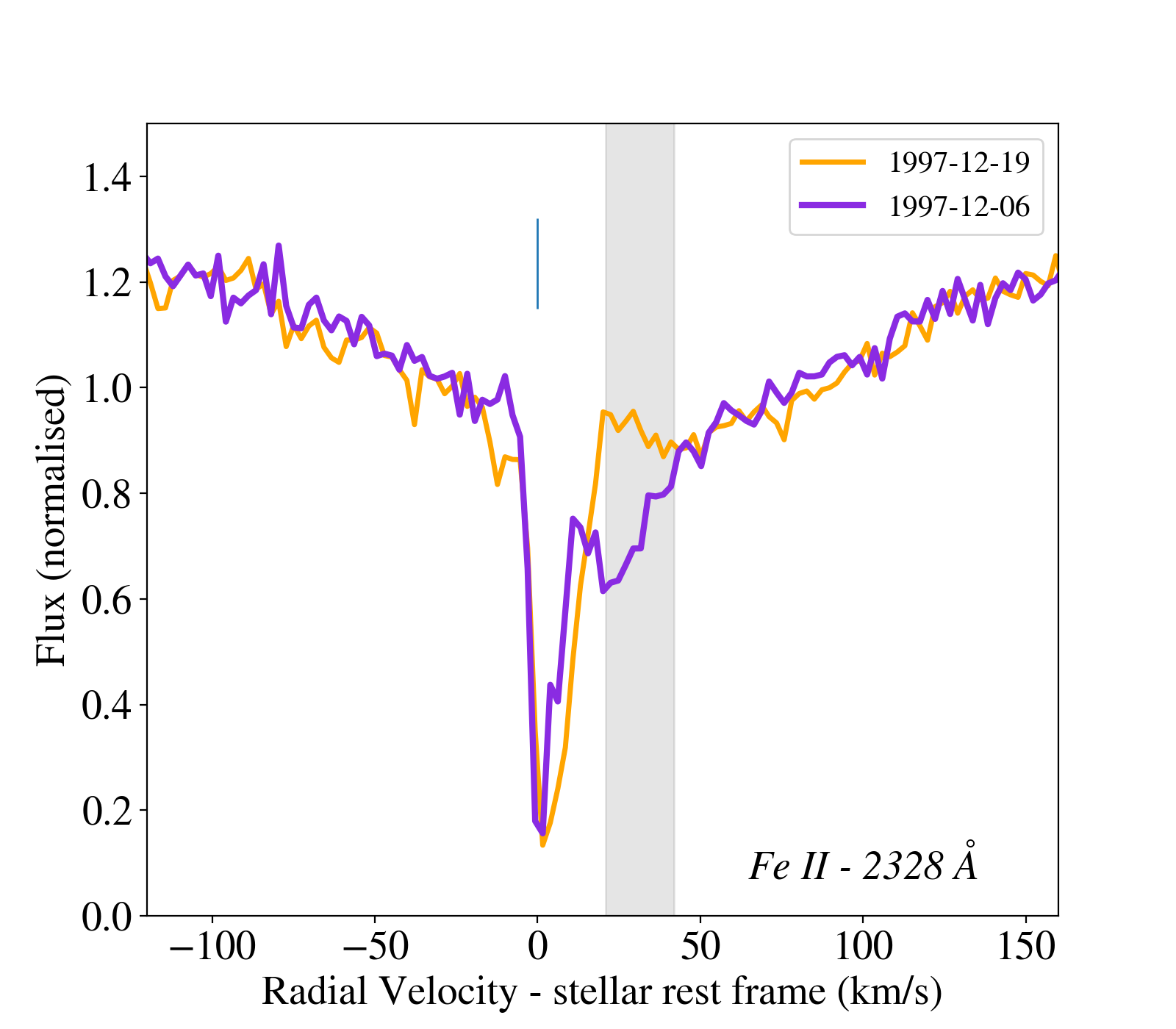}    
    \includegraphics[scale = 0.4,     trim = 8 0 0 30,clip]{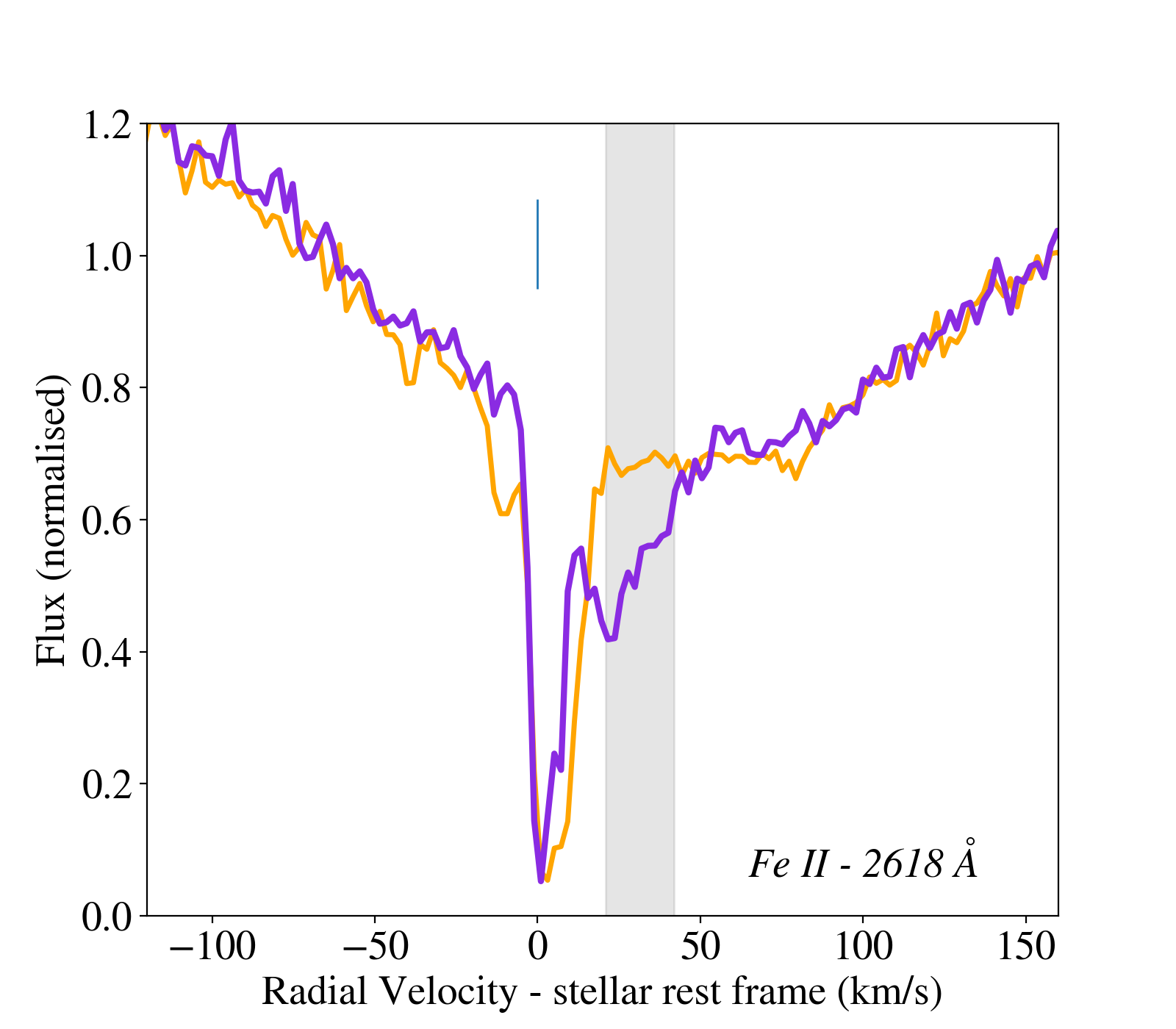}    
    
    \includegraphics[scale = 0.4,     trim = 8 0 0 15,clip]{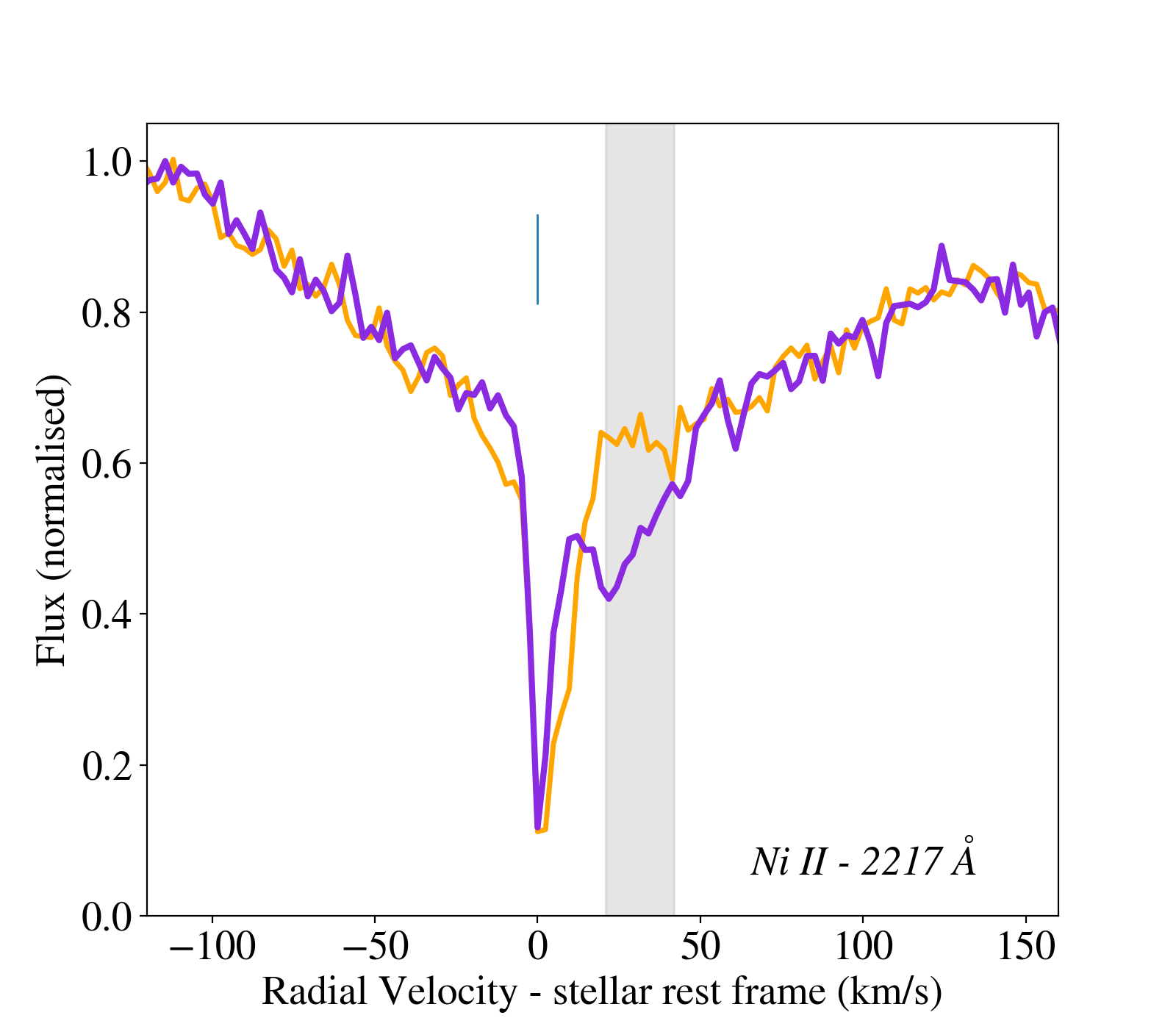}    
    \includegraphics[scale = 0.4,     trim = 8 0 0 15,clip]{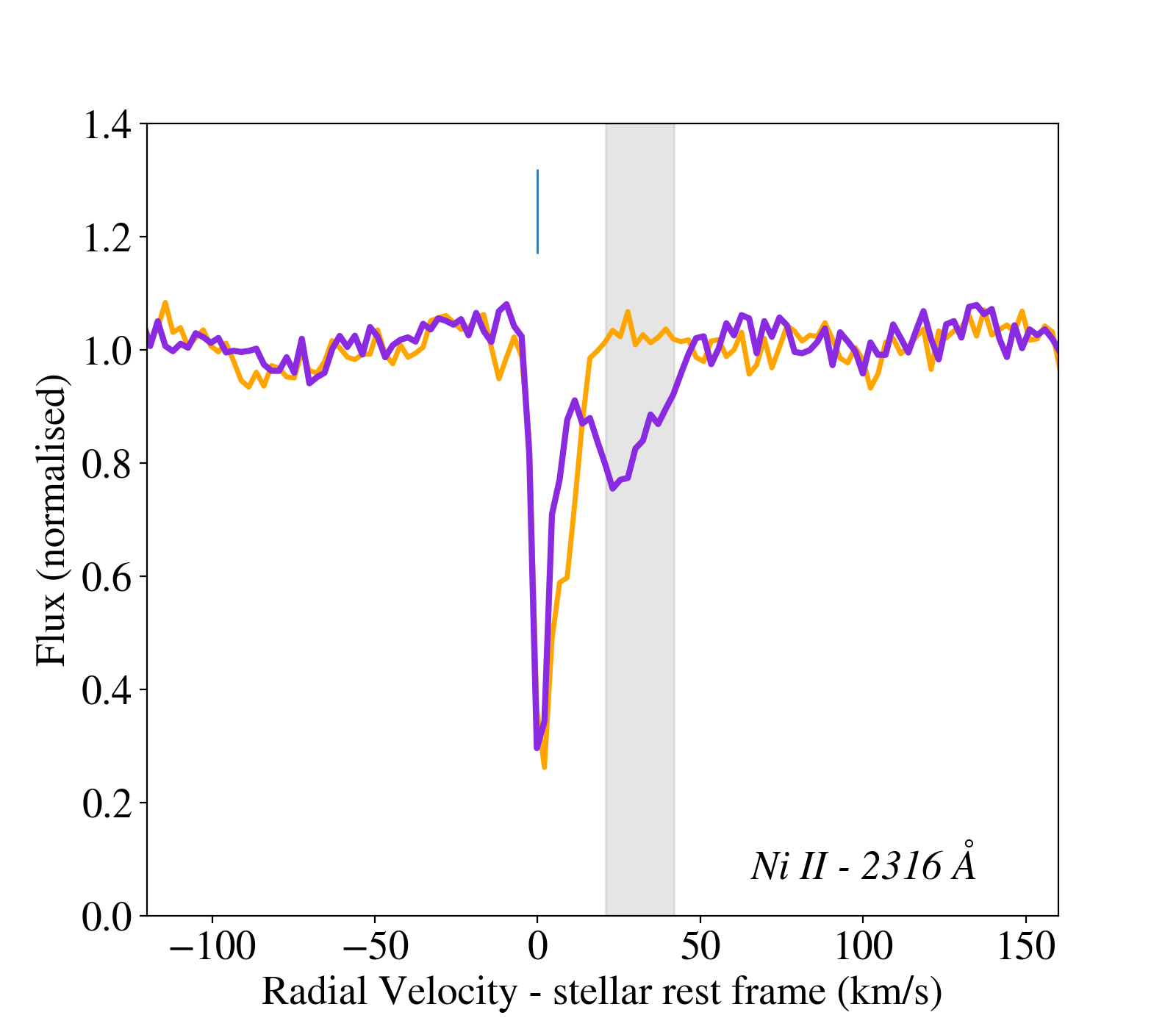}     

    \includegraphics[scale = 0.4,     trim = 8 0 0 15,clip]{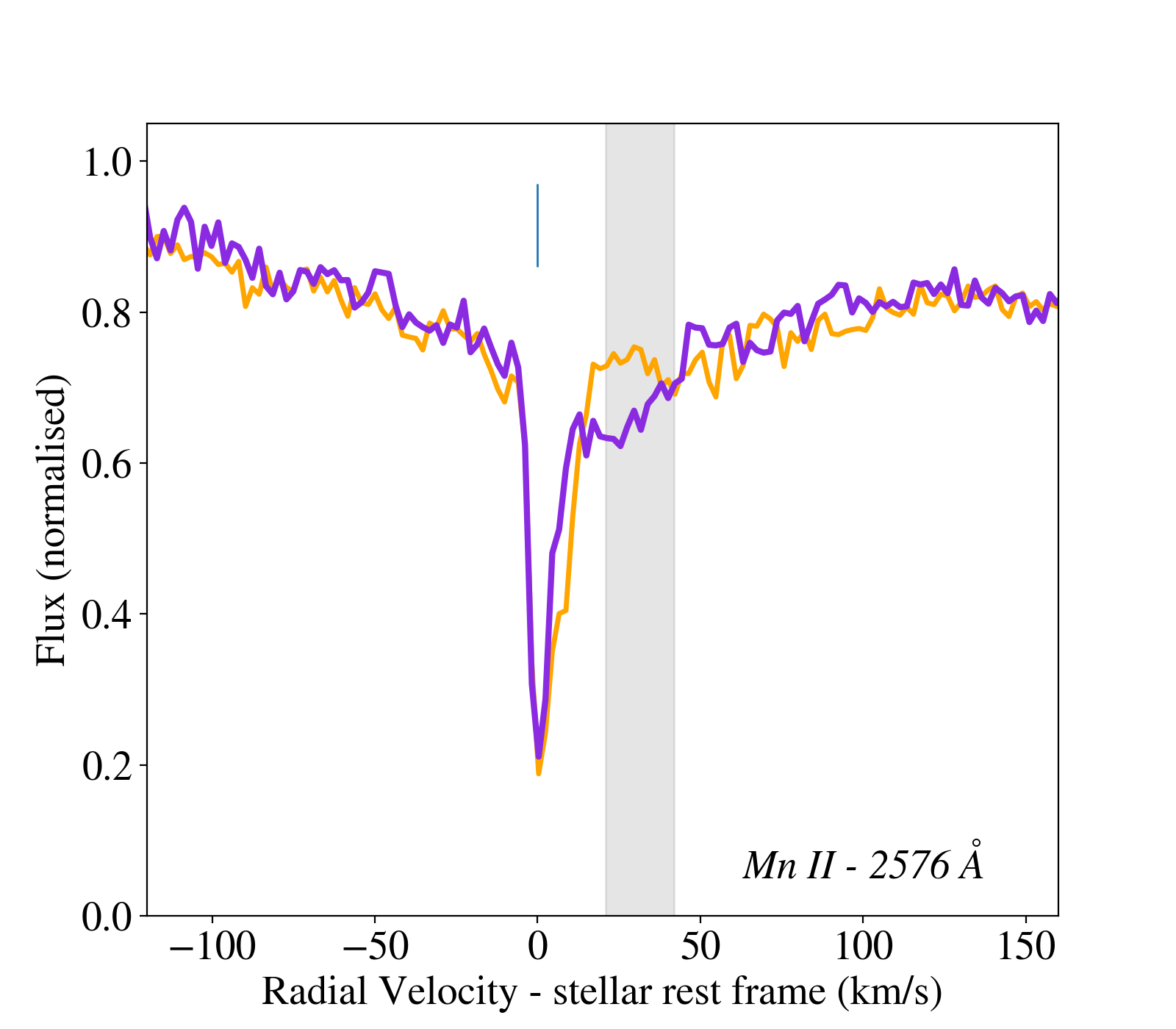}    
    \includegraphics[scale = 0.4,     trim = 8 0 0 15,clip]{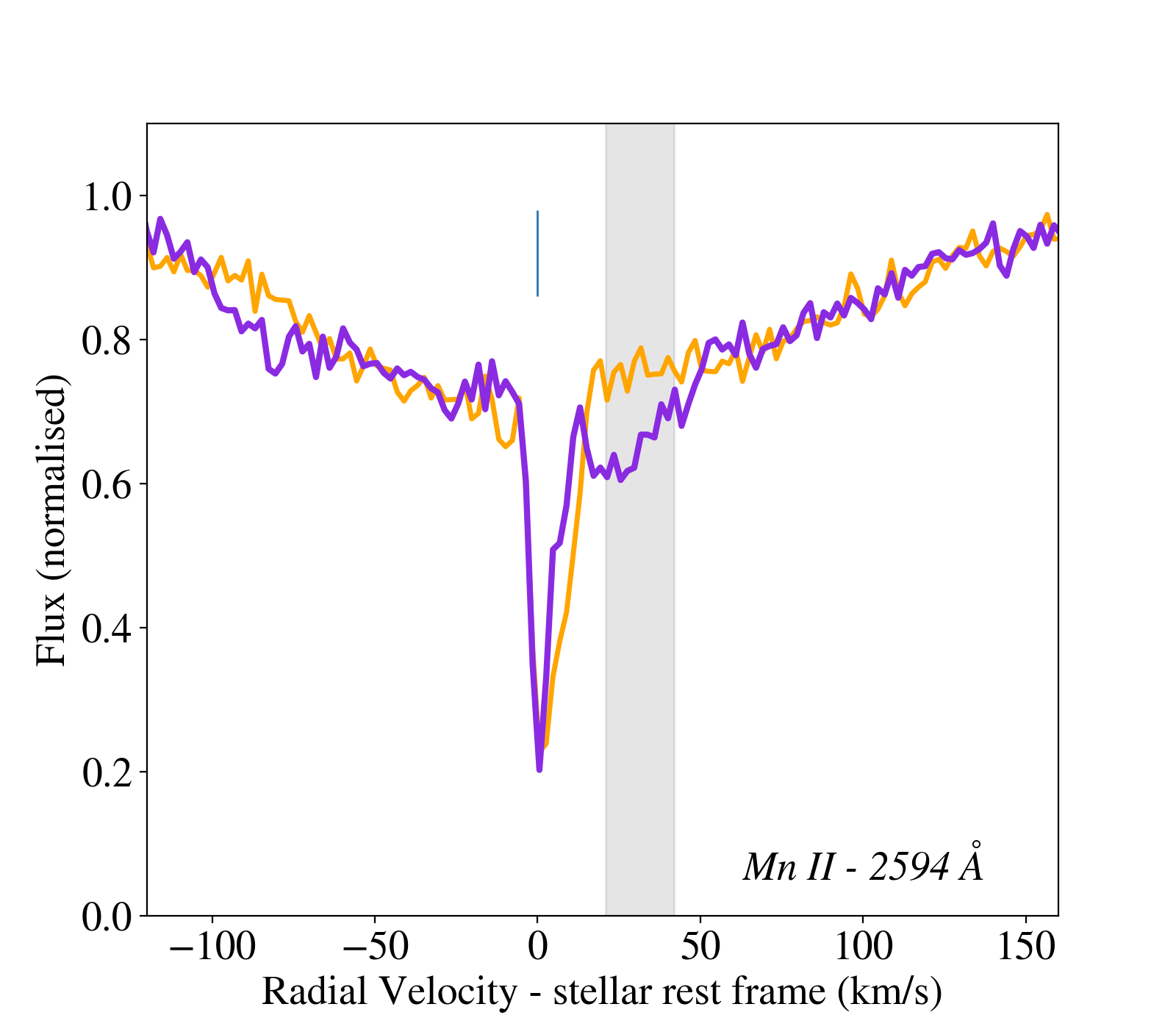}    

    \vspace{0.3 cm}
    \caption{Zoomed-in view of a few \feii, \niii\ and \mnii\ lines in \bp\ spectrum, at two different epochs (December 6 and 19, 1997). The grey area indicates the RV range (21-42 km/s) used to measure the December 6, 1997 comet's absorption depth throughout the study. For each epoch, all lines where observed within the same HST orbit, and are thus very close in time. The comet's absorption profile appears to be very similar from one species to another (same radial velocity, same width), although it seems slightly sharper in \feii\ and \niii\ than in \mnii. }
    \label{Fig. comparaison espèces I}
\end{figure*}

\newpage

\begin{figure*}[h!]
\centering
    \includegraphics[scale = 0.4,     trim = 8 0 0 30,clip]{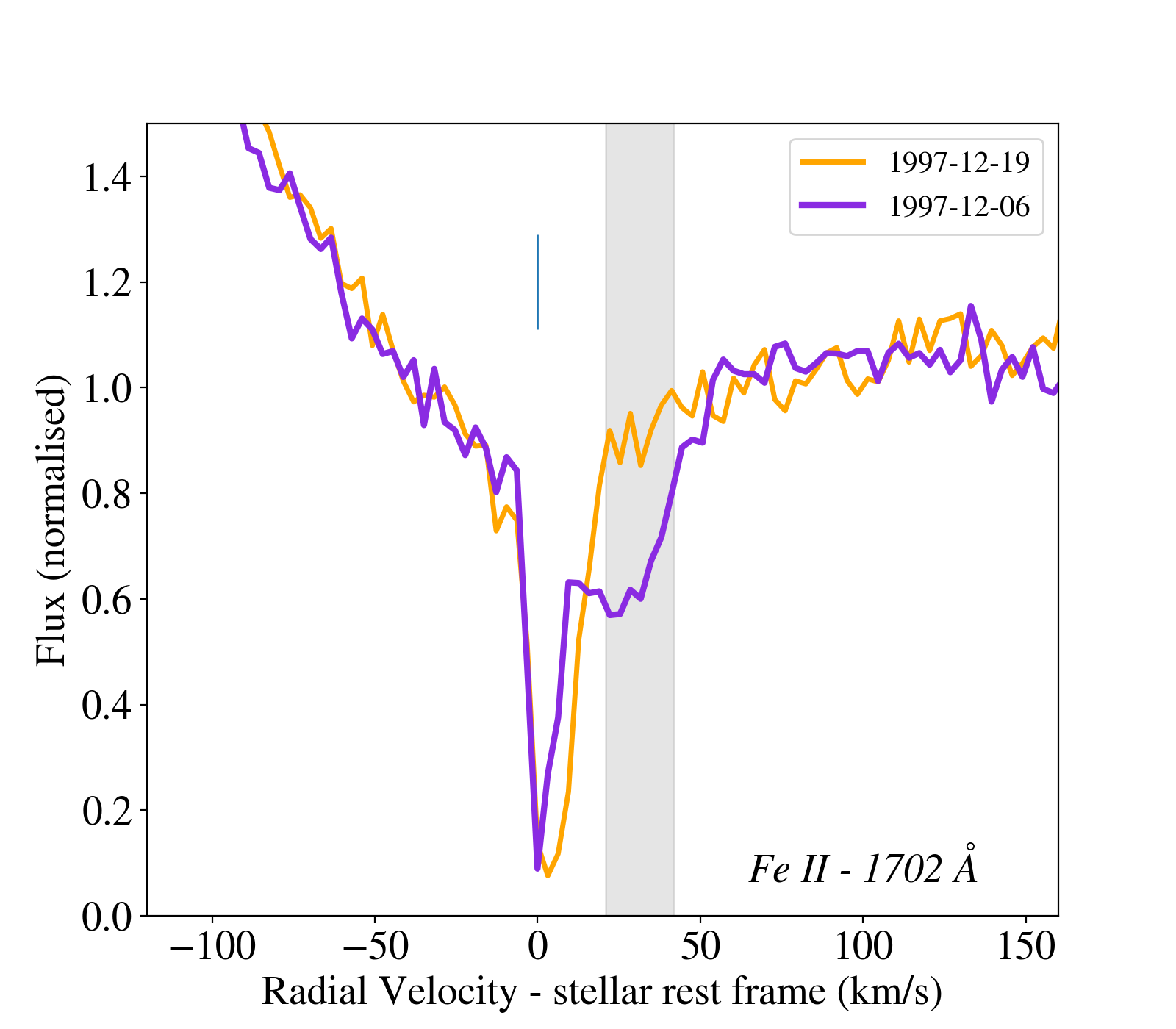}    
    \includegraphics[scale = 0.4,     trim = 8 0 0 30,clip]{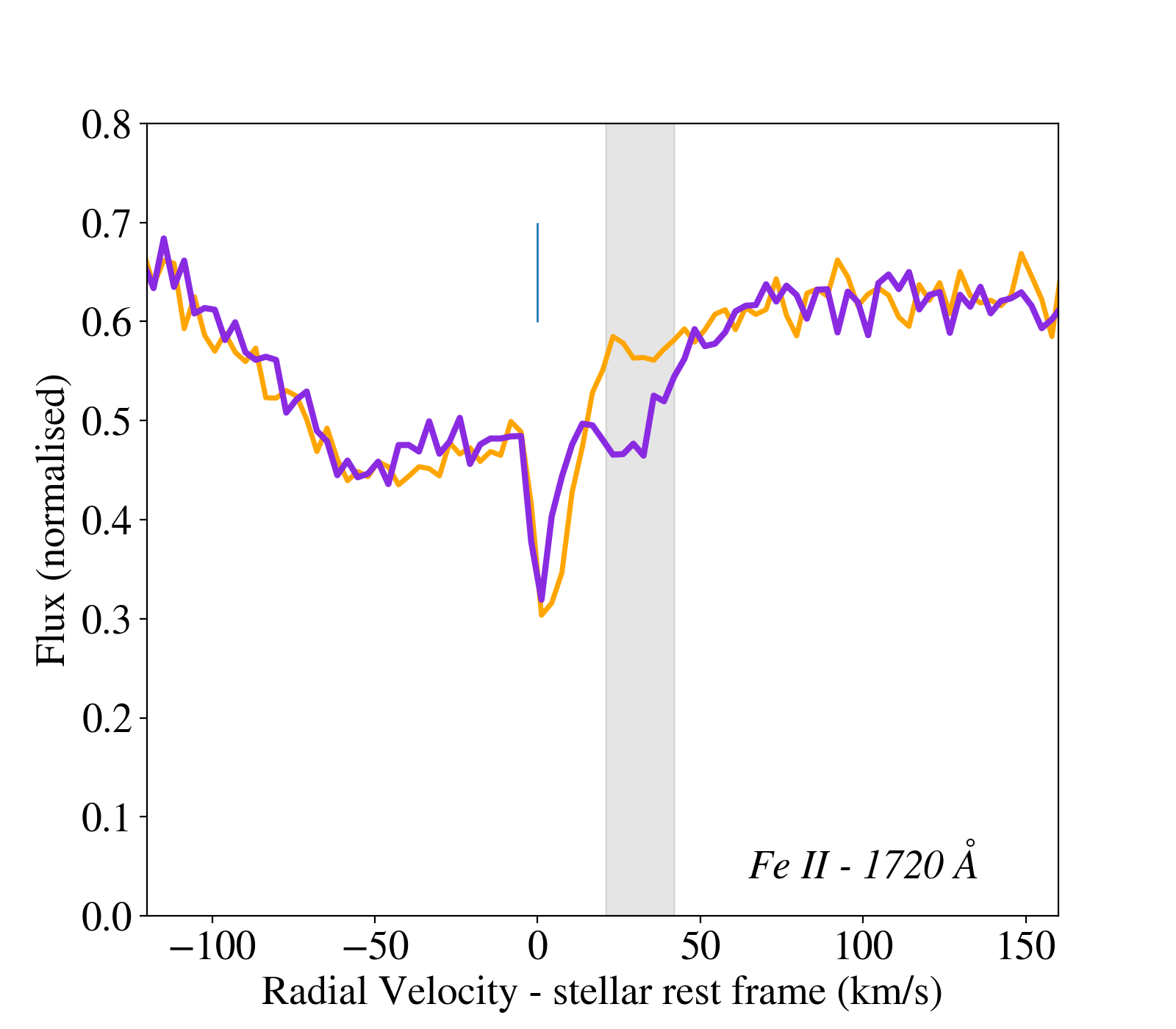}    
    
    \includegraphics[scale = 0.4,     trim = 8 0 0 15,clip]{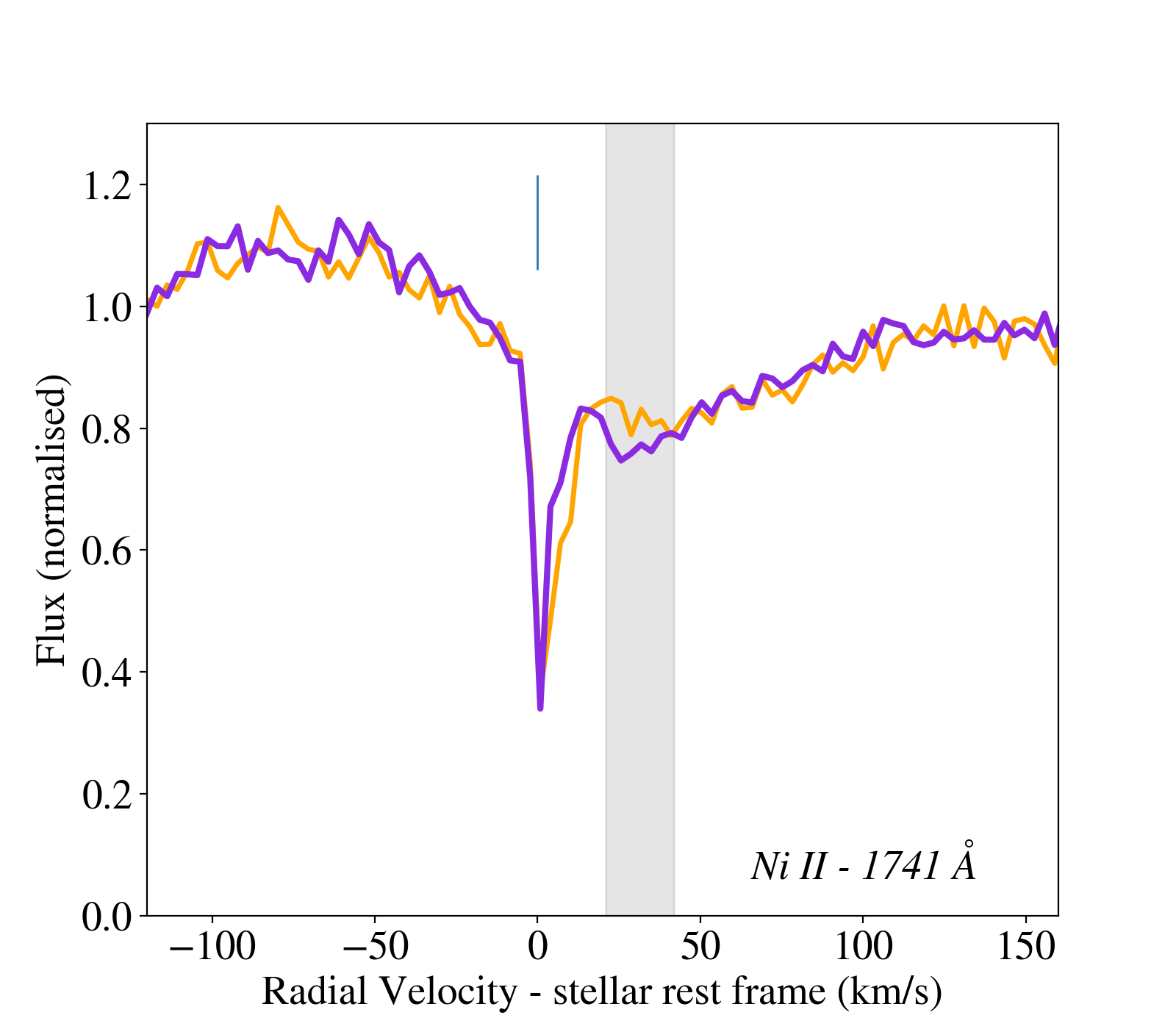}    
    \includegraphics[scale = 0.4,     trim = 8 0 0 15,clip]{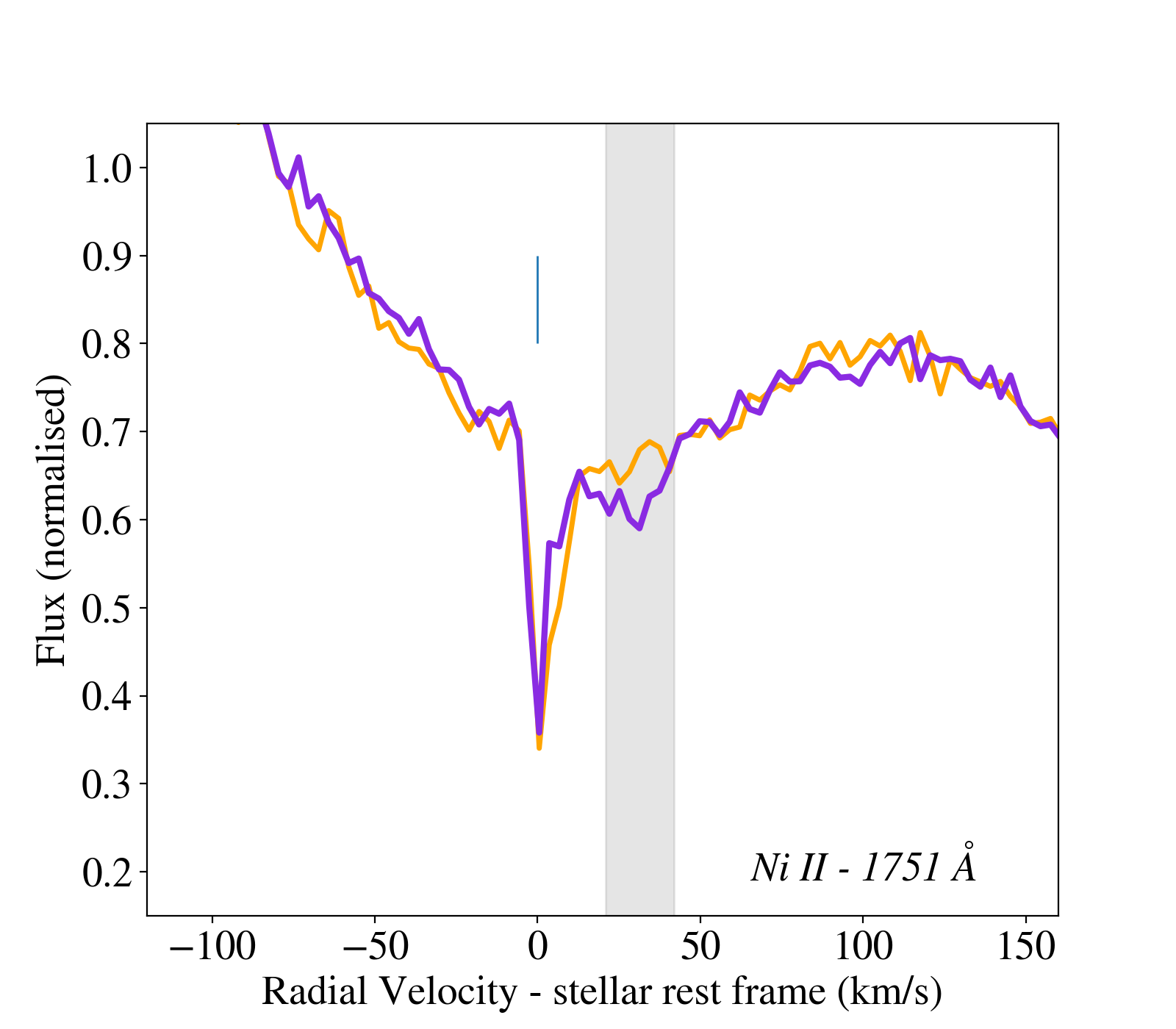}     

    \includegraphics[scale = 0.4,     trim = 8 0 0 15,clip]{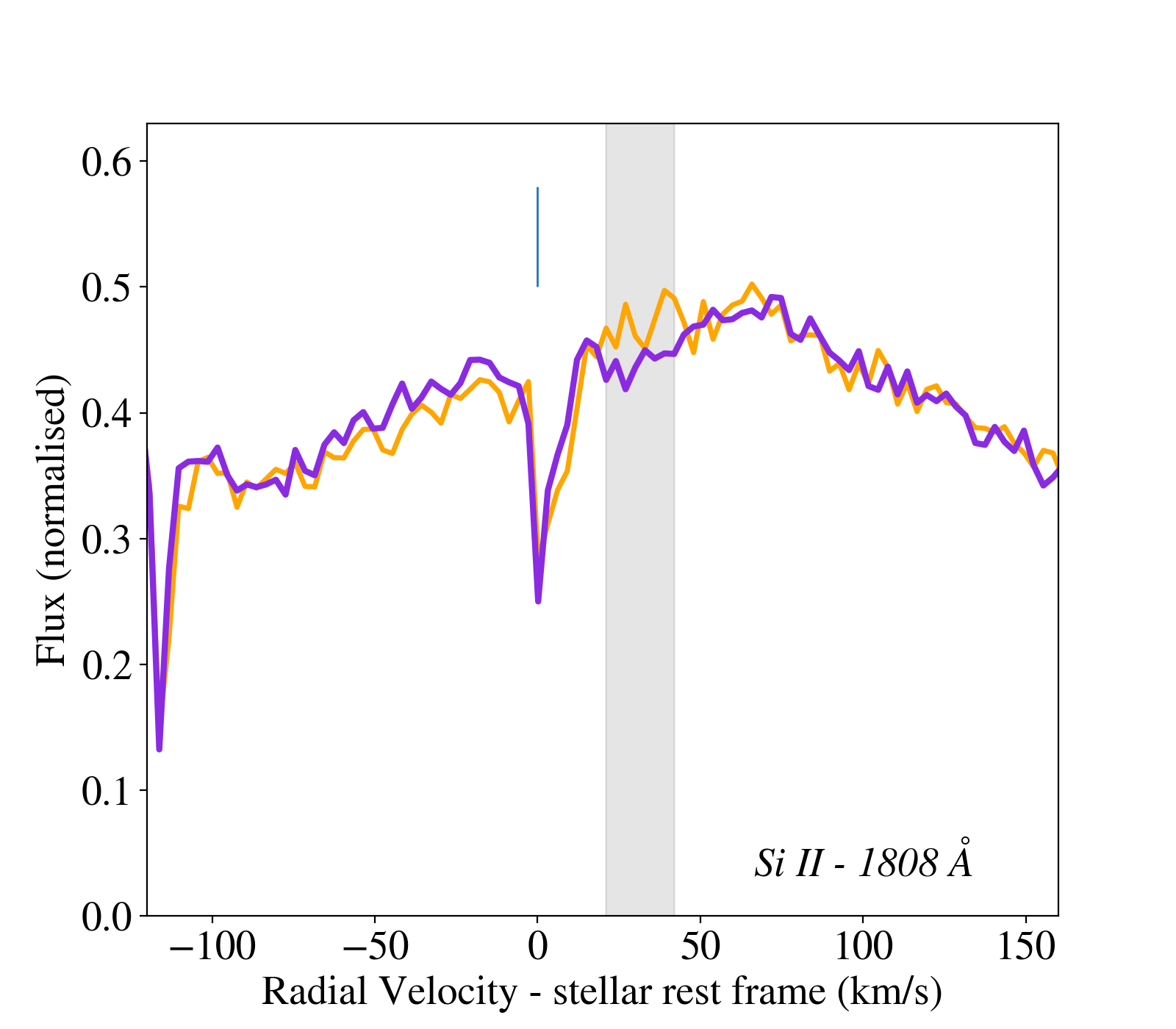}    
    \includegraphics[scale = 0.4,     trim = 8 0 0 15,clip]{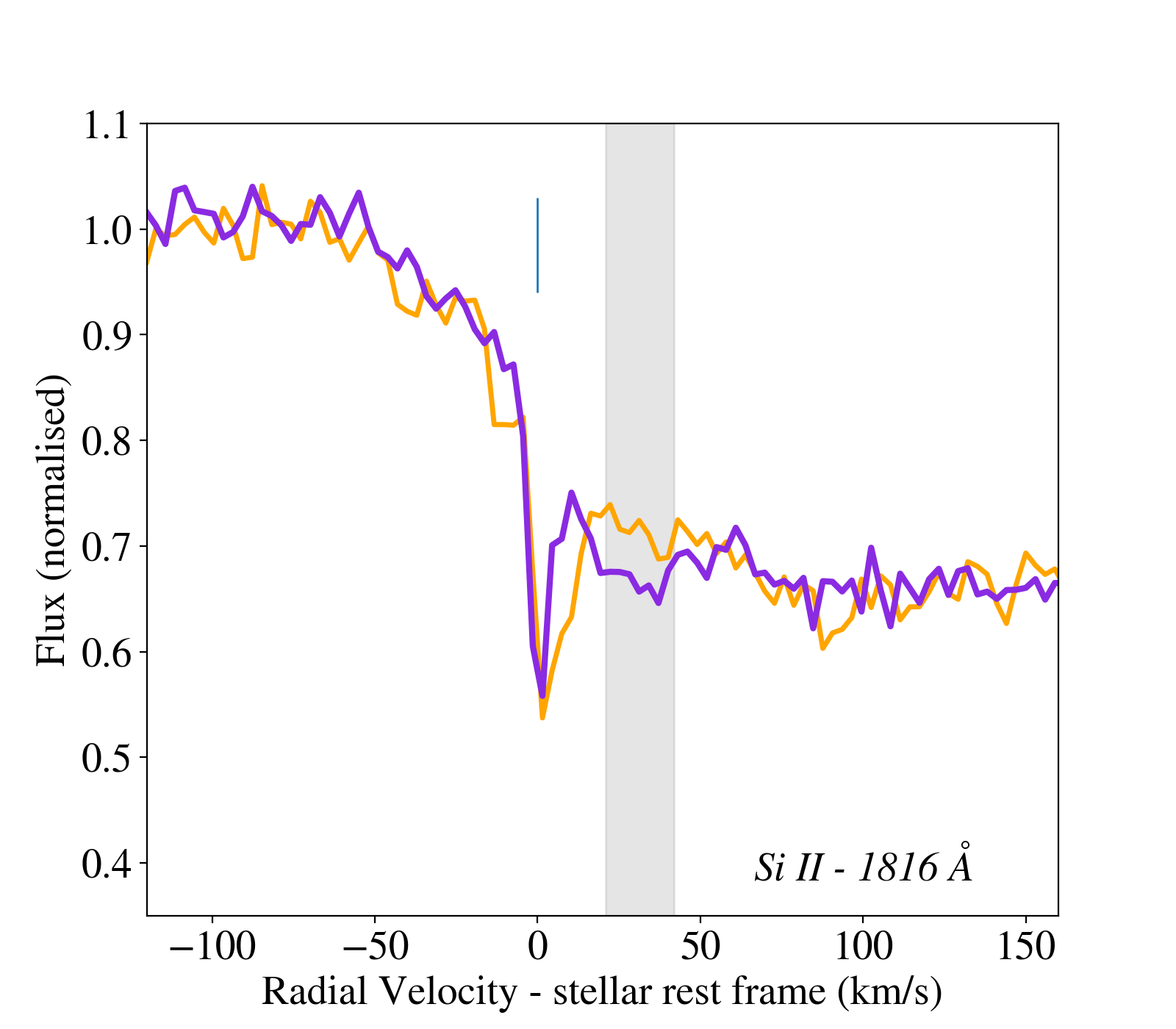}    

    \vspace{0.3 cm}
    \caption{Same as Fig. \ref{Fig. comparaison espèces I} for \feii, \niii\ and \siii\ lines. Again, the six lines were observed during the same HST orbit, for both visits.}
    \label{Fig. comparaison espèces II}
\end{figure*}

\newpage

\begin{figure*}[h!]
\centering
    \includegraphics[scale = 0.4,     trim = 8 0 0 30,clip]{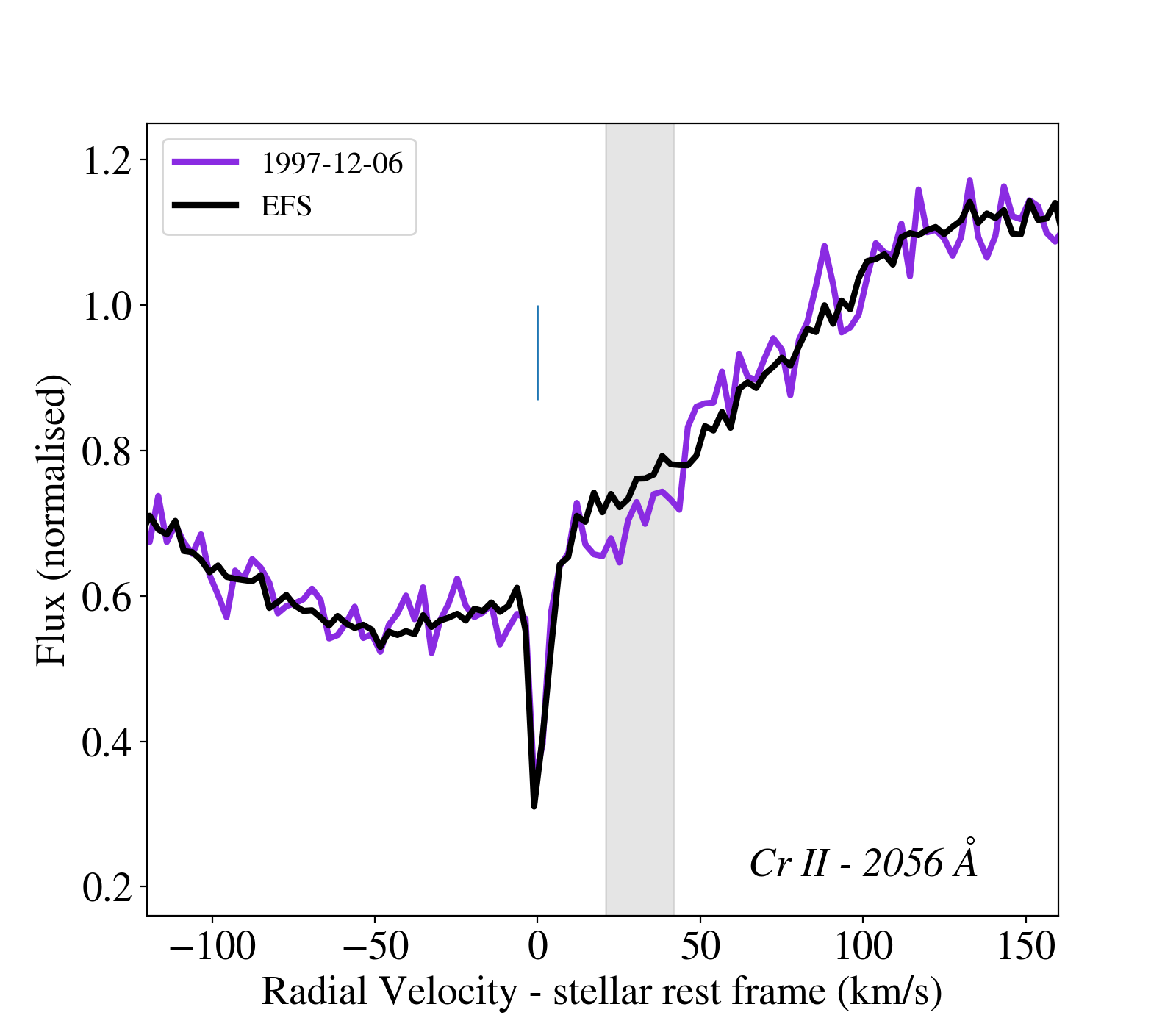}    
    \includegraphics[scale = 0.4,     trim = 8 0 0 30,clip]{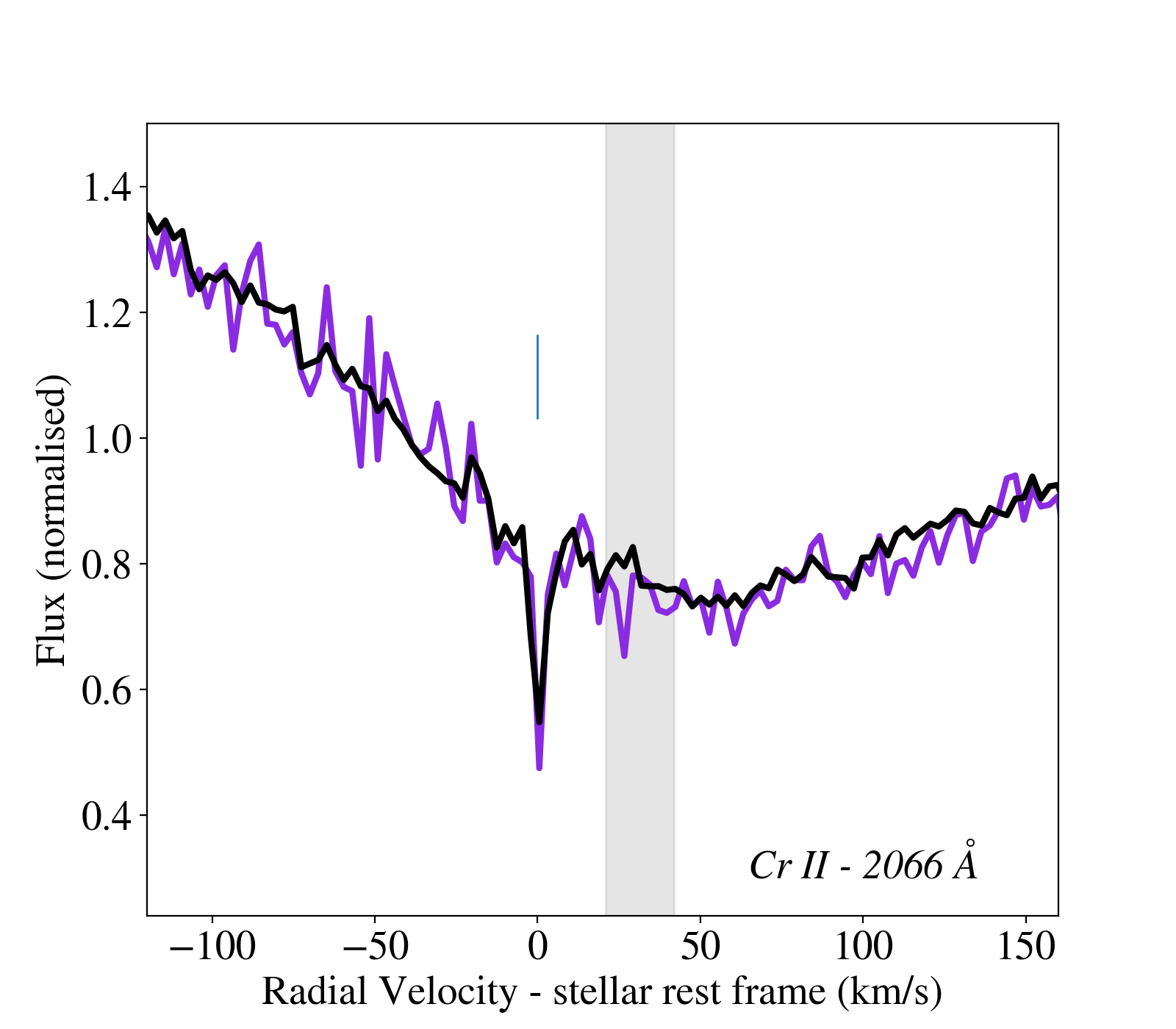}    
    
    \includegraphics[scale = 0.4,     trim = 8 0 0 15,clip]{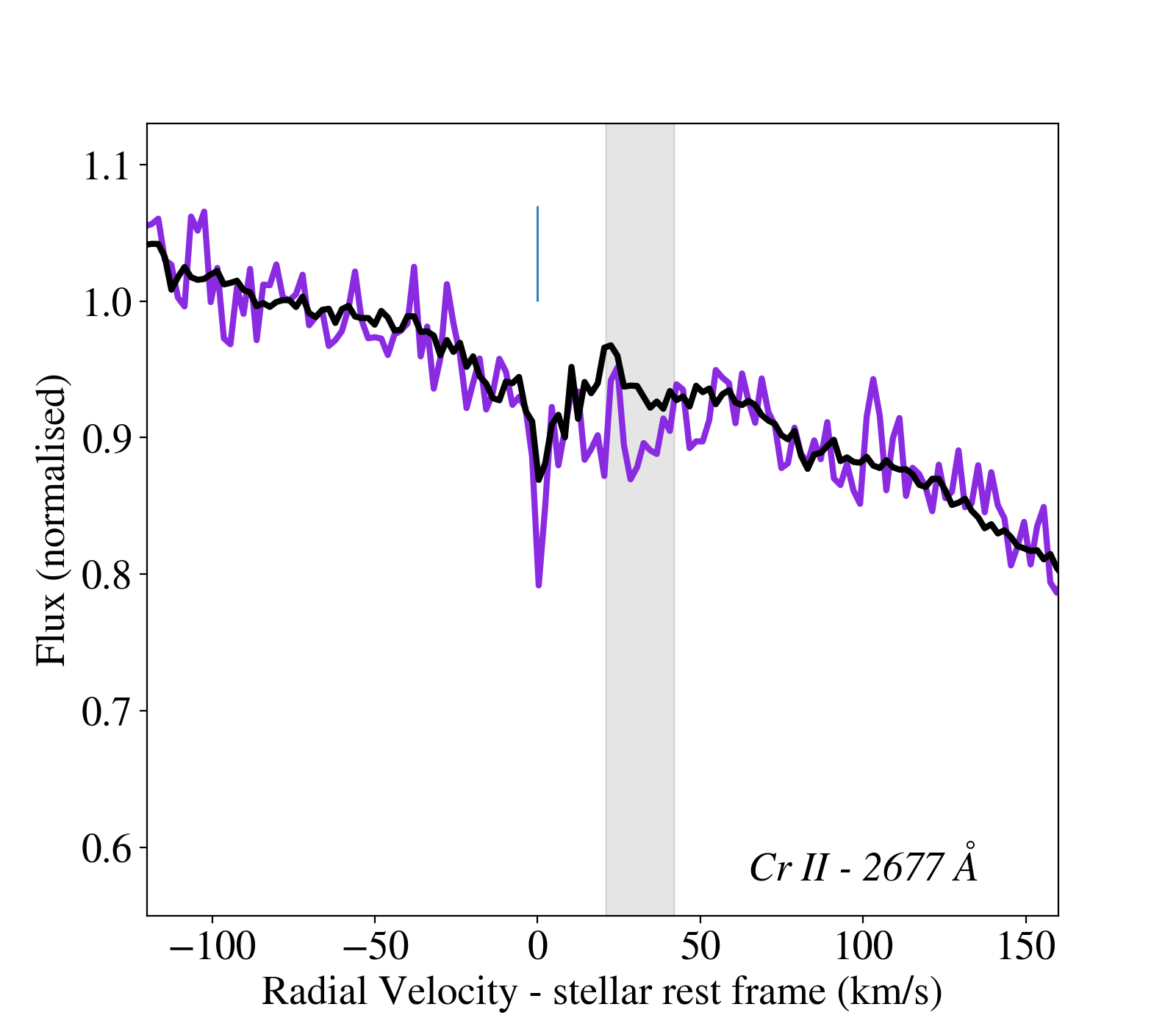}    
    \includegraphics[scale = 0.4,     trim = 8 0 0 15,clip]{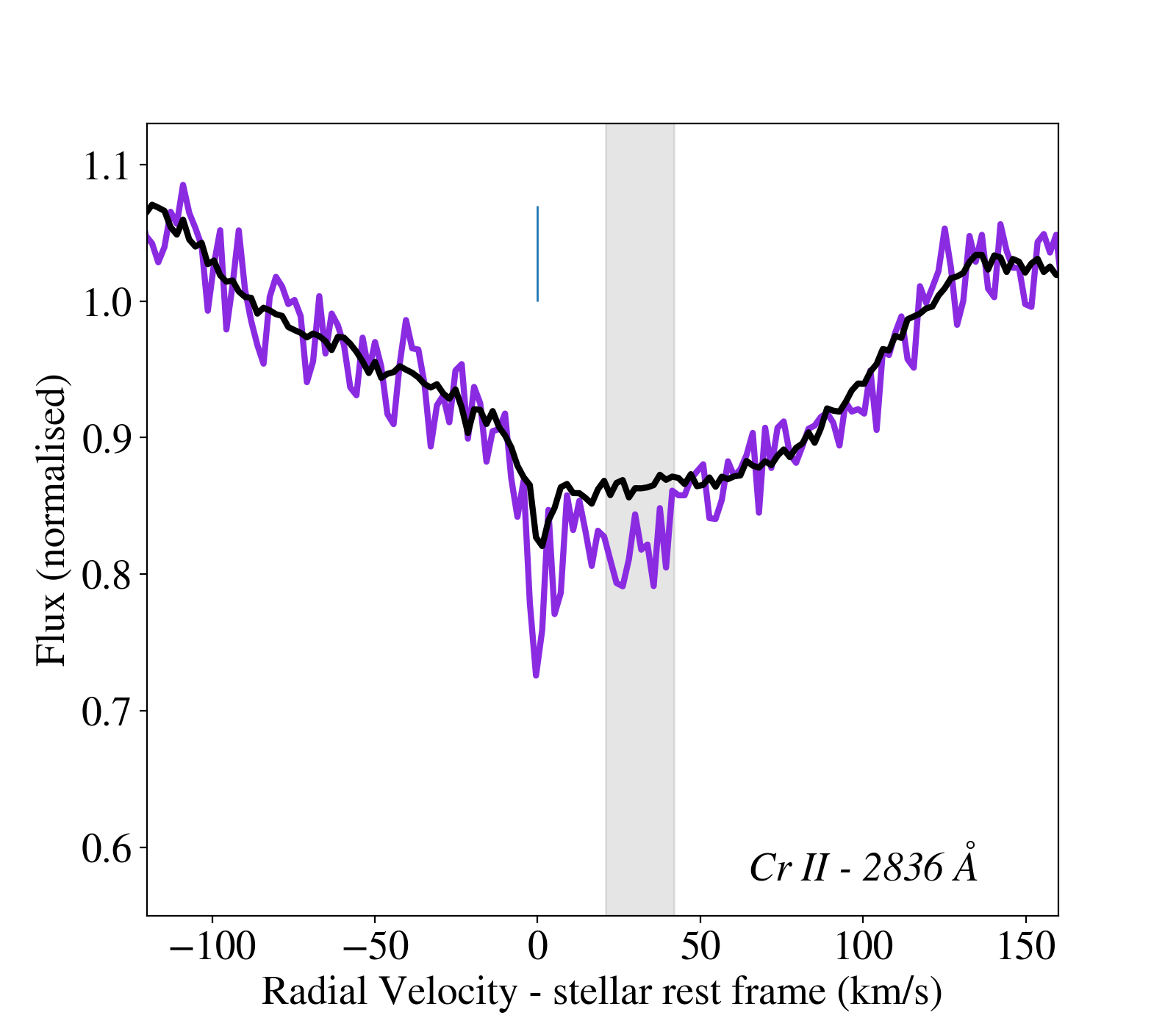}     

    \vspace{0.3 cm}
    \caption{Same as Fig. \ref{Fig. comparaison espèces I} for \crii\ lines, emphasising the December 6, 1997 observation (purple line). Contrary to Figs. \ref{Fig. comparaison espèces I} and \ref{Fig. comparaison espèces II}, the December 6, 1997 spectrum is compared to our reconstructed exocomet-free spectrum (EFS, black line), the December 19 observation being too noisy to allow a visual comparison.}
    \label{Fig. exemples Cr II}
\end{figure*}

\newpage

\FloatBarrier

\section{List of the studied lines}

\vspace{0.2 cm}

\begin{table}[h!]
    \centering 
    \begin{threeparttable}

    \caption{List of the \feii\ lines used in this study, sorted by lower level energy. The notations are the same as in Sect. 3. The three lowest terms are studied in Sect. 3 ($E_l = 0 - 8847 \ \si{cm^{-1}}$), while more excited levels are included in Sect. 5 ($E_l = 15845 - 32910 \ \si{cm^{-1}}$).}
    \begin{tabular}{ c c c c c c c c }          
            
            \cline{1-8}  
            \noalign{\smallskip}
            
            Configuration & Term & J & $E_l$\tnote{a} & $E_l/k_B$ & $\lambda_{lu}$\tnote{a} & $A_{ul}$\tnote{a} & $g_l f_{lu}$\tnote{a} \\
            & & & ($\si{cm^{-1}}$) & ($\si{K}$) & (\A) & ($10^7 \  \si{s^{-1}}$) &  \\
            
            \noalign{\smallskip}
            \cline{1-8} 
            \noalign{\smallskip}
            
            3d$^6$($^5$D)4s & a$^6$D & 9/2 & 0 & 0 & 2249.88 & 0.30 & 0.018 \\
            &&&&& 2260.78 & 0.32 & 0.024 \\
            &&&&& 2374.46 & 4.25 & 0.36 \\
            &&&&& 2382.76 & 31.3 & 3.20 \\
            &&&&& 2586.65 & 8.94 & 0.72 \\
            &&&&& 2600.17 & 23.5 & 2.38 \\
            
            \noalign{\medskip}

            && 7/2 & 385 & 554 & 2253.82 & 0.44 & 0.027 \\
            &&&&&                2280.62 & 0.45 & 0.035 \\
            &&&&&                2333.51 & 13.1 & 0.64 \\
            &&&&&                2365.55 & 5.90 & 0.40 \\
            &&&&&                2389.36 & 10.5 & 0.72 \\
            &&&&&                2396.35 & 25.9 & 2.23 \\        &&&&&                2612.65 & 12.0 & 0.98 \\
            &&&&&                2626.45 & 3.57 & 0.36 \\

            \noalign{\medskip}

            && 5/2 & 668 & 961 & 2251.63 & 0.32 & 0.015 \\
            &&&&&                2268.29 & 0.37 & 0.023 \\
            &&&&&                2328.11 & 6.60 & 0.21 \\
            &&&&&                2349.02 & 11.5 & 0.57 \\
            &&&&&                2381.49 & 3.10 & 0.21 \\
            &&&&&                2399.97 & 13.9 & 0.72 \\
            &&&&&                2405.62 & 19.6 & 1.36 \\
            &&&&&                2607.87 & 17.3 & 0.71 \\
            &&&&&                2618.40 & 4.88 & 0.30 \\
            &&&&&                2632.11 & 6.29 & 0.52 \\

            \noalign{\medskip}

            && 3/2 & 863 & 1241 & 2261.56 & 0.22 & 0.010 \\
            &&&&&                 2338.73 & 11.3 & 0.37 \\
            &&&&&                 2359.83 & 5.00 & 0.25 \\
            &&&&&                 2407.39 & 16.1 & 0.56 \\
            &&&&&                 2614.60 & 21.2 & 0.44 \\

            \noalign{\medskip}

            && 1/2 & 977 & 1406 & 2345.00 & 9.27 & 0.31 \\
            &&&&&                 2411.80 & 23.7 & 0.41 \\
            &&&&&                 2414.04 & 10.2 & 0.36 \\
            &&&&&                 2622.45 & 5.60 & 0.115 \\
            &&&&&                 2629.08 & 8.74 & 0.36 \\

            \noalign{\medskip}
            \cdashline{1-8}   
            \noalign{\medskip}

            3d$^7$ & a$^4$F & 9/2 & 1873 & 2694 & 2332.02 & 3.17 & 0.21 \\

            \noalign{\medskip}

            && 7/2 & 2430 & 3497 & 2361.02 & 6.20 & 0.31 \\
            &&&&&                  2362.74 & 1.41 & 0.094 \\
            &&&&&                  2380.00 & 2.73 & 0.185 \\
            &&&&&                  2392.21 & 0.38 & 0.032 \\
            
            \noalign{\medskip}

            && 5/2 & 2838 & 4083 & 2355.61 & 2.67 & 0.089 \\
            &&&&&                  2367.32 & 1.01 & 0.051 \\
            &&&&&                  2369.32 & 6.10 & 0.204 \\
            &&&&&                  2385.73 & 0.36 & 0.025 \\

            \noalign{\medskip}

            && 3/2 & 3117 & 4485 & 2371.22 & 1.73 & 0.058 \\
            &&&&&                  2375.92 & 9.80 & 0.166 \\
            &&&&&                  2385.11 & 3.20 & 0.109 \\
            
            \noalign{\smallskip}

            \cline{1-8}  

        \end{tabular} 
        
    \begin{tablenotes}
      \item[a] \small The level energies and line parameters ($\lambda_{lu}$, $A_{ul}$, $f_{lu}$) were all collected from the NIST database (Kramida et al. 2023). The uncertainties are typically of the order of 10 \%.
    \end{tablenotes}
    
\label{Tab. list lines Fe II}

\end{threeparttable}
\end{table}

\newpage

\onecolumn

\centering \small \textbf{Table D.1}, continued.

\begin{table}[h!]
    \centering 
    \begin{threeparttable}

    \begin{tabular}{ c c c c c c c c }  

            \cline{1-8}  
            \noalign{\smallskip}
            
            Configuration & Term & J & $E_l$ & $E_l/k_B$ & $\lambda_{lu}$ & $A_{ul}$ & $g_l f_{lu}$ \\
            & & & ($\si{cm^{-1}}$) & ($\si{K}$) & (\A) & ($10^7 \  \si{s^{-1}}$) &  \\
            
            \noalign{\smallskip}
            \cline{1-8} 
            \noalign{\smallskip}

            3d$^6$($^5$D)4s & a$^4$D & 7/2 & 7955 & 11446 & 2563.30 & 17.9 & 1.06 \\
            &&&&&                                           2715.22 & 5.70 & 0.38 \\            
            &&&&&                                           2740.36 & 22.1 & 1.99 \\
            &&&&&                                           2756.55 & 21.5 & 2.45 \\

            \noalign{\medskip}

            && 5/2 & 8392 & 12074 & 2564.27 & 1.73 & 0.058 \\
            &&&&&                   2592.31 & 5.70 & 0.35 \\
            &&&&&                   2725.69 & 0.96 & 0.064 \\
            &&&&&                   2728.35 & 9.38 & 0.42 \\
            &&&&&                   2747.79 & 16.9 & 1.15 \\

            \noalign{\medskip}

            && 3/2 & 8680 & 12490 & 2567.68 & 11.0 & 0.22 \\
            &&&&&                   2583.36 & 8.80 & 0.35 \\
            &&&&&                   2731.54 & 2.79 & 0.125 \\
            &&&&&                   2737.78 & 12.2 & 0.27 \\
            &&&&&                   2769.75 & 0.48 & 0.033 \\

            \noalign{\medskip}

            && 1/2 & 8847 & 12729 & 2578.69 & 12.0 & 0.24 \\
            &&&&&                   2744.01 & 19.7 & 0.89 \\
            &&&&&                   2762.66 & 1.38 & 0.063 \\

            \noalign{\medskip}
            \cdashline{1-8}   
            \noalign{\medskip}

            3d$^7$ & a$^2$G & 9/2 & 15845 & 22797 & 1746.82 & 12.1 & 0.55 \\

            \noalign{\medskip}

            && 7/2 & 16369 & 23553 & 1761.37 & 14.2 & 0.53 \\
            
            \noalign{\medskip}
            \cdashline{1-8}   
            \noalign{\medskip}

            3d$^6$($^3$P2)4s & b$^4$P & 5/2 & 20831 & 29971 & 2445.26 & 27.8 & 1.99 \\
            &&&&&                                             2527.05 & 24.7 & 1.42 \\           
            \noalign{\medskip}

            && 3/2 & 21812 & 31383 & 2446.31 & 20.7 & 1.11 \\
            
            \noalign{\medskip}
            \cdashline{1-8}   
            \noalign{\medskip}

            3d$^6$($^3$H)4s & a$^4$H & 13/2 & 21251 & 30577 & 2526.15 & 19.1 & 2.56 \\

            \noalign{\medskip}

            && 11/2 & 21430 & 30834 & 2534.39 & 19.2 & 2.22 \\

            \noalign{\medskip}

            && 9/2 & 21582 & 31052 & 2499.65 & 21.2 & 2.38 \\

            \noalign{\medskip}

            && 7/2 & 21712 & 31239 & 2512.52 & 23.0 & 2.18 \\
            &&&&&                    2535.18 & 18.3 & 1.41 \\

            \noalign{\medskip}
            \cdashline{1-8}   
            \noalign{\medskip}

            3d$^6$($^3$F2)4s & b$^4$F & 9/2 & 22637 & 32571 & 2424.88 & 22.1 & 2.34 \\
            &&&&&                                             2480.91 & 15.5 & 1.14 \\
            &&&&&                                             2530.31 & 22.0 & 2.11 \\
            
            \noalign{\medskip}

            && 7/2 & 22810 & 32820 & 2430.82 & 19.1 & 1.69 \\
            &&&&&                    2471.42 & 15.4 & 0.85 \\

            \noalign{\medskip}
            \cdashline{1-8}   
            \noalign{\medskip}

            3d$^5$4s$^2$ & a$^6$S & 5/2 & 23317 & 33550 & 1785.28 & 109 & 4.17 \\
            &&&&&                                         1786.76 & 110 & 3.16 \\
            
            \noalign{\medskip}
            \cdashline{1-8}   
            \noalign{\medskip}

            3d$^6$($^3$G)4s & a$^4$G & 11/2 & 25429 & 36587 & 2440.04 & 22.5 & 2.81 \\
            &&&&&                                             2490.58 & 19.4 & 2.17 \\
            
            \noalign{\medskip}

            && 9/2 & 25805 & 37129 & 2459.53 & 23.1 & 2.51 \\
        
            \noalign{\medskip}

            && 7/2 & 25982 & 37383 & 2462.60 & 24.3 & 2.21 \\

            \noalign{\medskip}
            \cline{1-8}  
            
        \end{tabular} 
        
    \begin{tablenotes}
      \item[a] \small The 2530.31 \A\ line is blended with another \feii\ transition at 2530.32 \A\ ($E_l\,=\,21812 \ \si{cm^{-1}}$, $gf = 0.78$). An effective $gf$-value of 2.89 was thus used for this line.
    \end{tablenotes}
    
\end{threeparttable}
\end{table}

\newpage

\small \textbf{Table D.1}, continued.

\begin{table}[h!]
    \centering 
    \begin{threeparttable}

    \begin{tabular}{ c c c c c c c c }  

            \cline{1-8}  
            \noalign{\smallskip}
            
            Configuration & Term & J & $E_l$ & $E_l/k_B$ & $\lambda_{lu}$ & $A_{ul}$ & $g_l f_{lu}$ \\
            & & & ($\si{cm^{-1}}$) & ($\si{K}$) & (\A) & ($10^7 \  \si{s^{-1}}$) &  \\
            
            \noalign{\smallskip}
            \cline{1-8} 
            \noalign{\smallskip}

            3d$^6$($^3$H)4s & b$^2$H & 11/2 & 26170 & 37654 & 2490.58  & 19.4 & 2.17 \\
            &&&&&                                             2768.32  & 15.8 & 2.55 \\
                        
            \noalign{\medskip}

            && 9/2 & 26353 & 37917 & 2550.79 & 17.9 & 1.75 \\
            &&&&&                    2754.10 & 18.9 & 2.58 \\
            
            \noalign{\medskip}
            \cdashline{1-8}   
            \noalign{\medskip}

            3d$^6$($^3$G)4s & b$^2$G & 9/2 & 30389 & 43723 & 2504.63 & 22.3 & 2.10 \\
            &&&&&                                            2693.40 & 14.0 & 1.83 \\
                        
            \noalign{\medskip}

            && 7/2 & 30764 & 44264 & 2515.14 & 21.1 & 1.60 \\
            &&&&&                    2685.55 & 15.7 & 1.70 \\

            \noalign{\medskip}
            \cdashline{1-8}   
            \noalign{\medskip}

            d$^6$($^1$I)4s & a$^2$I & 13/2 & 32876 & 47302 & 2433.61 & 28.6 & 3.56 \\
            &&&&&                                            2593.56 & 27.4 & 4.42 \\
                        
            \noalign{\medskip}

            && 11/2 & 32910 & 47351 & 2435.47 & 27.9 & 1.98 \\
            &&&&&                     2448.50 & 19.7 & 1.77 \\

            \noalign{\smallskip}
            \cline{1-8}

        \end{tabular}

\end{threeparttable}
\end{table}

\vspace{1.5 cm}

\begin{table}[h!]
    \centering 
    \begin{threeparttable}

    \caption{List of the \crii\ lines used in this study, sorted by lower level energy.}

    \begin{tabular}{ c c c c c c c c }          
            
            \cline{1-8}  
            \noalign{\smallskip}
            
            Configuration & Term & J & $E_l$\tnote{a} & $E_l/k_B$ & $\lambda_{lu}$\tnote{a} & $A_{ul}$\tnote{a} & $g_l f_{lu}$\tnote{a} \\
            & & & ($\si{cm^{-1}}$) & ($\si{K}$) & (\A) & ($10^7 \  \si{s^{-1}}$) &  \\
            
            \noalign{\smallskip}
            \cline{1-8} 
            \noalign{\smallskip}
            
            3d$^5$  & a$^6$S & 5/2 & 0 & 0 & 2056.26 & 1.22 & 0.62 \\
            &&&&&                            2062.24 & 1.19 & 0.46 \\
            &&&&&                            2066.16 & 1.20 & 0.31 \\

            \noalign{\medskip}
            \cdashline{1-8}   
            \noalign{\medskip}

            3d$^4$($^5$D)4s & a$^6$D & 3/2 & 12033 & 17313 & 2669.50 & 10.0 & 0.85 \\
            &&&&&                                            2679.59 & 8.02 & 0.52 \\
            &&&&&                                            2856.51 & 11.3 & 0.83 \\
            &&&&&                                            2867.58 & 13.6 & 0.67 \\

            \noalign{\medskip}

            && 5/2 & 12148 & 17478 & 2672.60 & 10.9 & 0.47 \\
            &&&&&                    2751.54 & 9.56 & 0.65 \\
            &&&&&                    2850.67 & 15.2 & 1.48 \\
            &&&&&                    2865.95 & 11.1 & 0.82 \\

            \noalign{\medskip}

            && 7/2 & 12304 & 17703 & 2673.62 & 5.74 & 0.37 \\
            &&&&&                    2844.08 & 18.9 & 2.29 \\
            &&&&&                    2863.41 & 8.66 & 0.85 \\
            
            \noalign{\medskip}

            && 9/2 & 12496 & 17980 & 2677.95\tnote{b} & 20.9 & 2.25 \\
            &&     &       &       & 2767.35          & 22.3 & 2.05 \\
            &&&&&                    2836.47          & 25.0 & 3.62 \\

            \noalign{\smallskip}

            \cline{1-8}  

        \end{tabular} 
        
    \begin{tablenotes}
      \item[a] \small The level energies of \crii\ were obtained from the NIST database, as well as the line parameters for the ground level. For the lines rising from the a$^6$D term, the NIST database appears to be very incomplete. We thus used the values from Nilsson et al. (2006).
      \item[b] \small The 2677.95 \A\ line is strongly blended with another \crii\ line ($E_l = 12304 \ \si{cm^{-1}}$, $gf = 1.02$), with $\Delta$RV $\sim 200 \ \si{m/s}$. A corrected $gf$-value of 3.28 was thus used for this doublet line.
      \end{tablenotes}
    
\label{Tab. list lines Cr II}

\end{threeparttable}
\end{table}

\newpage

\begin{table}[h!]
    \centering 
    \begin{threeparttable}

    \caption{List of the \niii\ lines used in this study, sorted by lower level energy.}

    \begin{tabular}{ c c c c c c c c }          
            
            \cline{1-8}  
            \noalign{\smallskip}
            
            Configuration & Term & J & $E_l$\tnote{a} & $E_l/k_B$ & $\lambda_{lu}$\tnote{a} & $A_{ul}$\tnote{a} & $g_l f_{lu}$\tnote{a} \\
            & & & ($\si{cm^{-1}}$) & ($\si{K}$) & (\A) & ($10^7 \  \si{s^{-1}}$) &  \\
            
            \noalign{\smallskip}
            \cline{1-8} 
            \noalign{\smallskip}
            
            3p$^6$3d$^9$  & $^2$D & 5/2 & 0 & 0 & 1703.41 & 1.86 & 0.032 \\
            &&&&&                                 1709.61 & 7.25 & 0.191 \\
            &&&&&                                 1741.56 & 9.48 & 0.259 \\
            &&&&&                                 1751.92 & 4.55 & 0.168 \\
            &&&&&                                 1773.95 & 0.67 & 0.025 \\
            &&&&&                                 1804.47 & 0.71 & 0.028 \\
            
            \noalign{\medskip}

            && 3/2 & 1507 & 2168 & 1754.81 & 2.30 & 0.064 \\
            &&&&&                  1788.49 & 3.50 & 0.101 \\

            \noalign{\medskip}
            \cdashline{1-8}   
            \noalign{\medskip}

            3p$^6$3d$^8$($^3$F)4s & $^4$F & 9/2 & 8394 & 12077 & 2166.23 & 2.40 & 1.69 \\
            &&&&&                                         2217.17 & 3.40 & 3.01 \\
            &&&&&                                         2223.64 & 9.80 & 0.73 \\
            &&&&&                                         2316.75 & 2.88 & 1.85 \\

            \noalign{\medskip}

            && 7/2 & 9330 & 13424 & 2169.77 & 15.8 & 0.89 \\
            &&&&&                   2211.07 & 3.90 & 0.29 \\
            &&&&&                   2225.55 & 16.5 & 0.98 \\
            &&&&&                   2270.91 & 15.6 & 1.21 \\
            &&&&&                   2303.70 & 29.0 & 1.38 \\

            \noalign{\medskip}

            && 5/2 & 10116 & 14555 & 2175.82 & 17.7 & 0.75 \\
            &&&&&                    2207.41 & 16.6 & 0.97 \\
            &&&&&                    2227.02 & 13.0 & 0.58 \\
            &&&&&                    2265.16 & 14.3 & 0.88 \\

            \noalign{\medskip}

            && 3/2 & 10664 & 15343 & 2185.29 & 29.0 & 0.83 \\
            &&&&&                    2202.09 & 13.0 & 0.57 \\
            &&&&&                    2254.55 & 19.8 & 0.91 \\

            \noalign{\medskip}
            \cdashline{1-8}   
            \noalign{\medskip}

            3p$^6$3d$^8$($^3$F)4s & $^2$F & 7/2 & 13550 & 19496 & 2279.47 & 28.0 & 1.31 \\
            &&&&&                                         2297.26 & 19.8 & 1.25 \\
            &&&&&                                         2395.25 & 17.0 & 1.46 \\
            &&&&&                                         2438.63 & 5.40 & 0.48 \\
            &&&&&                                         2511.63 & 5.80 & 0.55 \\
            \noalign{\medskip}

            && 5/2 & 14996 & 21576 & 2287.79 & 28.0 & 0.88 \\
            &&&&&                    2298.98 & 28.0 & 1.33 \\
            &&&&&                    2416.87 & 21.0 & 1.47 \\
            &&&&&                    2546.66 & 1.56 & 0.12 \\

            \noalign{\smallskip}

            \cline{1-8}  

        \end{tabular} 
        
    \begin{tablenotes}
      \item[a] \small The level energies of \niii\ were obtained from the NIST database. For the line parameters ($\lambda_{lu}$, $\A_{lu}$, $f_{lu}$), we used the values from Boissé \& Bergeron (2019) for the ground state, Fedchak \& Lawler (1999) for the $1507 \ \si{cm^{-1}}$ state, and the NIST database for higher levels.
    \end{tablenotes}
    
\label{Tab. list lines Ni II}

\end{threeparttable}
\end{table}

\newpage
\section{Energy distribution of \feplus\ under various physical conditions}

\label{Appendix - excitation diagram of Fe II}

\begin{figure*}[h!]
\centering
    \includegraphics[scale = 0.45,     trim = 10 0 30 10,clip]{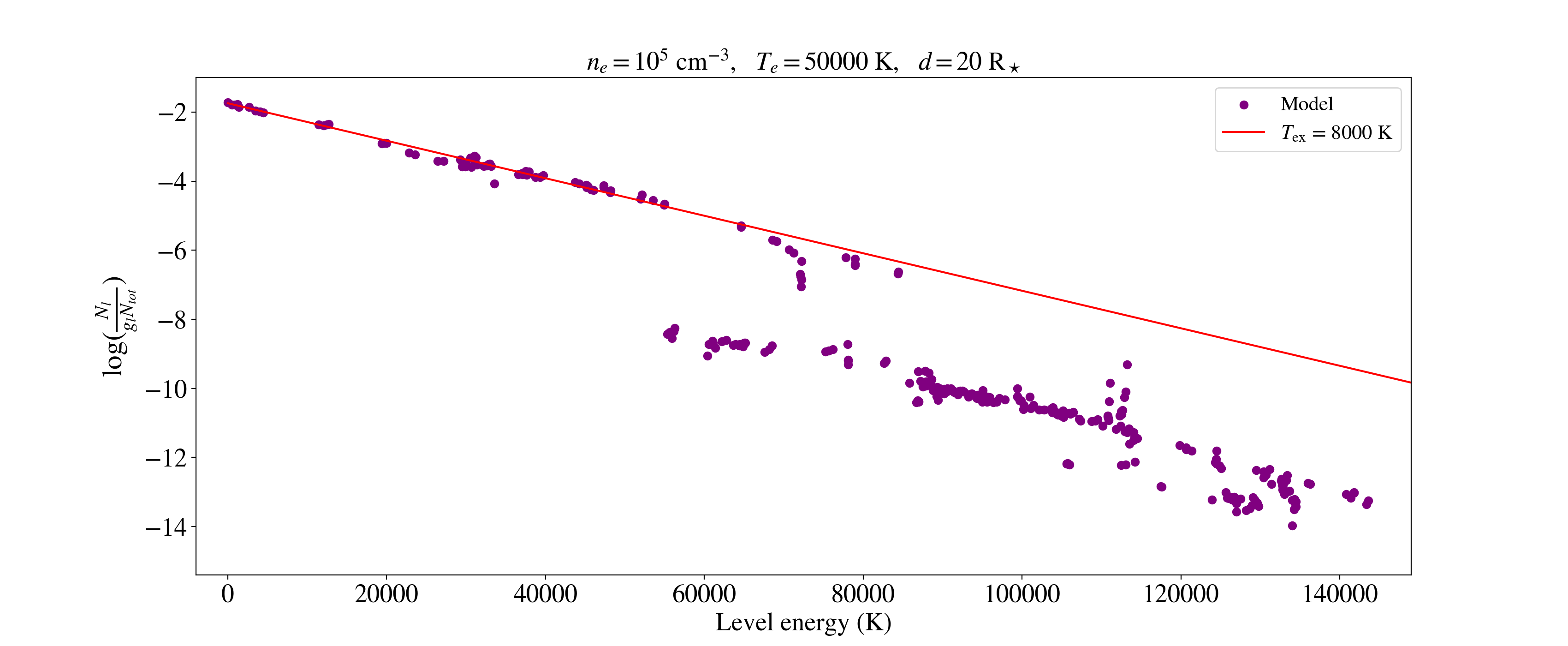}  
    
    \includegraphics[scale = 0.45,     trim = 10 0 30 10,clip]{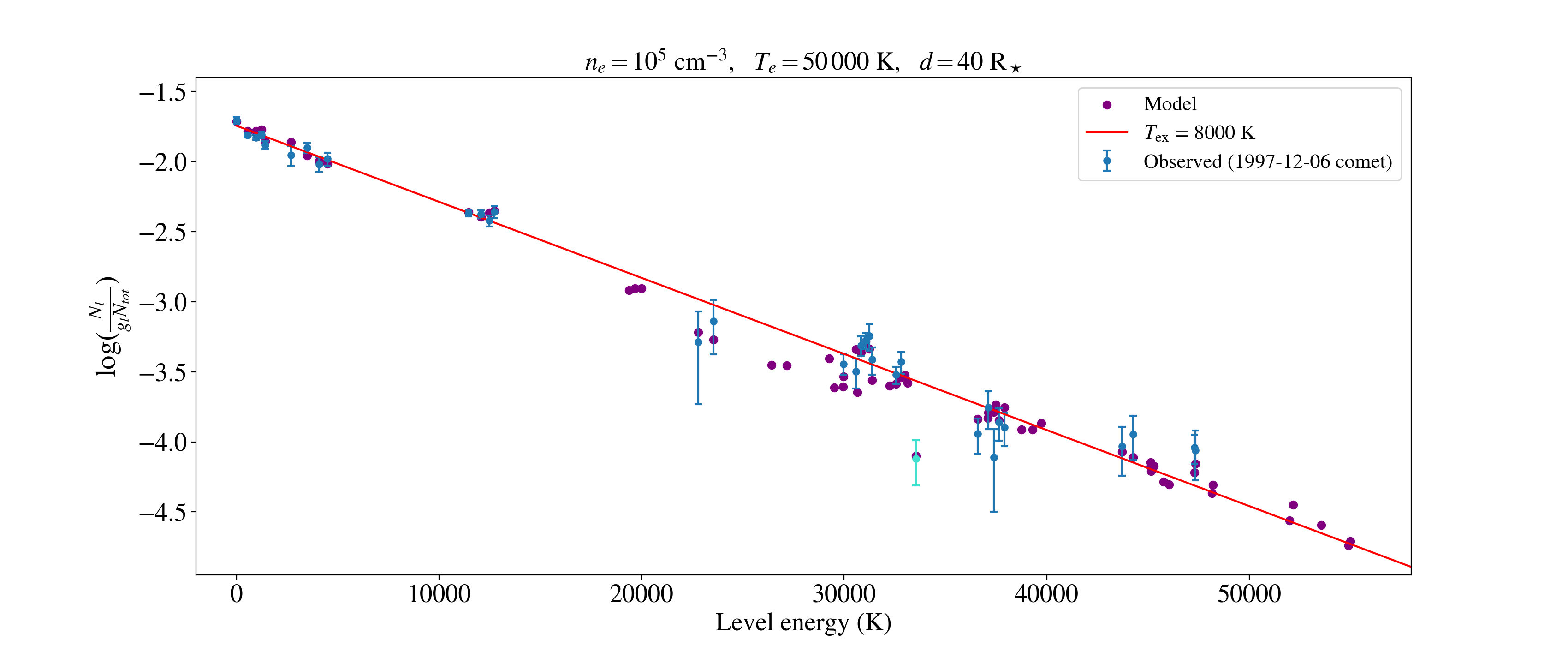}  
    
    \caption{Results from our \feplus\ excitation model in the radiative regime (Sect. \ref{Sect. Radiative regime}), for $n_e = 10^5 \ \si{cm^{-3}}$, $T_e = 50\,000 \ \si{K}$, $d = 20$ R$_\star$. \textbf{Top}: Calculated relative abundances for all 340 \feii\ levels. The two branches are clearly visible: the less excited branch corresponds to low excitation energies ($\leq 80\,000 \ \si{K}$) and high abundances ($\frac{N_l}{g_l N_{\rm tot}} \sim 10^{-6}$ to $10^{-2}$), while the excited branch is characterised by high energies ($\geq 55\,000 \ \si{K}$) and strong depletion ($\frac{N_l}{g_l N_{\rm tot}} \leq 10^{-8}$). Here, the effect of collisions is negligible, due to the low electronic density. The calculated energy distribution is thus not impacted by the electronic temperature. 
    \textbf{Bottom}: Comparison between calculated abundances and measurements from various \feplus\ excitation energy levels observed in the December~6, 1997 comet. In the radiative regime, we obtain a fairly good agreement between the theoretical results and the measurements. The peculiar a$^6$S$_{5/2}$ level (33\,550 K) is highlighted in turquoise colour.}
    \label{Fig. Result excitation Radiative}
\end{figure*}

\begin{figure*}[h!]
\centering
    \includegraphics[scale = 0.45,     trim = 10 0 30 10,clip]{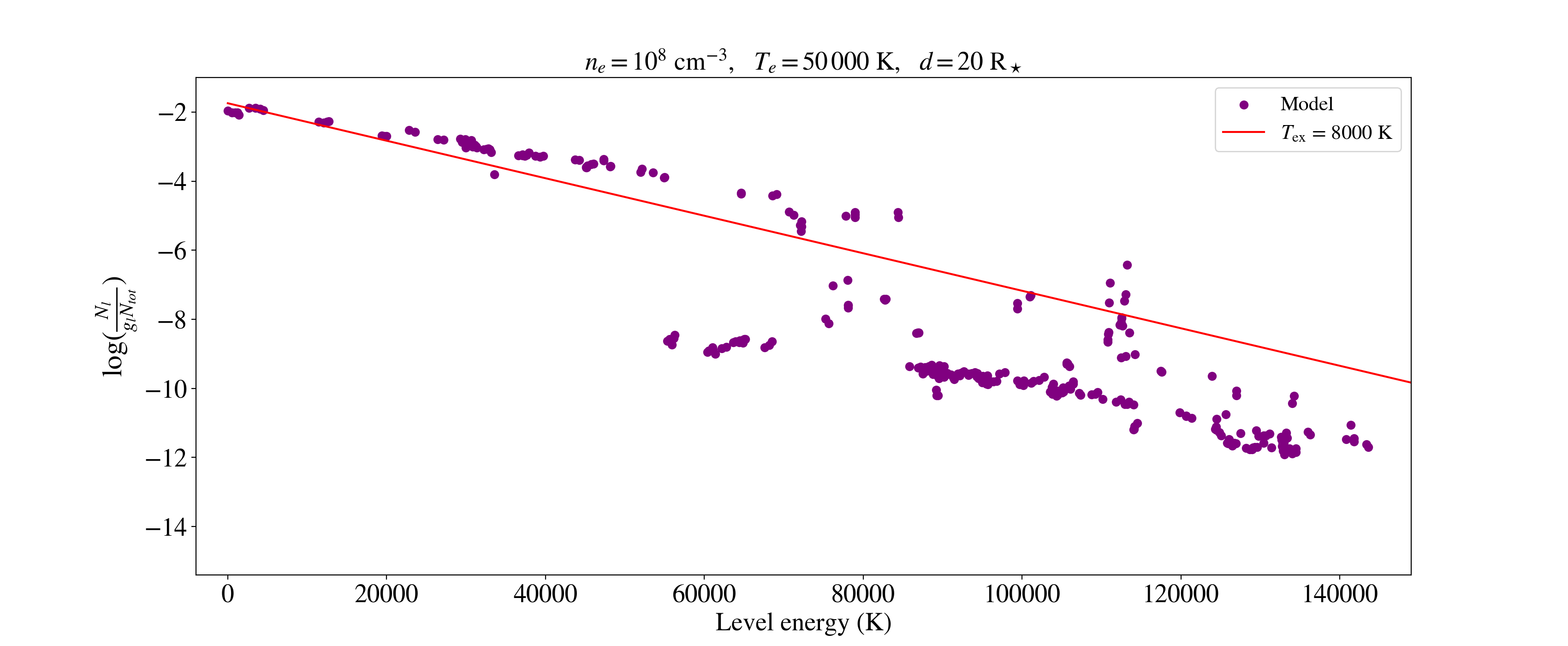}  
    
    \includegraphics[scale = 0.45,     trim = 10 0 30 10,clip]{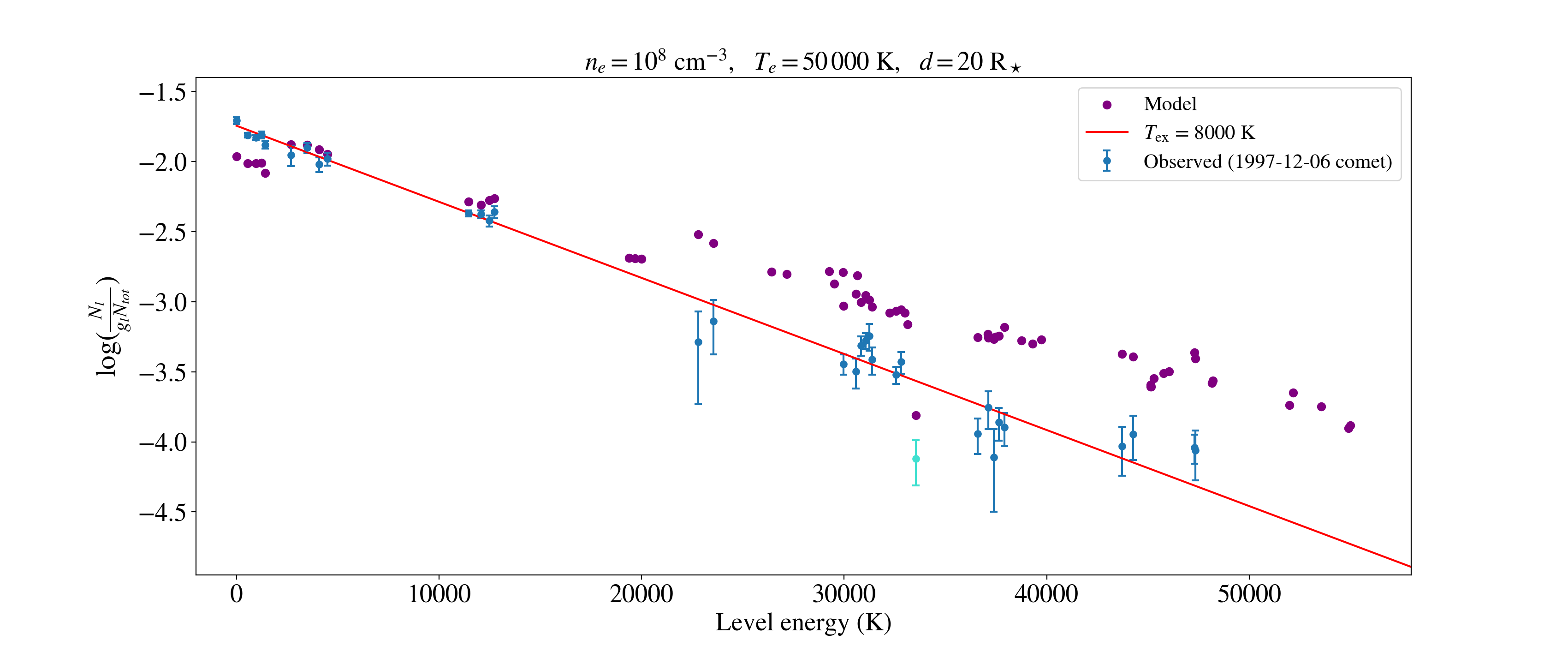}  
    
    \caption{Results from our \feplus\ excitation model in the semi-collisional regime (Sect. \ref{Sect. Semi-collisional regime}), for $n_e = 10^8 \, \si{cm^{-3}}$, $T_e = 50\,000 \ \si{K}$, $d = 20$ R$_\star$. \textbf{Top}: Calculated relative abundances for all 340 \feii\ levels. Due to electronic collisions, the excitation temperature for low-lying levels is increased in comparison to the lower density case with $n_e = 10^5 \, \si{cm^{-3}}$ (Fig. \ref{Fig. Result excitation Radiative}).
    \textbf{Bottom}: Comparison between calculated abundances and measurements from various \feplus\ excitation energy levels observed in the December~6, 1997 comet. In this case, the outcome of the model is highly inconsistent with our measurements. }
    \label{Fig. Result excitation Semi-col 1}
\end{figure*}

\begin{figure*}[h!]
\centering
    \includegraphics[scale = 0.45,     trim = 10 0 30 10,clip]{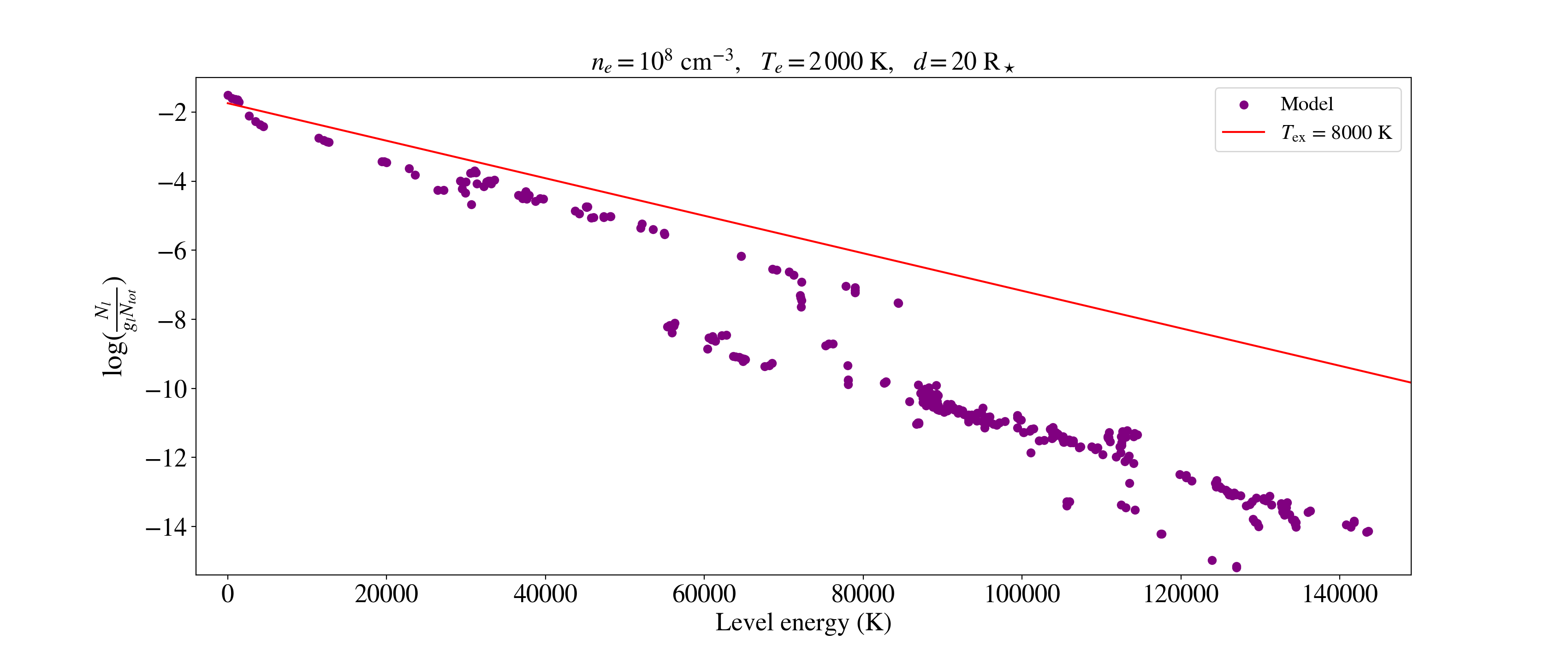}  
    
    \includegraphics[scale = 0.45,     trim = 10 0 30 10,clip]{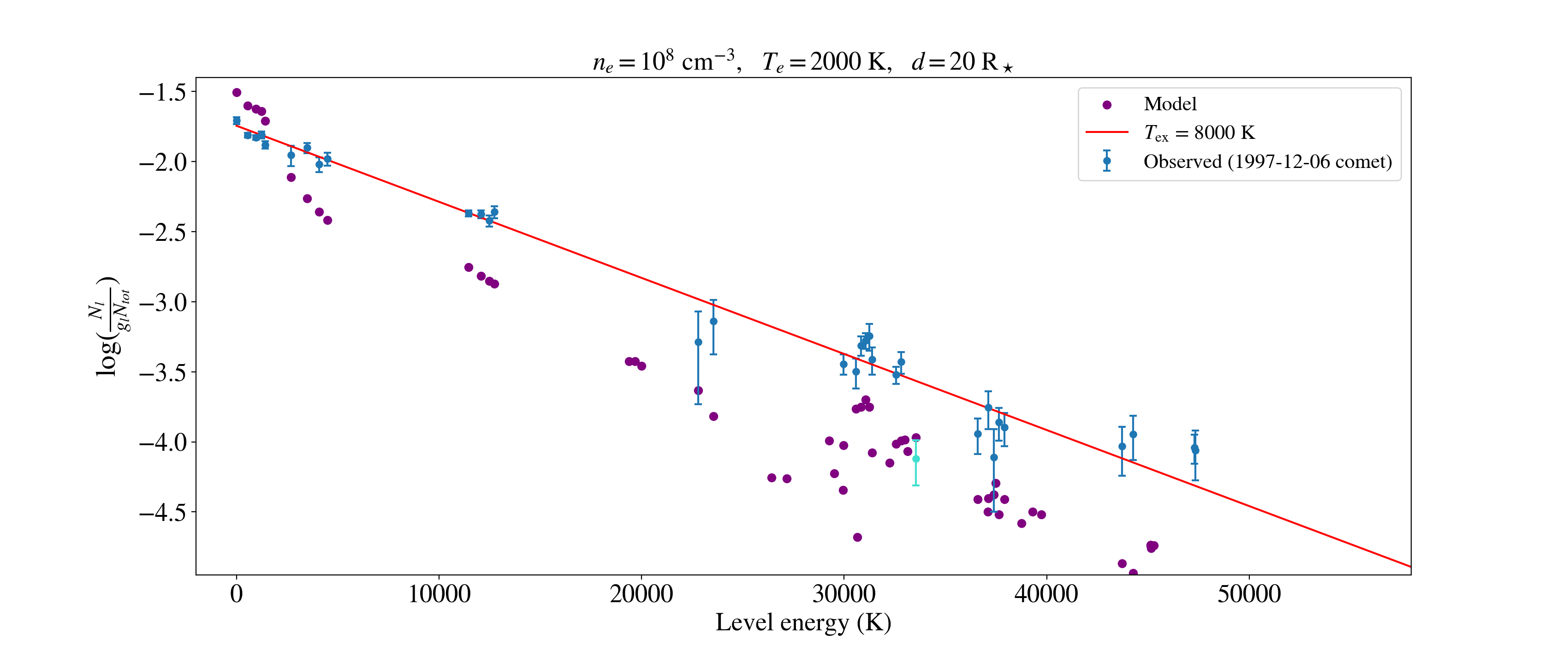}

    \caption{Results from our \feplus\ excitation model in the semi-collisional regime (Sect. \ref{Sect. Semi-collisional regime}), for $n_e = 10^8 \ \si{cm^{-3}}$, $T_e = 2\,000 \ \si{K}$, $d = 20$ R$_\star$. \textbf{Top}: Calculated relative abundances for all 340 \feii\ levels. Here, contrary to Fig. \ref{Fig. Result excitation Semi-col 1}, collisions with electrons tend to reduce the excitation temperature of the low-excitation branch, as the kinetic temperature ($2000 \ \si{K}$) is lower than the stellar effective temperature ($8000 \ \si{K}$). 
    \textbf{Bottom}: Comparison between calculated abundances and measurements from various \feplus\ excitation energy levels observed in the December~6, 1997 comet. Similarly to Fig. \ref{Fig. Result excitation Semi-col 1}, the agreement between our calculations and the observed \feii\ energy distribution is very poor. }
    \label{Fig. Result excitation Semi-col 2}
\end{figure*}

\begin{figure*}[h!]
\centering
    \includegraphics[scale = 0.45,     trim = 10 0 30 10,clip]{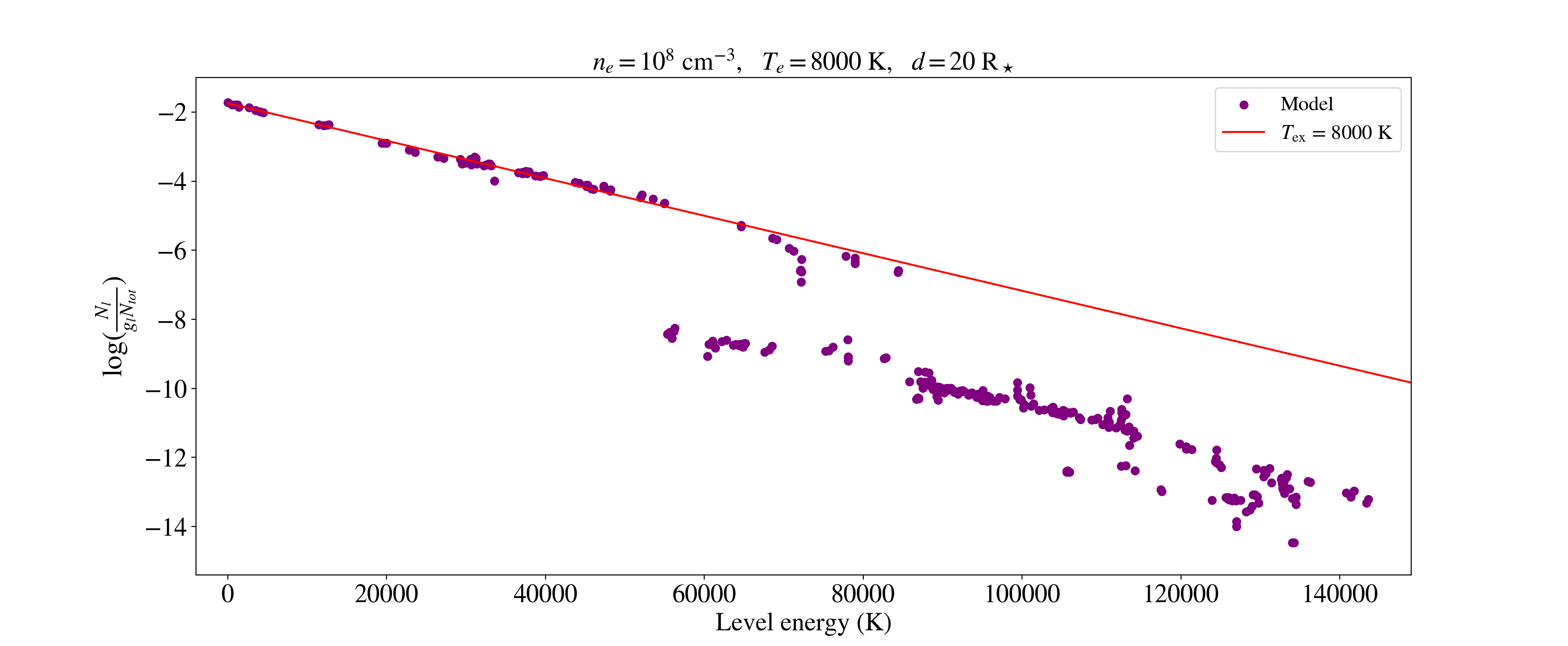}  
    
    \includegraphics[scale = 0.45,     trim = 10 0 30 10,clip]{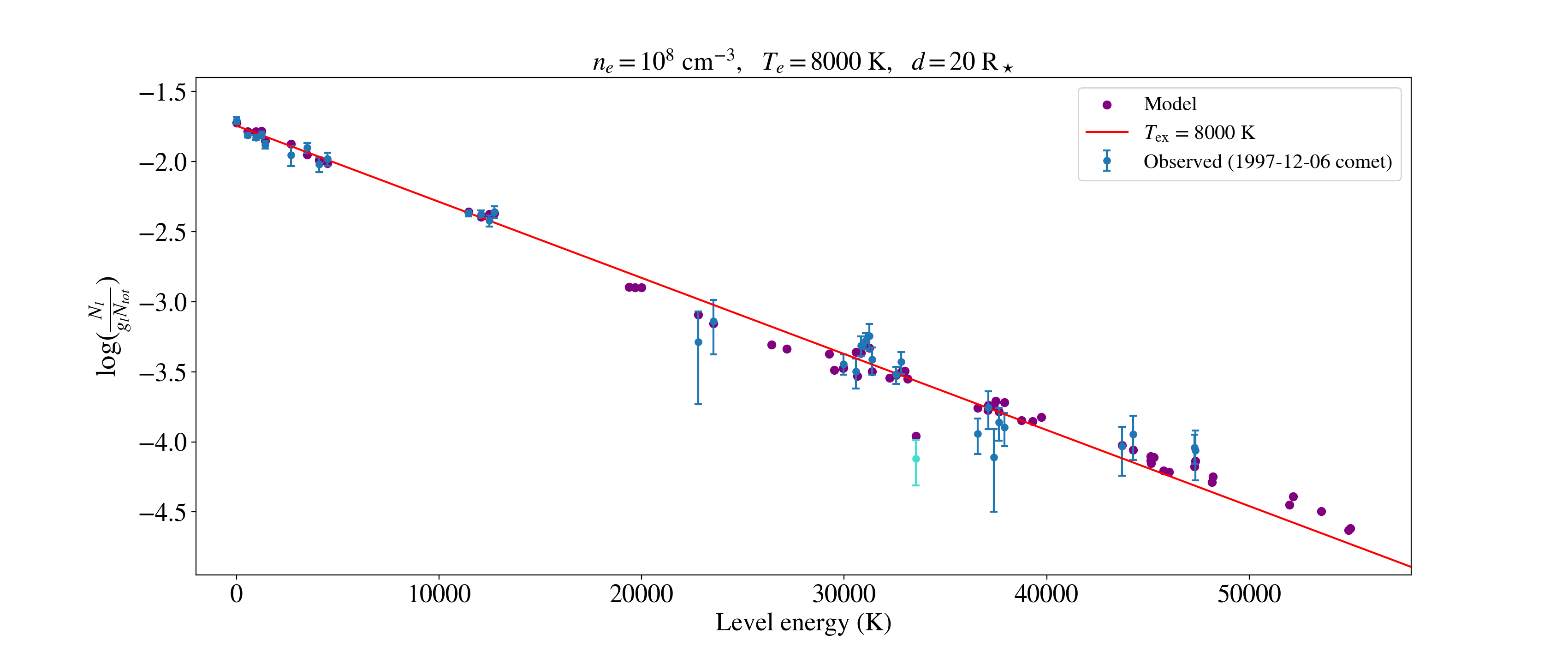}  
    
    \caption{Results from our \feplus\ excitation model in the semi-collisional regime (Sect. \ref{Sect. Semi-collisional regime}), for $n_e = 10^8 \ \si{cm^{-3}}$, $T_e = 8000 \ \si{K}$, $d = 20$ R$_\star$. \textbf{Top}: Calculated relative abundances for all 340 \feii\ levels. Here, the electronic temperature ($8000 \ \si{K}$) is equal to the stellar effective temperature, leading to a peculiar regime: both radiation and collision tend to populate \feplus\ at the same excitation temperature. The effect of collision is thus here barely noticeable, contrary to the previous semi-collisional examples (Fig. \ref{Fig. Result excitation Semi-col 1} and \ref{Fig. Result excitation Semi-col 2}).
    \textbf{Bottom}:  Comparison between calculated abundances and measurements from various \feplus\ excitation energy levels observed in the December~6, 1997 comet. In this peculiar case, we obtain a fairly good agreement between the theoretical results and the measurements.}
    \label{Fig. Result excitation Semi-col 3}
\end{figure*}

\begin{figure*}[h!]
\centering
    \includegraphics[scale = 0.45,     trim = 10 0 30 10,clip]{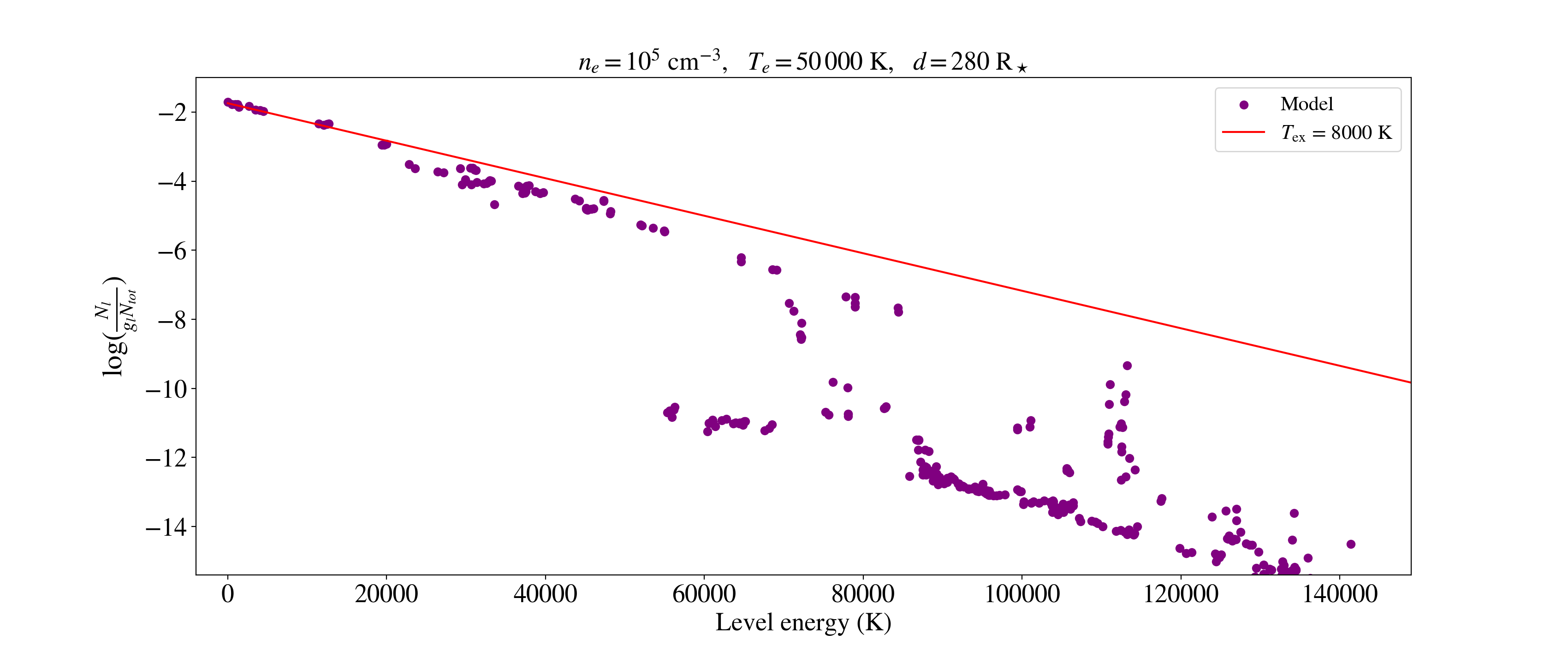}  
    
    \includegraphics[scale = 0.45,     trim = 10 0 30 10,clip]{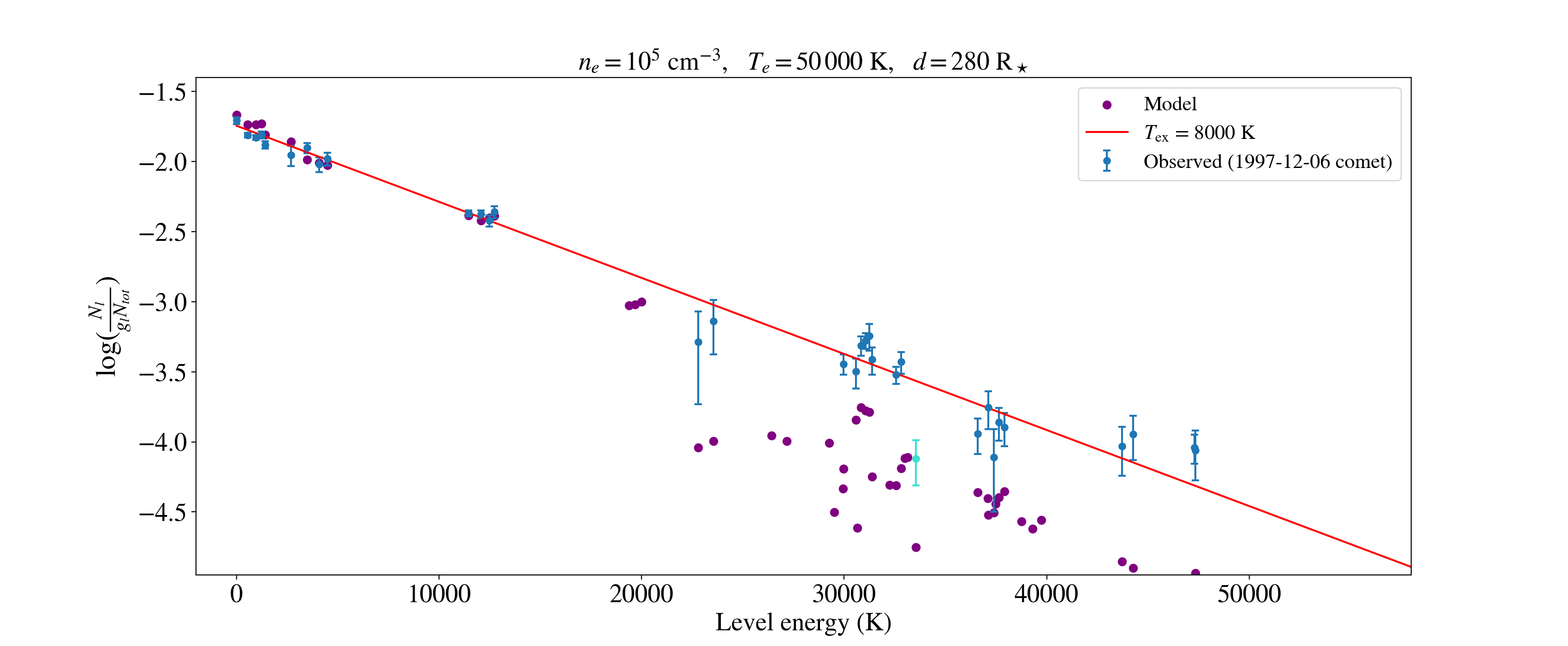}  

    \caption{Results from our \feplus\ excitation model in the large distance regime (Sect. \ref{Sect. Large distance regime}), for $n_e = 10^5 \ \si{cm^{-3}}$, $T_e = 50\,000 \ \si{K}$, $d = 280$ R$_\star = 2$ au. \textbf{Top}: Calculated abundances for all 340 \feii\ levels. The result is somehow similar to the radiative regime (Fig. \ref{Fig. Result excitation Radiative}), but here spontaneous emission in forbidden transitions start to play a significant role, depleting all excitation level to the benefit of the ground state. 
    \textbf{Bottom}: Comparison between calculated abundances and measurements from the December 6, 1997 comet. The model agreement is poorer than in the radiative regime.}
    \label{Fig. Result excitation large distance}
\end{figure*}

\FloatBarrier
\newpage
\section{$\rchi^2$ maps for various comet-to-star distances}

\begin{figure*}[h!]
\centering
    \includegraphics[scale = 0.34,     trim = 50 0 60 30,clip]{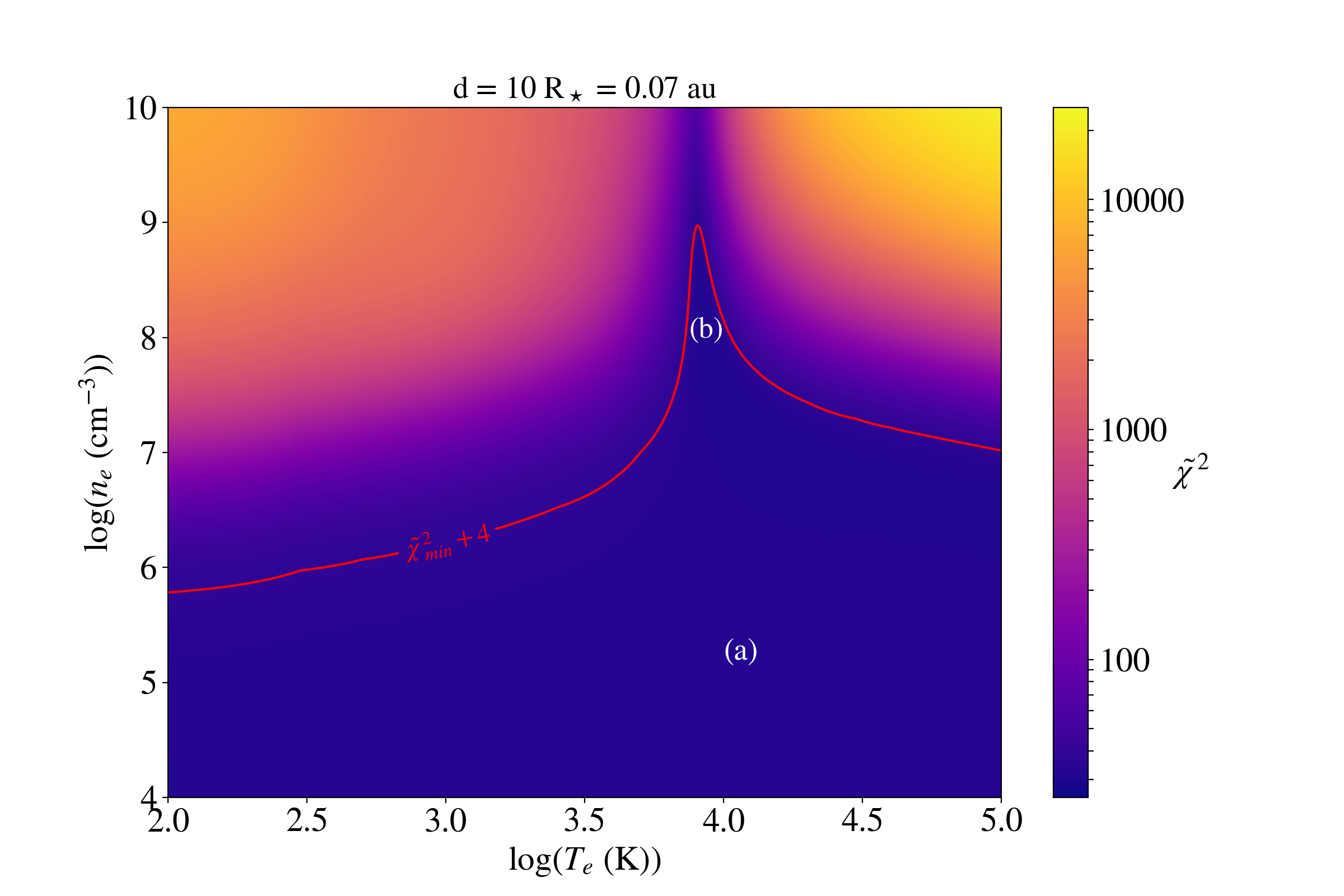}    
    \includegraphics[scale = 0.34,     trim = 50 0 60 30,clip]{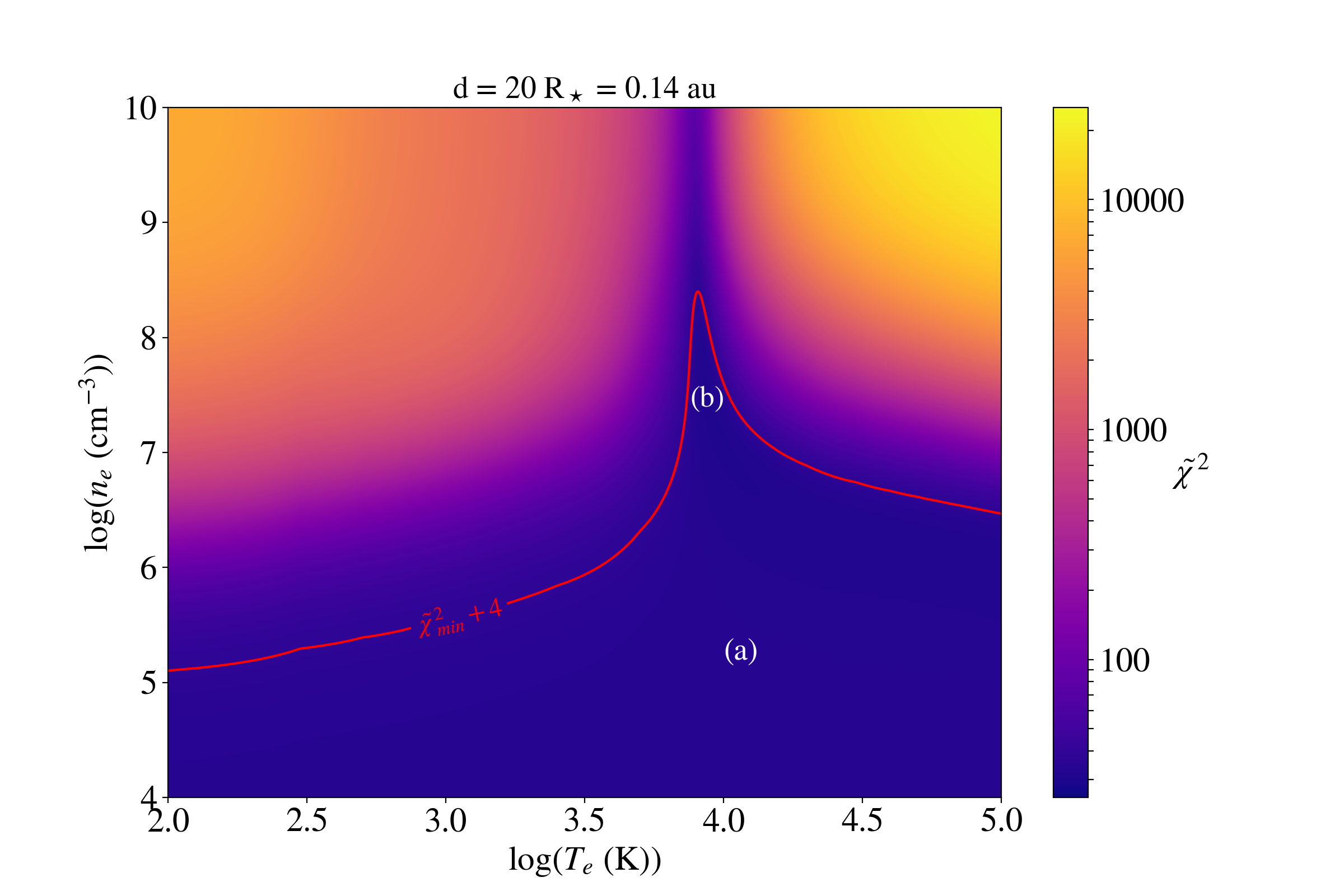}   

    \includegraphics[scale = 0.34,     trim = 50 0 60 30,clip]{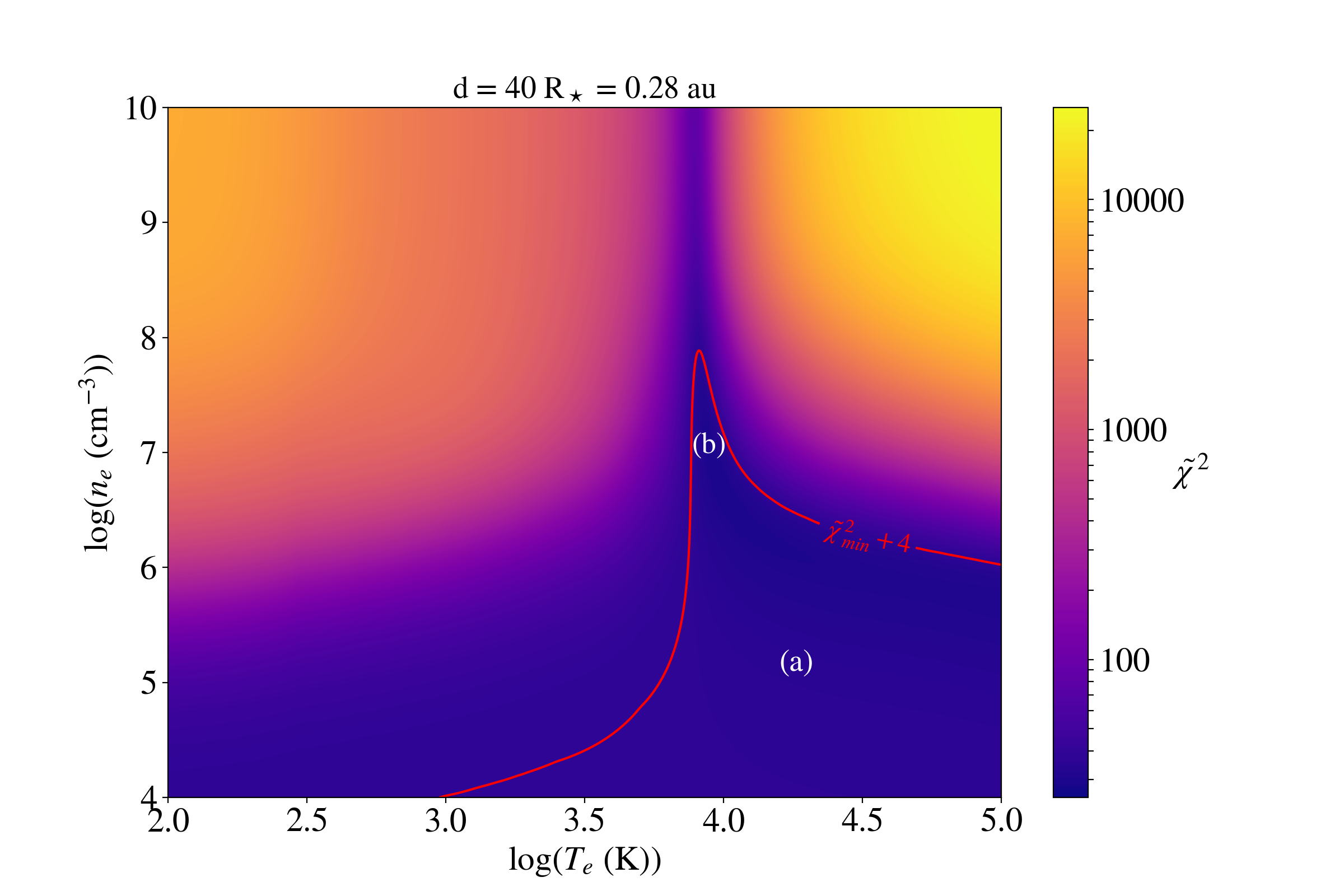}     
    \includegraphics[scale = 0.34,     trim = 50 0 60 30,clip]{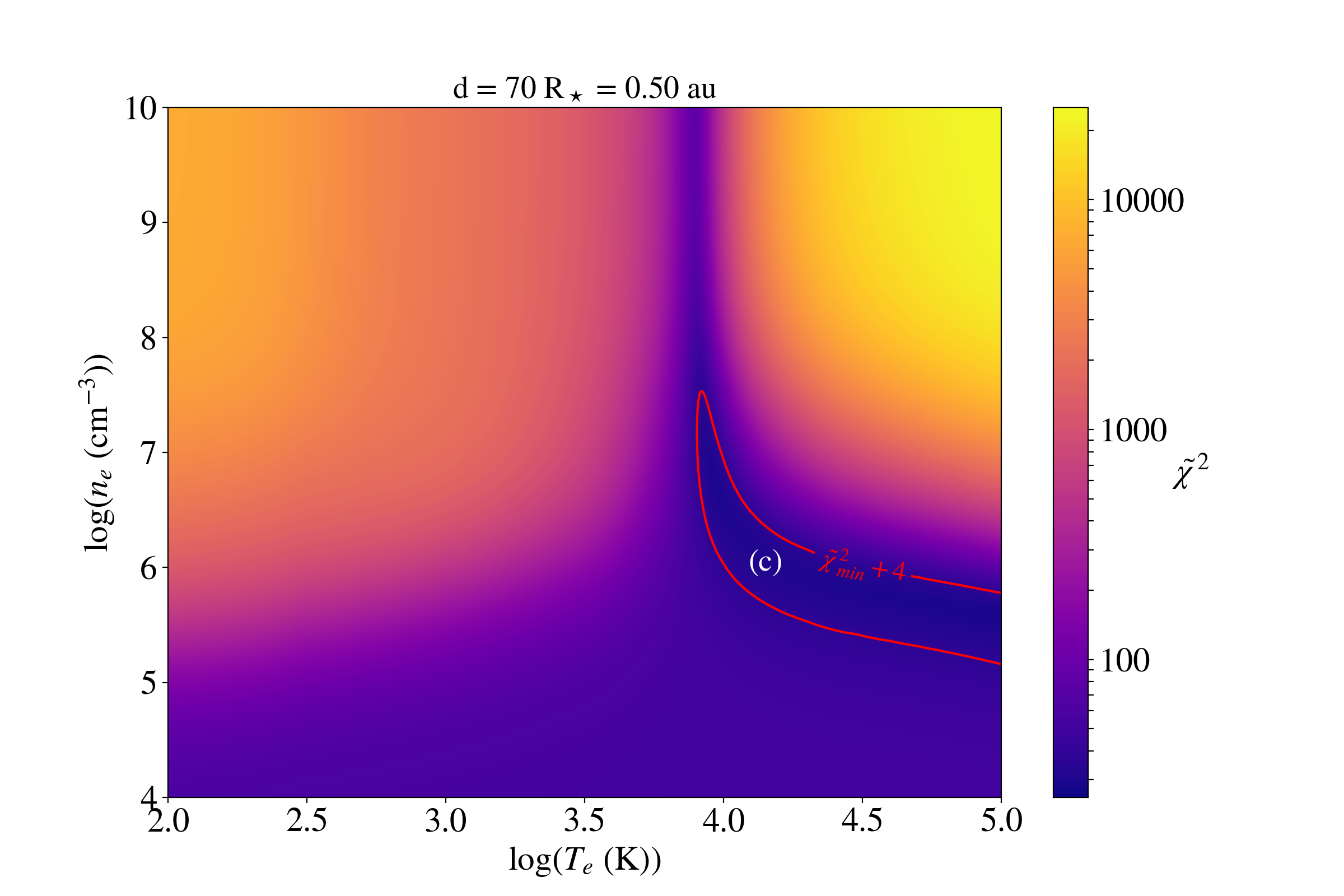}      

    \includegraphics[scale = 0.34,     trim = 50 0 60 30,clip]{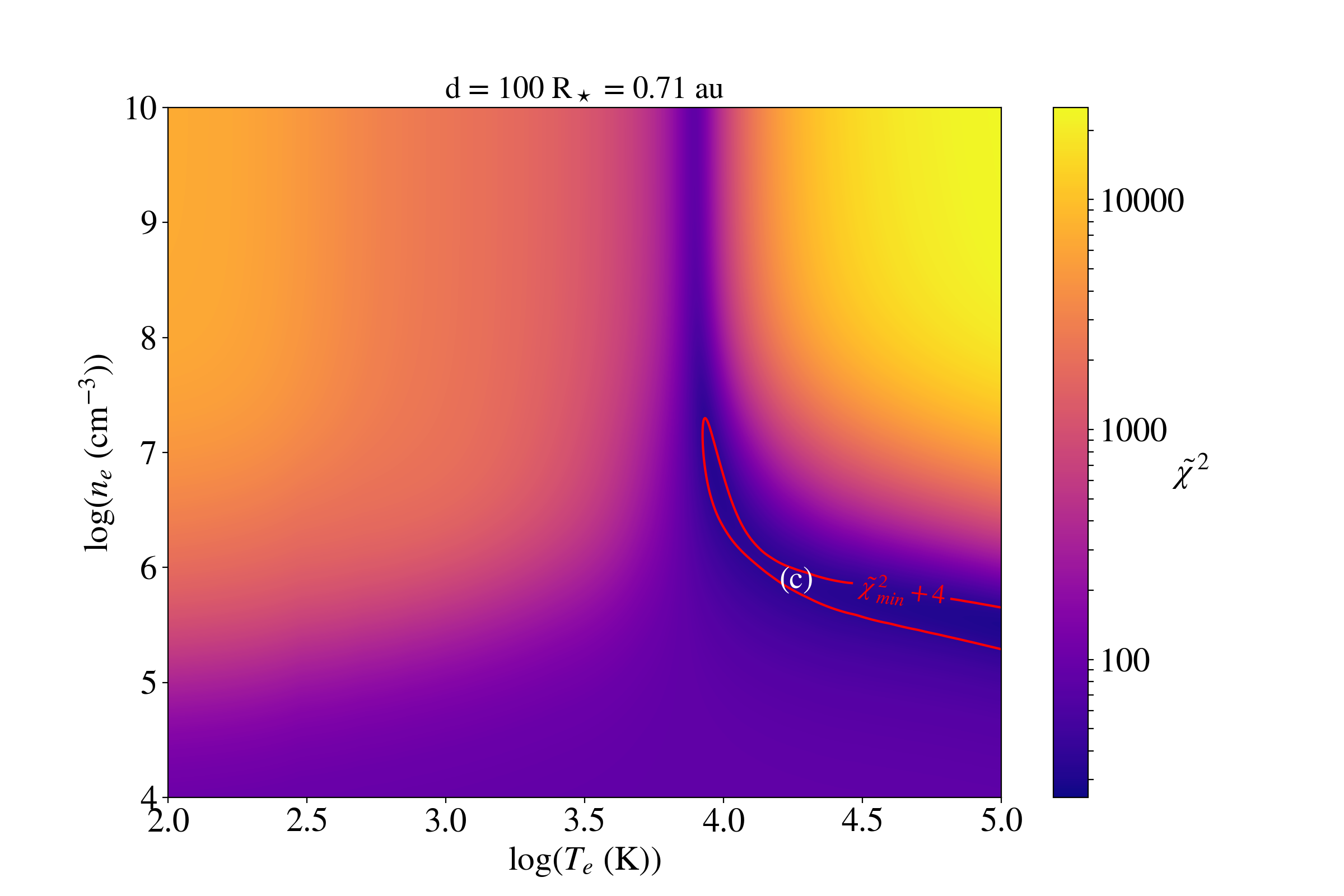}      
    \includegraphics[scale = 0.34,     trim = 50 0 60 30,clip]{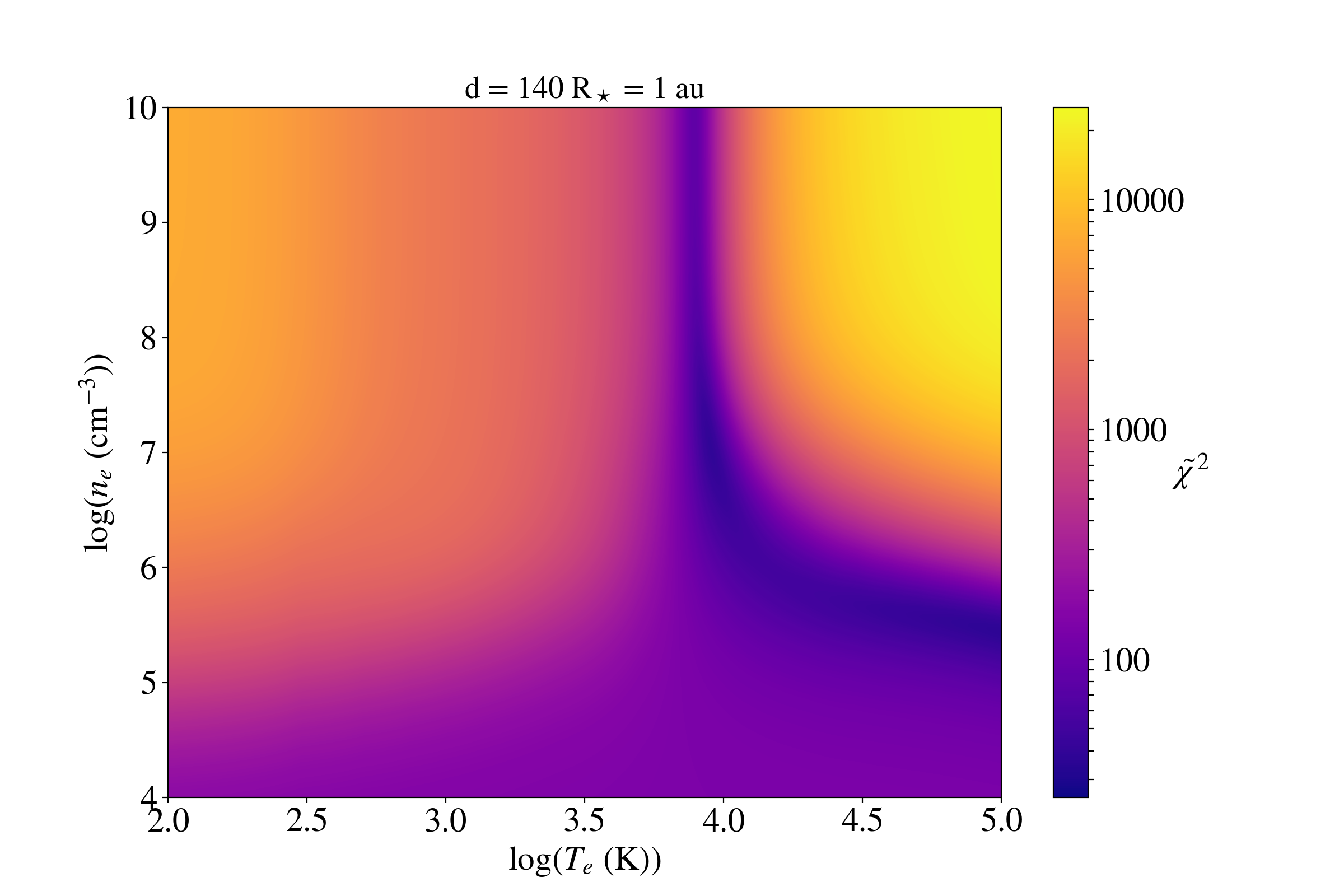}      

    \vspace{0.3 cm}
    \caption{$\rchi^2$ maps calculated in Sect. \ref{Sect. Comparison with data}, characterising the agreement between the observed \feplus\ excitation diagram in the December 6, 1997 comet, and the results from our excitation model. The six maps were calculated at different comet-to-star distances ($d$). The x-axis shows the electronic temperature within the comet ($T_e$), the y-axis shows its electronic density ($n_e$), and  different colours indicate the $\chi^2$ value for the corresponding parameters ($n_e$, $T_e$, $d$). The acceptable parameter space in shown with deep blue colour; the red line shows the $\tilde \rchi_{min}^2 + 4$ contour, associated with a 95 \% confidence. Letters indicate the different regimes compatible with our data (see Sect. \ref{Sect. Comparison with data}): a radiative regime associated with a low electronic density (a), a semi-collisional regime at an electronic temperature close to 8000 K (b), and a large distance regime with $n_e \simeq 10^6 \ \si{cm^{-3}}$ and $T_e \geq 10^4$~K (c).}
    \label{Fig. Chi2 maps}
\end{figure*}

\FloatBarrier
\newpage
\onecolumn

\section{View of the studied spectra}

\begin{figure*}[h!]
\centering
    \includegraphics[scale = 0.28,     trim = 80 27 80 30,clip]{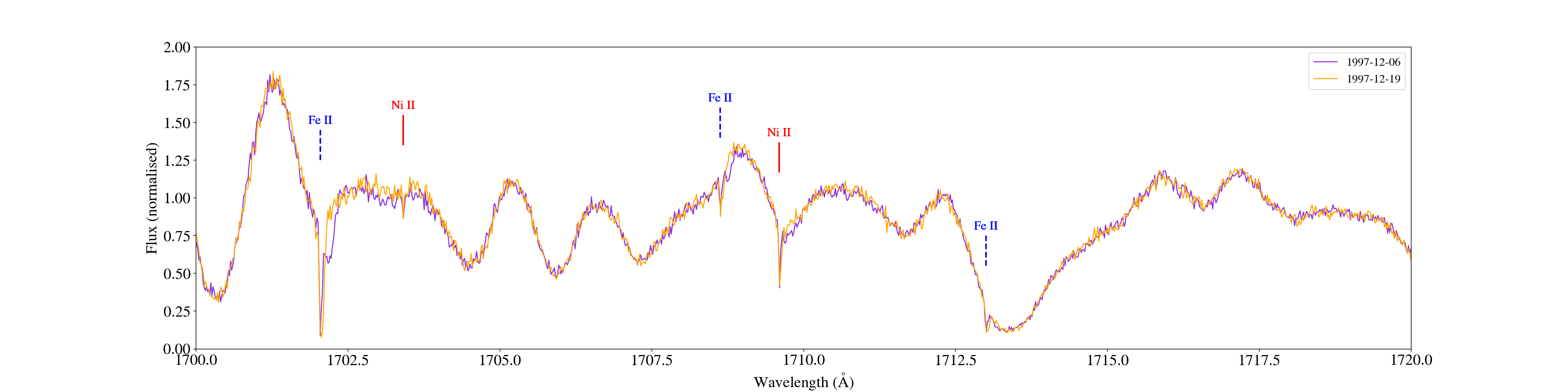}    
    \includegraphics[scale = 0.28,     trim = 80 27 80 30,clip]{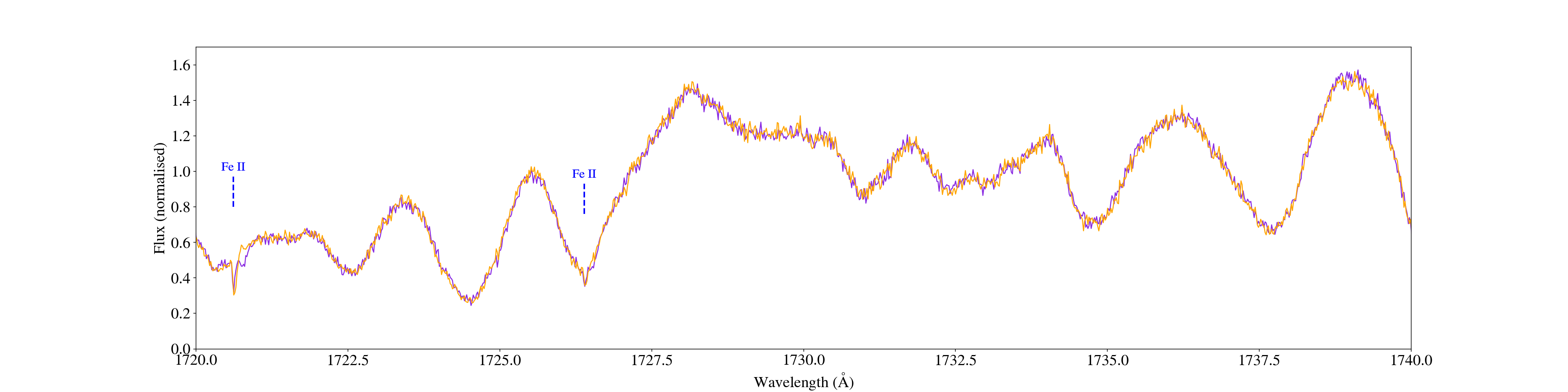}    
    \includegraphics[scale = 0.28,     trim = 80 27 80 30,clip]{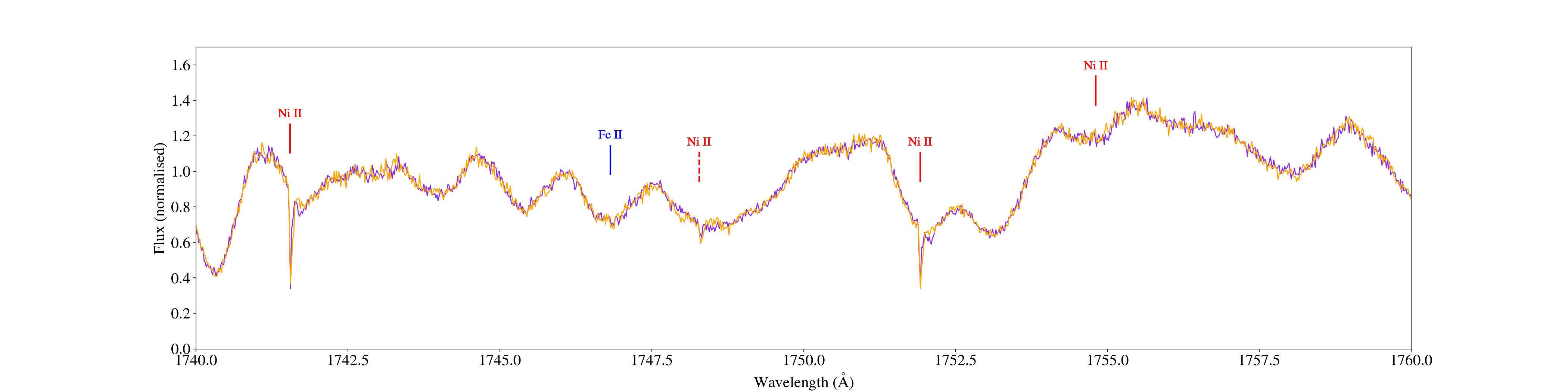}    
    \includegraphics[scale = 0.28,     trim = 80 0 80 30,clip]{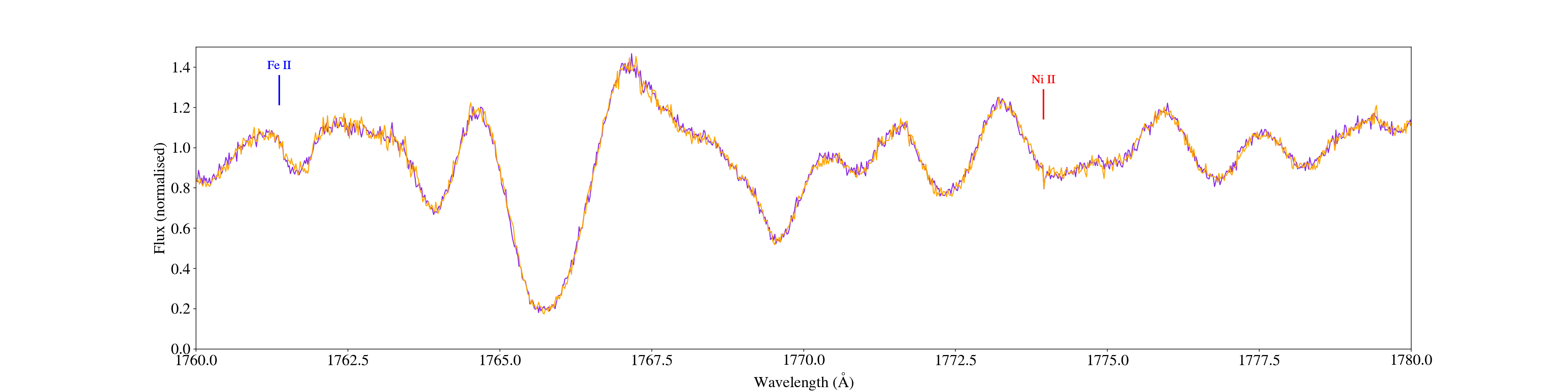}    
    \caption{Extended view of the two \bp\ spectra obtained in 1997 (December 6 and December 19). The \feii, \niii\ and \crii\ lines used in the present study are indicated with solid lines and different colours. Dotted lines indicate lines not included in our study, associated with various species (\feii, \fei, \mnii...). The redshifted absorption of the December 6, 1997 comet is clearly visible in most of the spectral lines}
    \label{Fig. Full spectrum}
\end{figure*}

\newpage

\begin{figure*}[h!]
\centering
    \includegraphics[scale = 0.28,     trim = 80 27 80 30,clip]{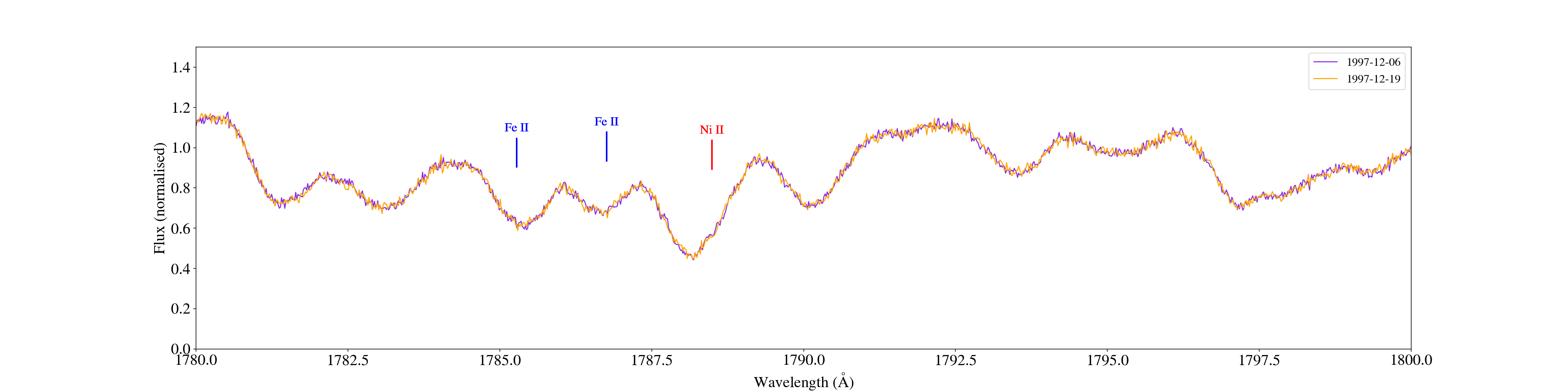}    
    \includegraphics[scale = 0.28,     trim = 80 27 80 30,clip]{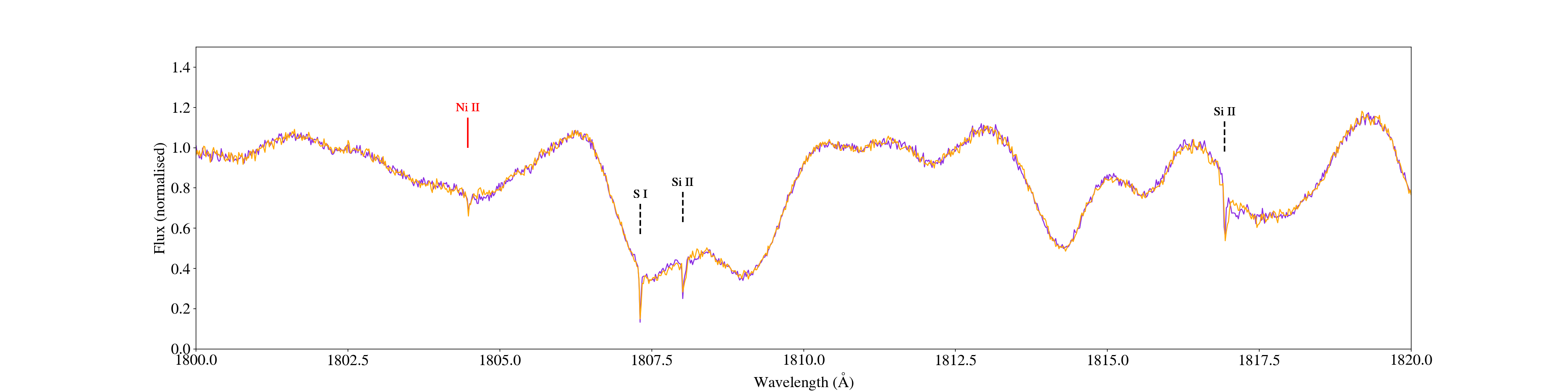}    
    \includegraphics[scale = 0.28,     trim = 80 27 80 30,clip]{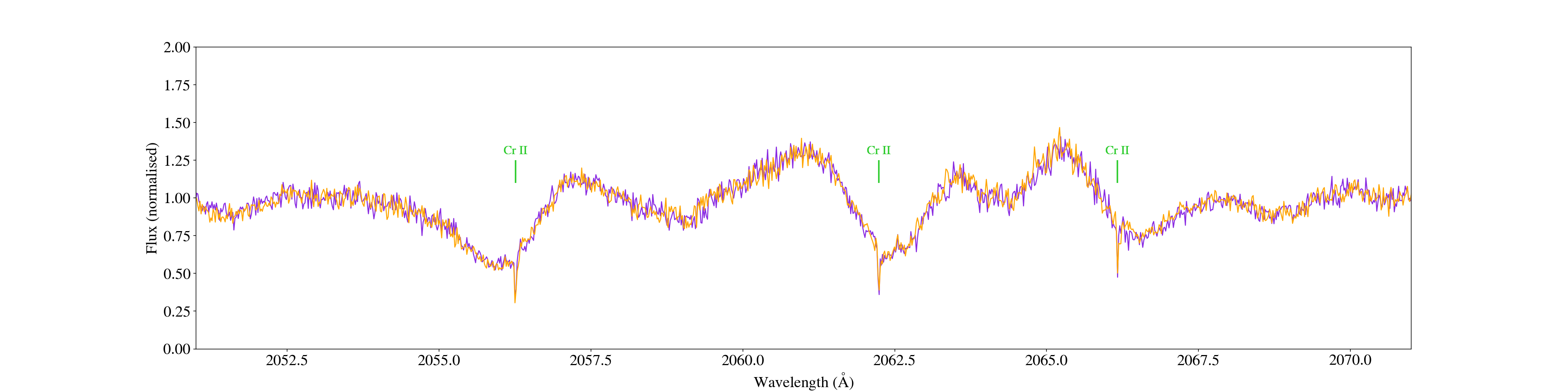}    
    \includegraphics[scale = 0.28,     trim = 80 27 80 30,clip]{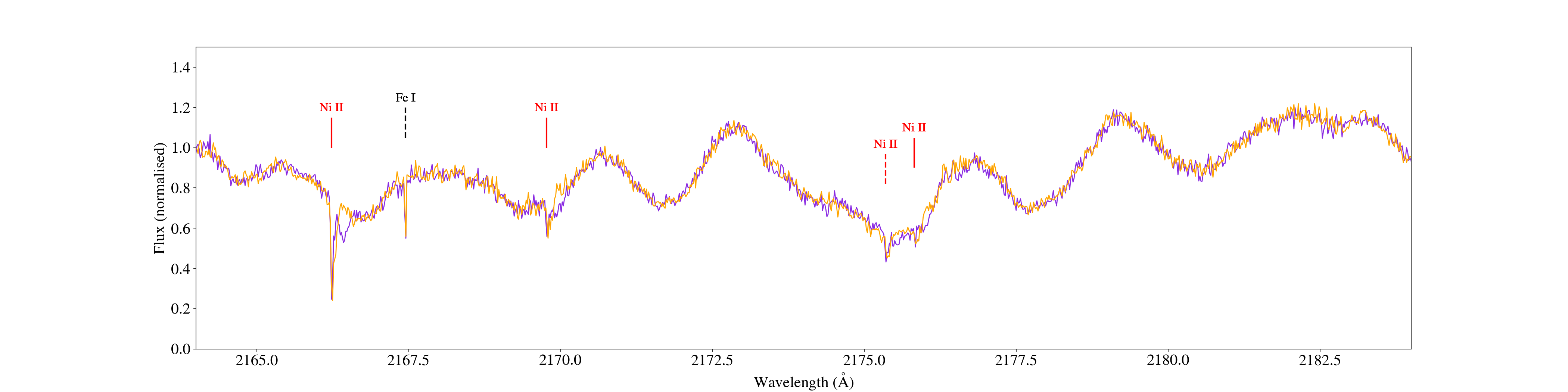}    
    \includegraphics[scale = 0.28,     trim = 80 0 80 30,clip]{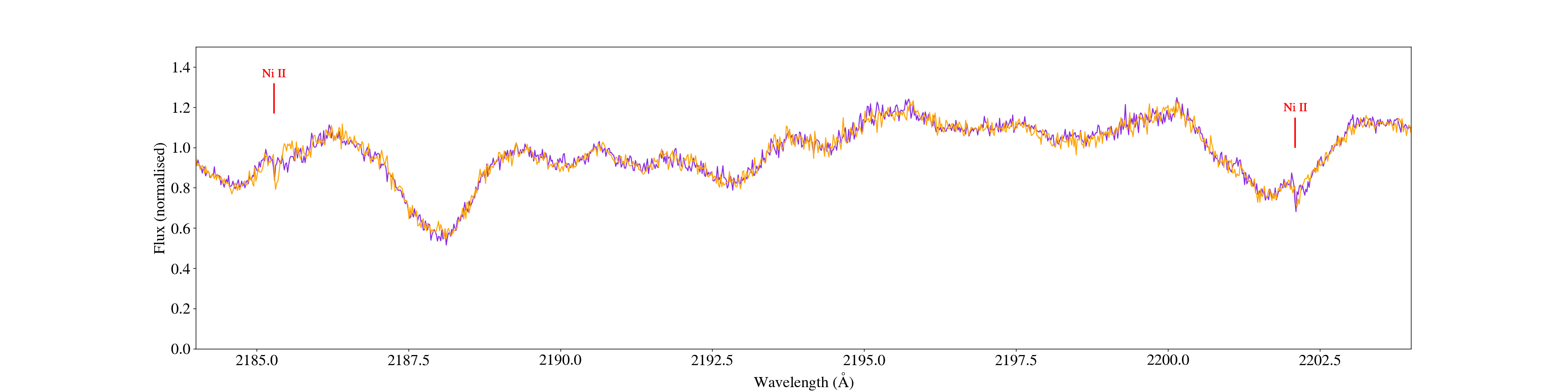}    
\end{figure*}
\small \textbf{Figure G.1}, continued.

\newpage

\begin{figure*}[h!]
\centering
    \includegraphics[scale = 0.28,     trim = 80 27 80 30,clip]{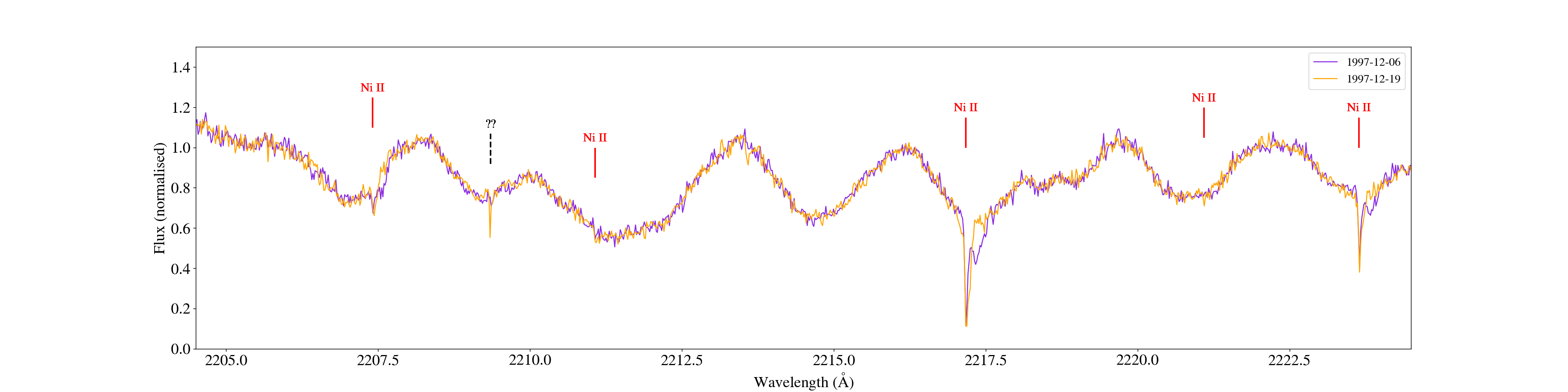}    
    \includegraphics[scale = 0.28,     trim = 80 27 80 30,clip]{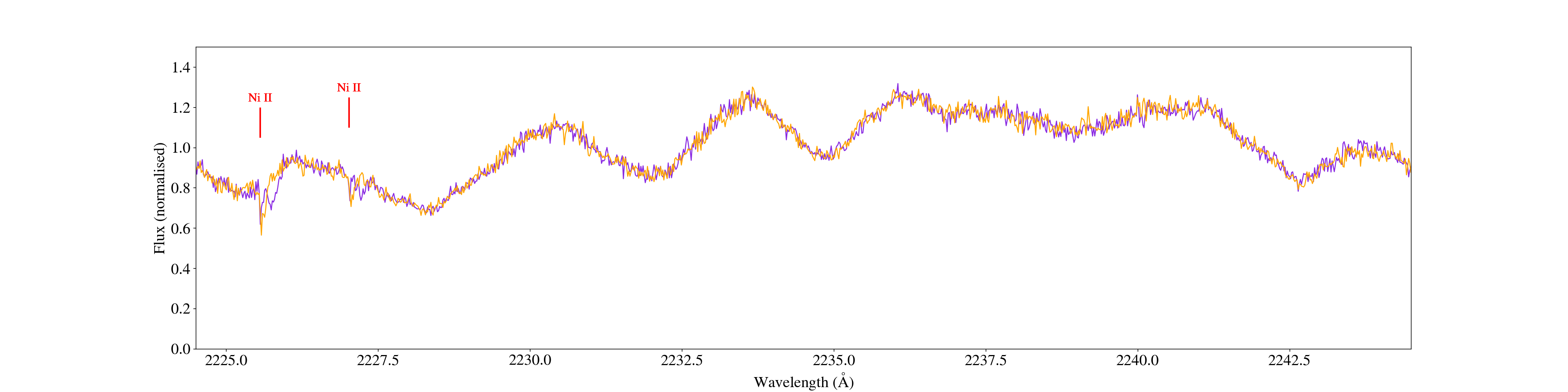}    
    \includegraphics[scale = 0.28,     trim = 80 27 80 30,clip]{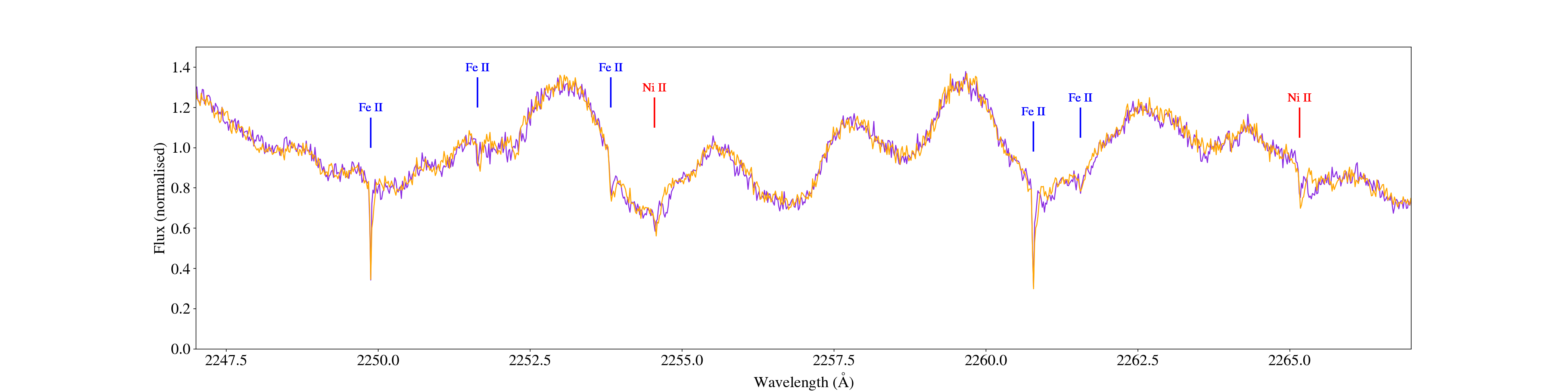}    
    \includegraphics[scale = 0.28,     trim = 80 27 80 30,clip]{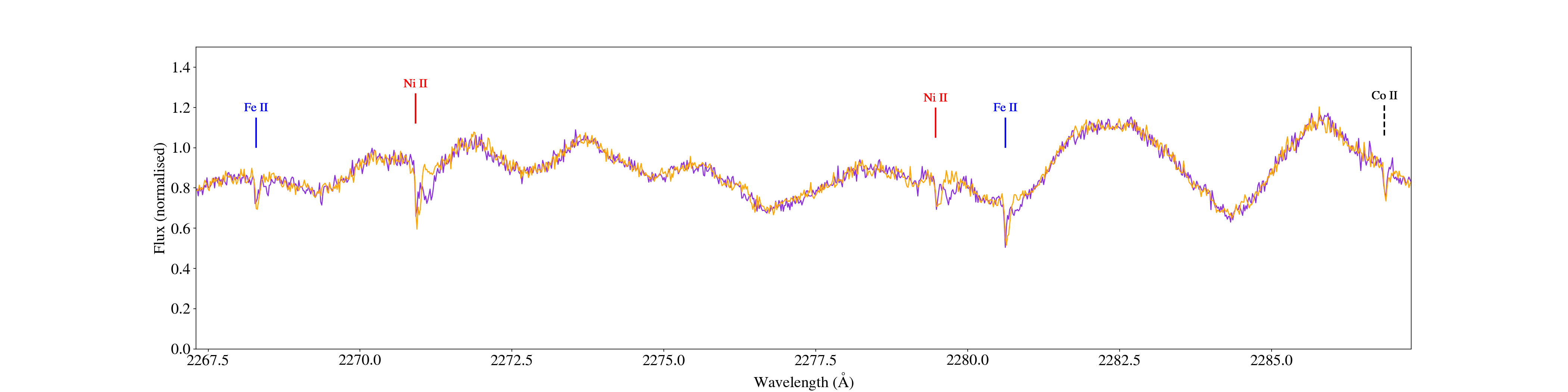}    
    \includegraphics[scale = 0.28,     trim = 80 0 80 30,clip]{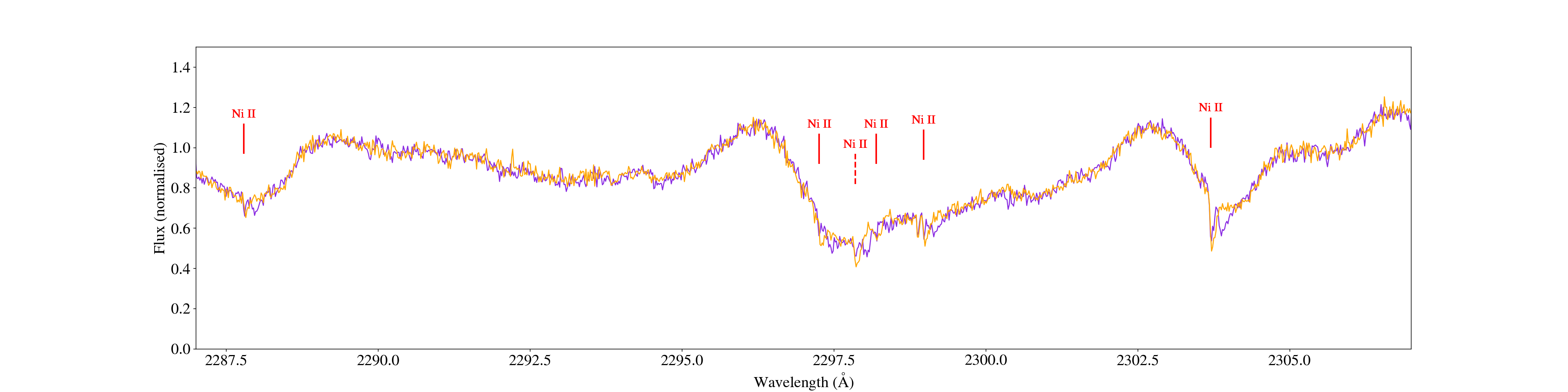}    
\end{figure*}
\small \textbf{Figure G.1}, continued.

\newpage

\begin{figure*}[h!]
\centering
    \includegraphics[scale = 0.28,     trim = 80 27 80 30,clip]{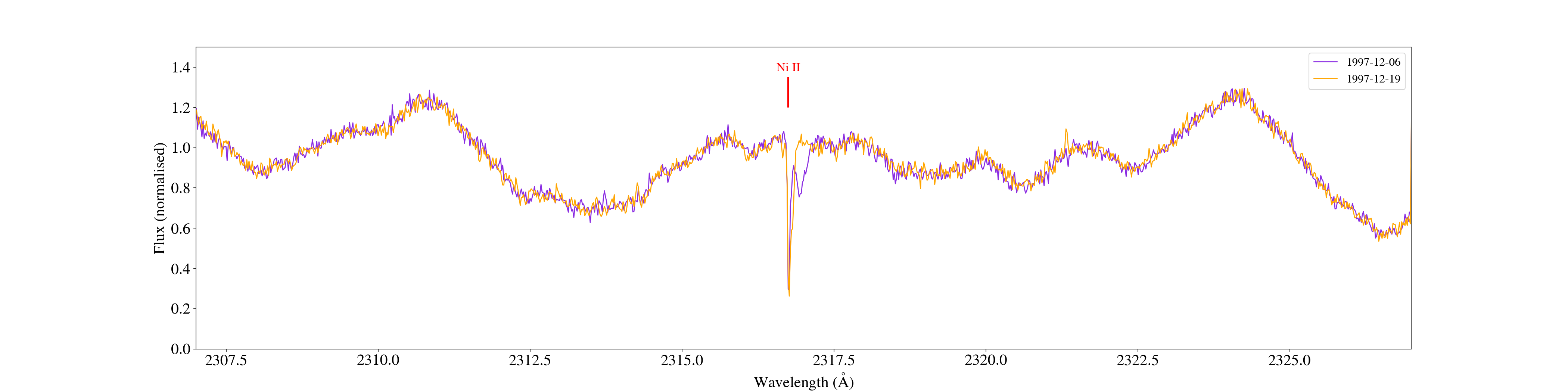}    
    \includegraphics[scale = 0.28,     trim = 80 27 80 30,clip]{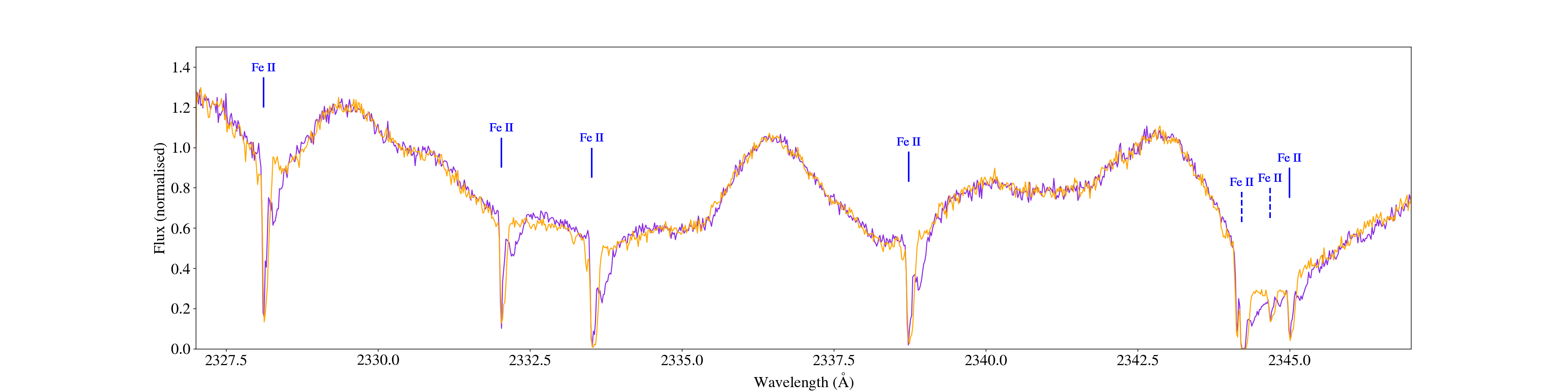}    
    \includegraphics[scale = 0.28,     trim = 80 27 80 30,clip]{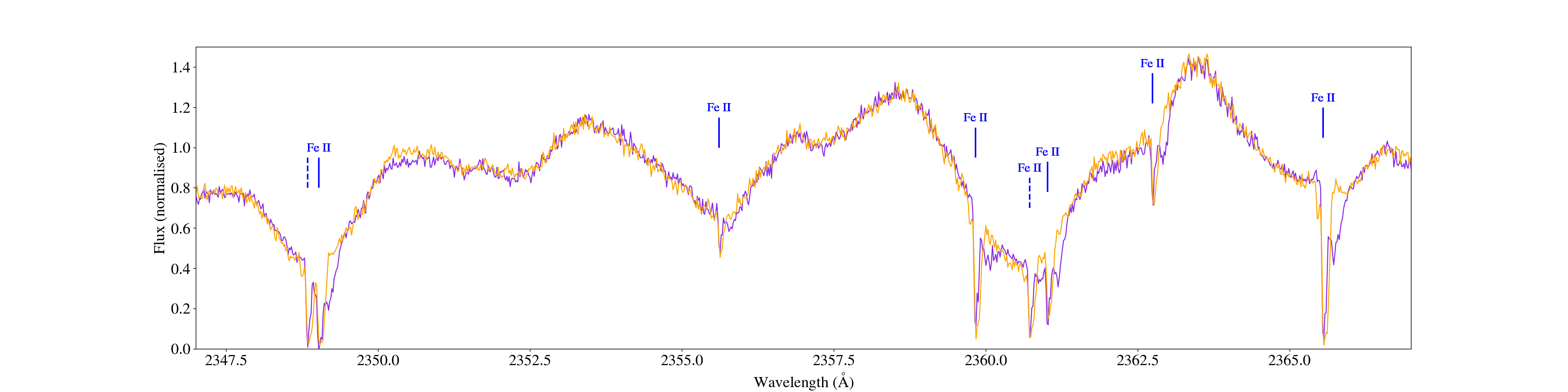}    
    \includegraphics[scale = 0.28,     trim = 80 27 80 30,clip]{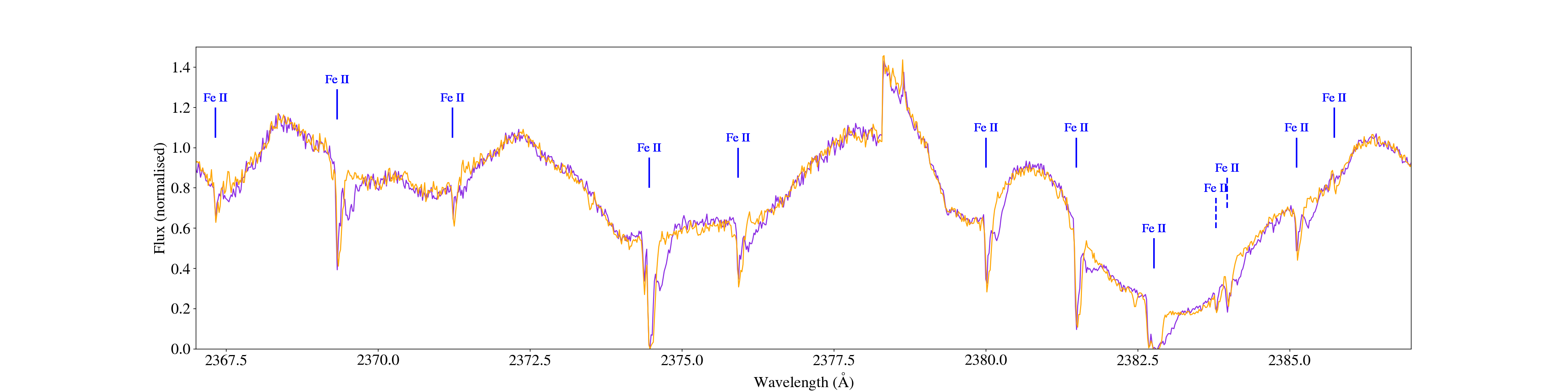}    
    \includegraphics[scale = 0.28,     trim = 80 0 80 30,clip]{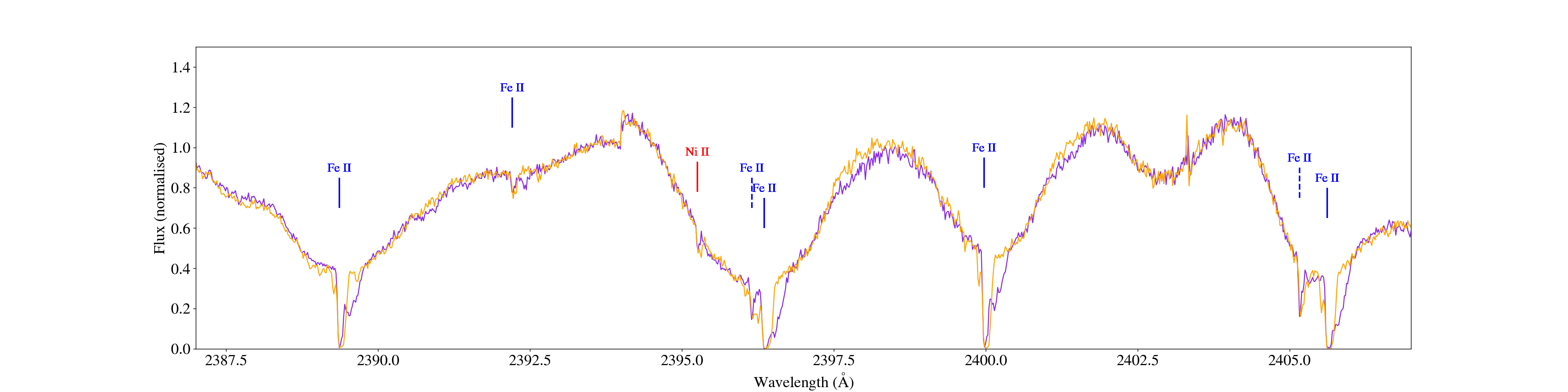}    
    \caption{}
\end{figure*}
\small \textbf{Figure G.1}, continued.

\newpage

\begin{figure*}[h!]
\centering
    \includegraphics[scale = 0.28,     trim = 80 27 80 30,clip]{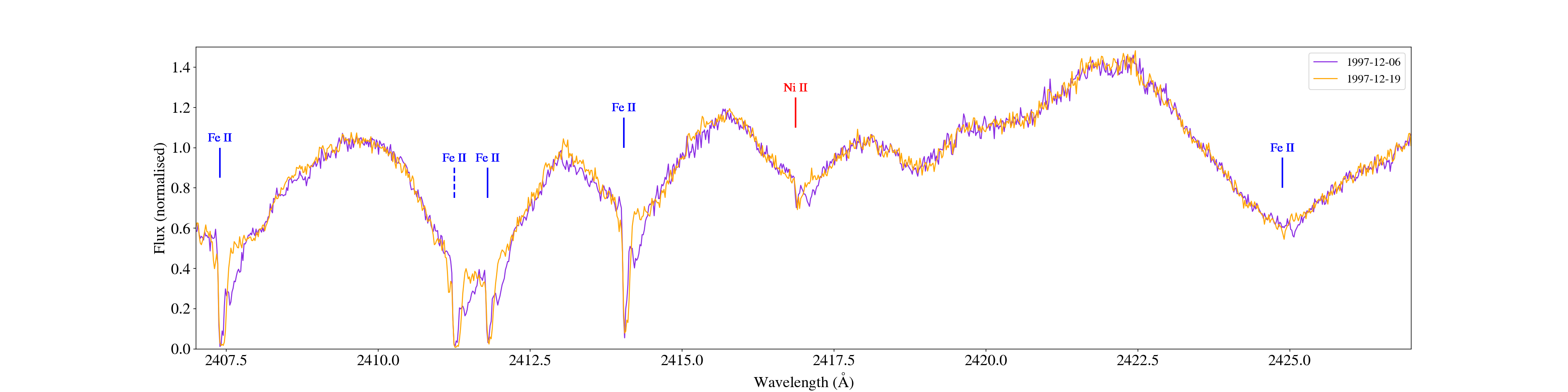}    
    \includegraphics[scale = 0.28,     trim = 80 27 80 40,clip]{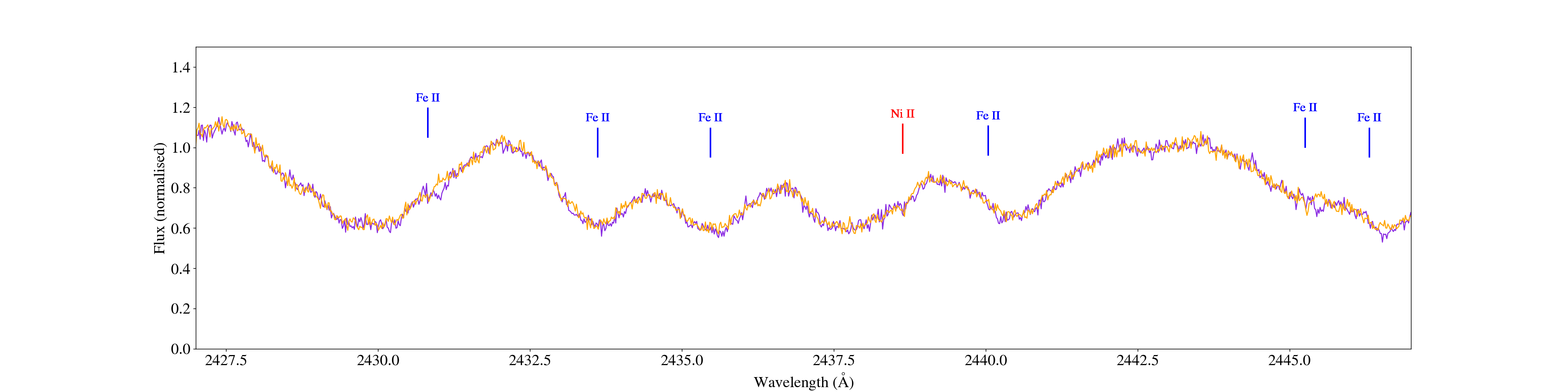}    
    \includegraphics[scale = 0.28,     trim = 80 27 80 40,clip]{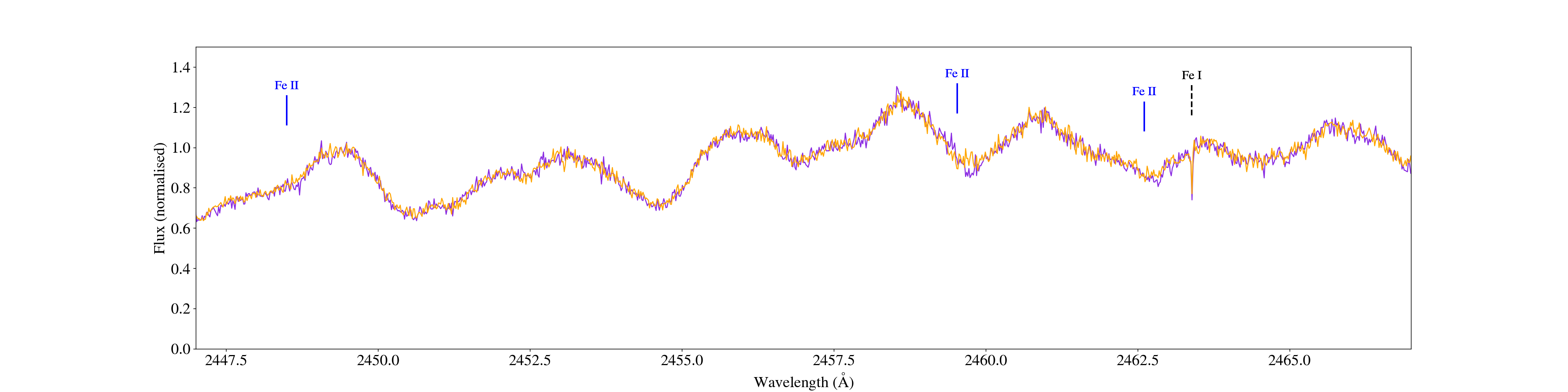}    
    \includegraphics[scale = 0.28,     trim = 80 27 80 40,clip]{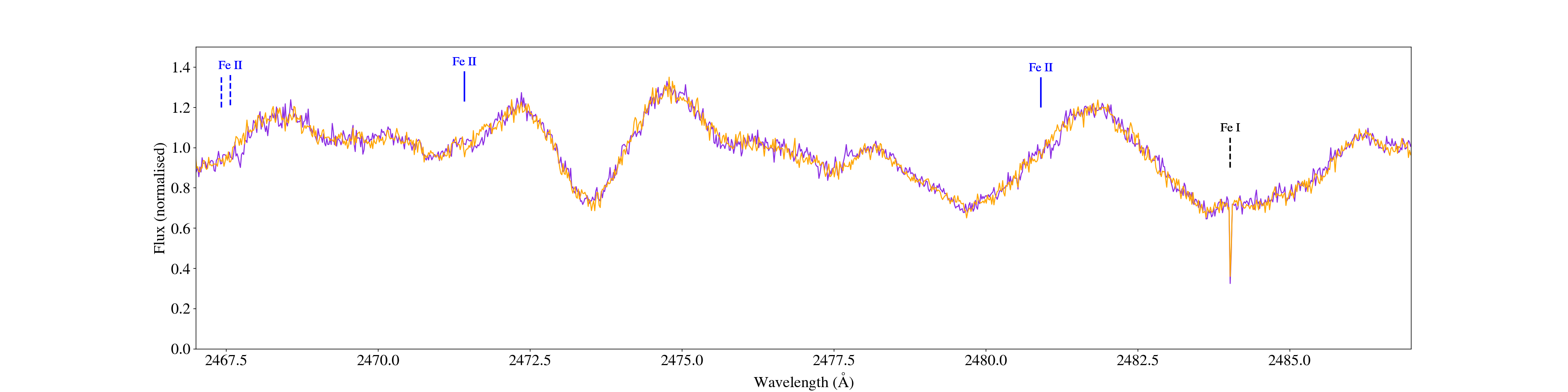}    
    \includegraphics[scale = 0.28,     trim = 80 0 80 40,clip]{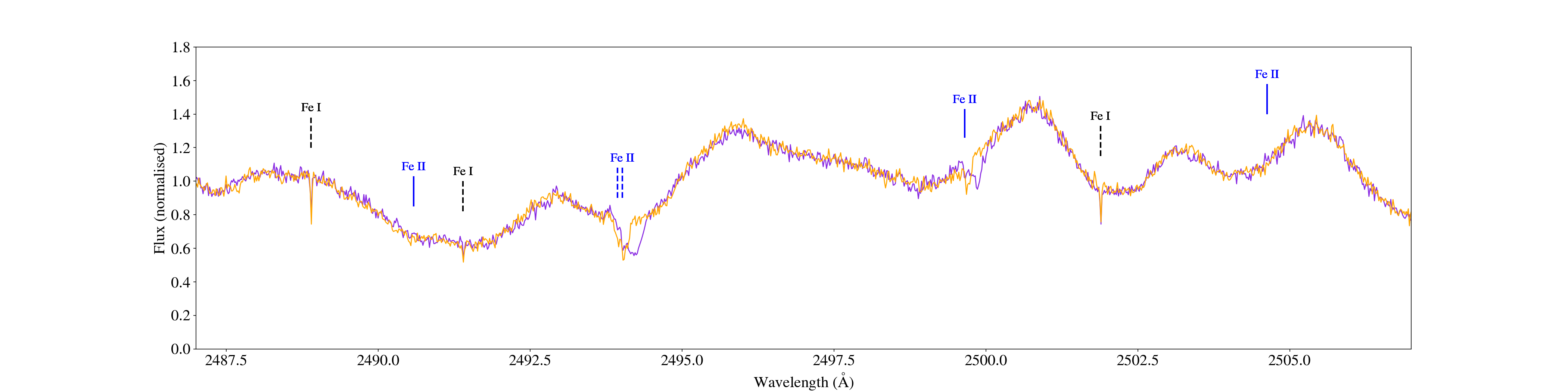}  
\end{figure*}
\small \textbf{Figure G.1}, continued.

\newpage

\begin{figure*}[h!]
\centering
    \includegraphics[scale = 0.28,     trim = 80 27 80 40,clip]{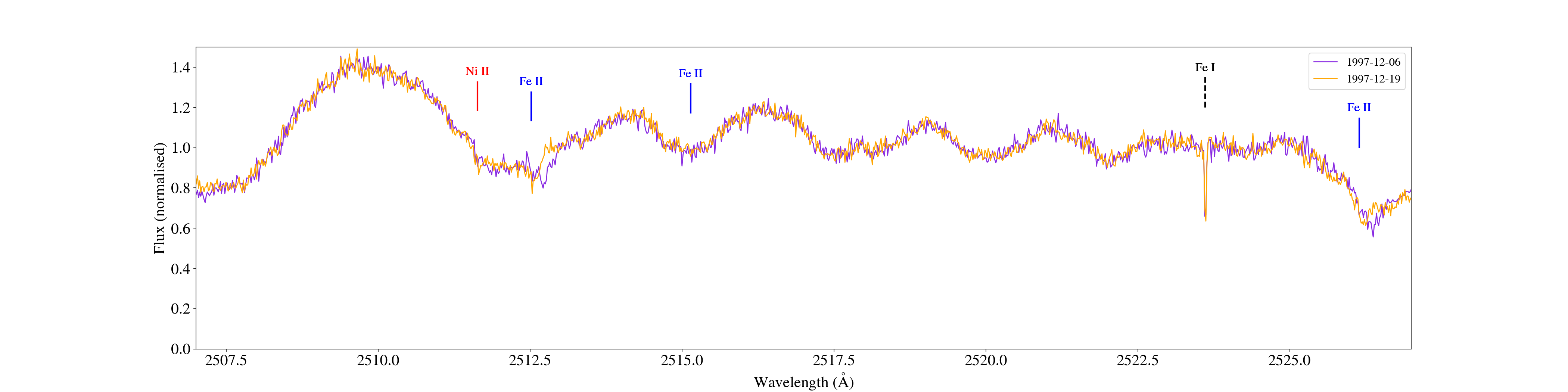}    
    \includegraphics[scale = 0.28,     trim = 80 27 80 40,clip]{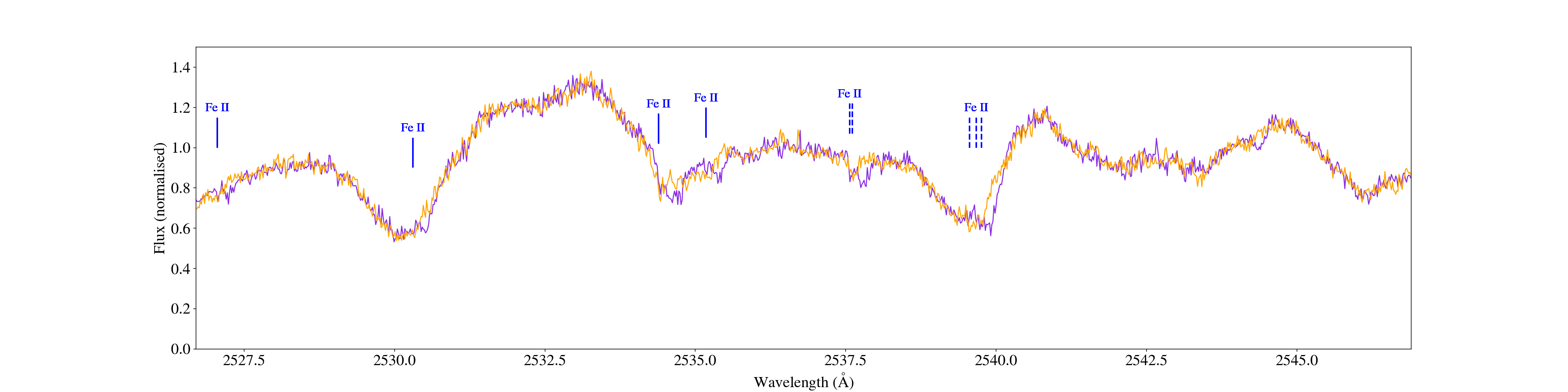}    
    \includegraphics[scale = 0.28,     trim = 80 27 80 40,clip]{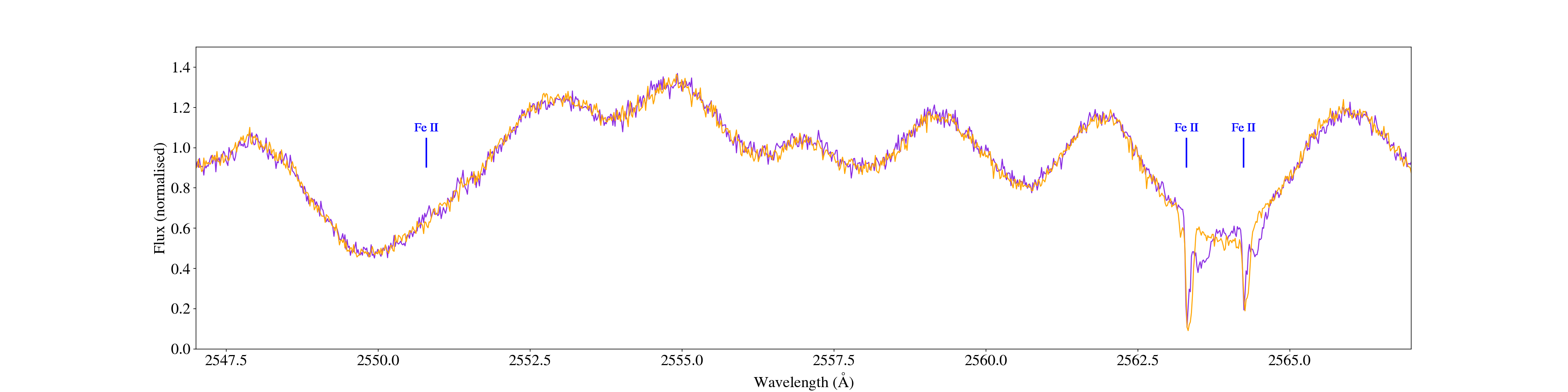}    
    \includegraphics[scale = 0.28,     trim = 80 27 80 40,clip]{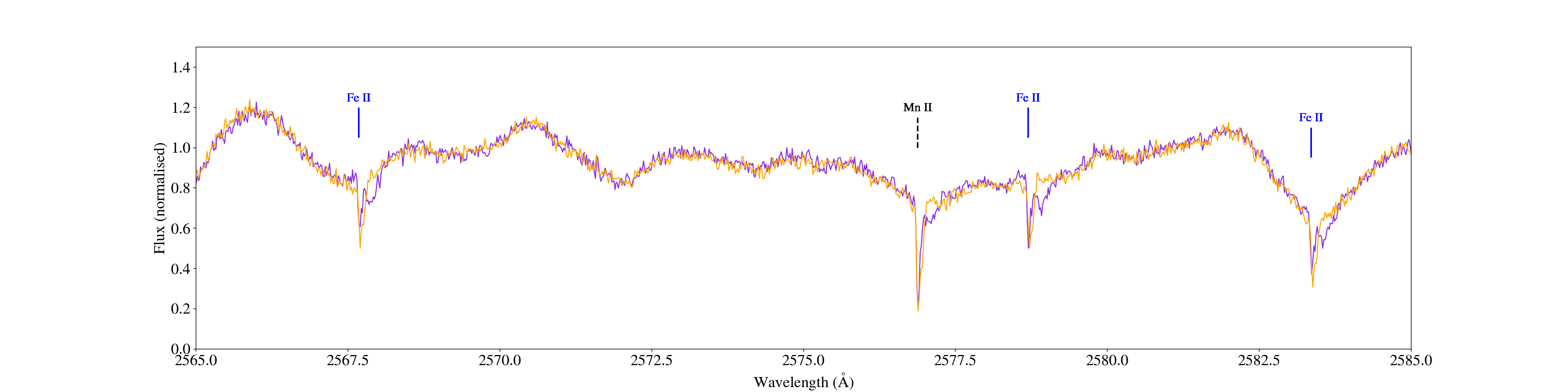}    
    \includegraphics[scale = 0.28,     trim = 80 0 80 40,clip]{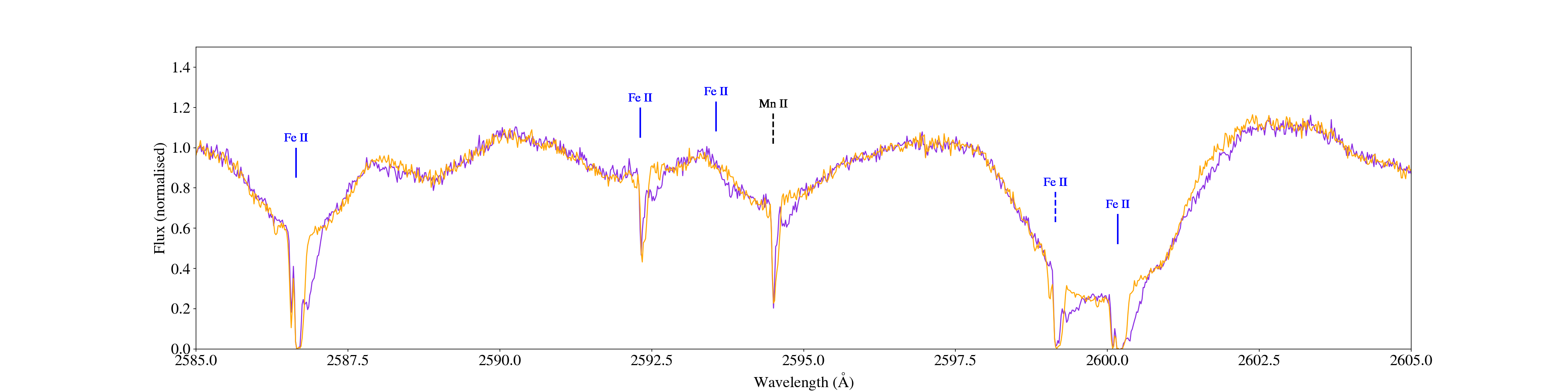}    
\end{figure*}
\small \textbf{Figure G.1}, continued.

\newpage

\begin{figure*}[h!]
\centering
    \includegraphics[scale = 0.28,     trim = 80 27 80 40,clip]{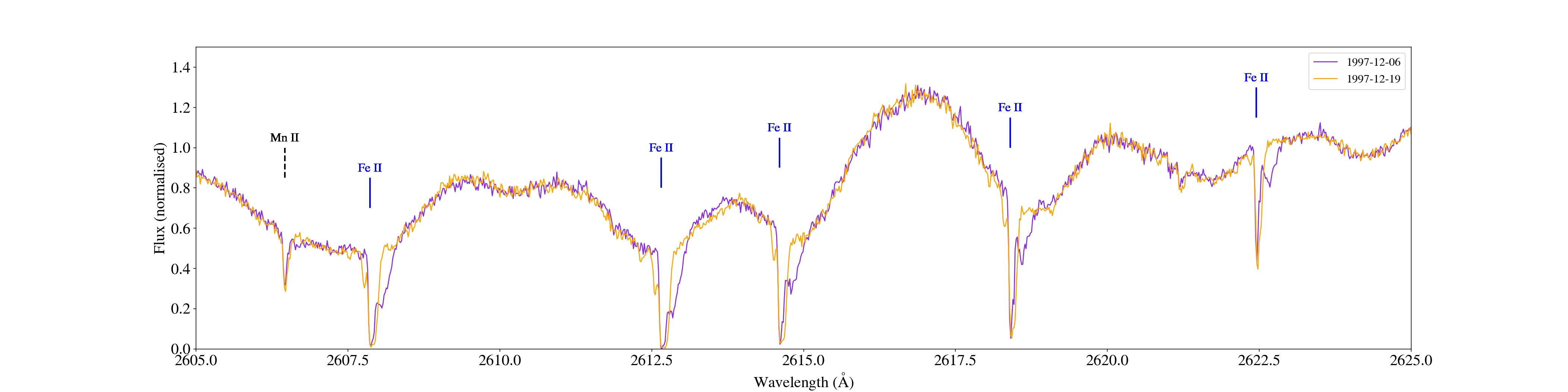}    
    \includegraphics[scale = 0.28,     trim = 80 27 80 40,clip]{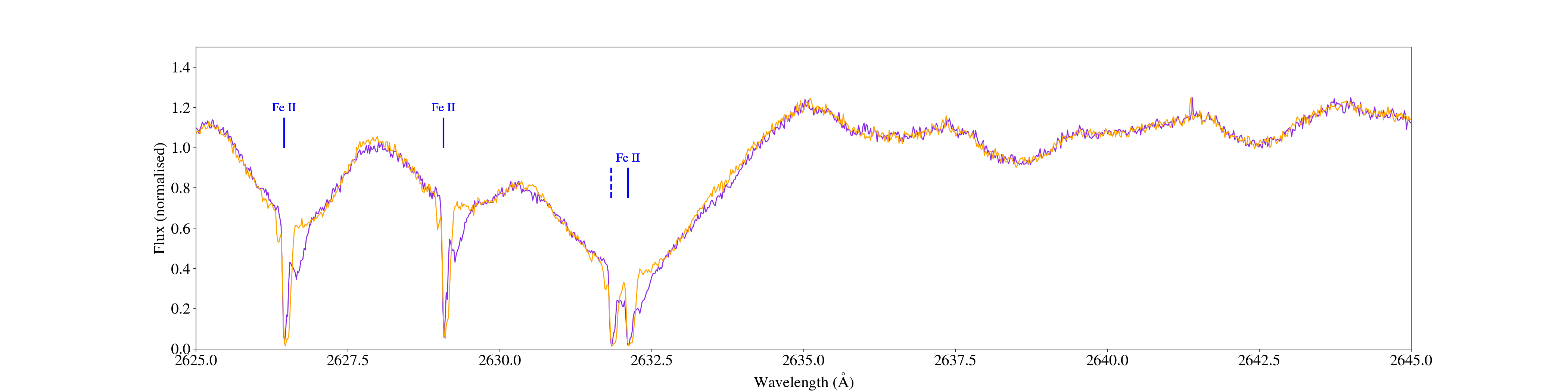}    
    \includegraphics[scale = 0.28,     trim = 80 27 80 40,clip]{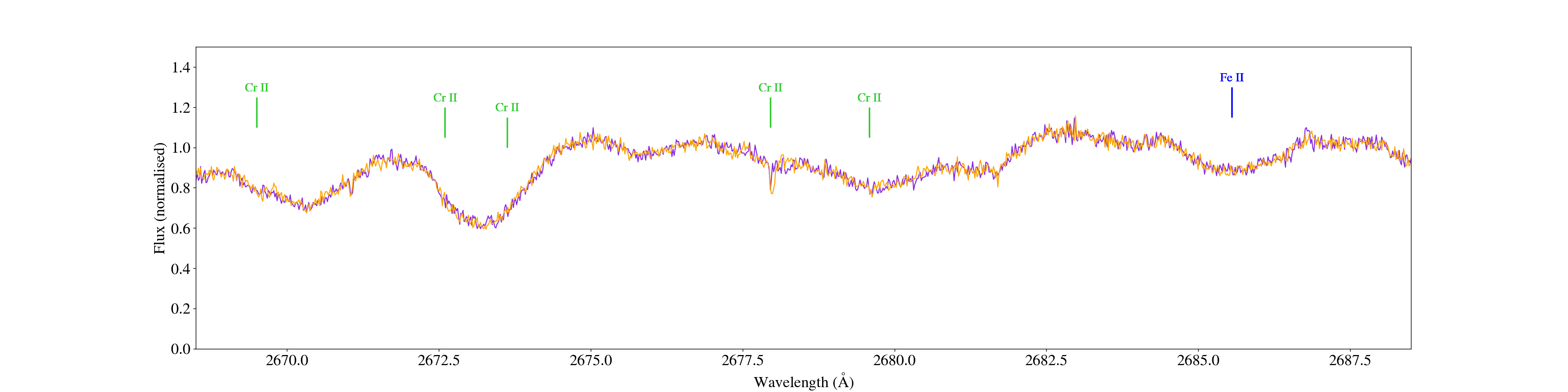}    
    \includegraphics[scale = 0.28,     trim = 80 27 80 40,clip]{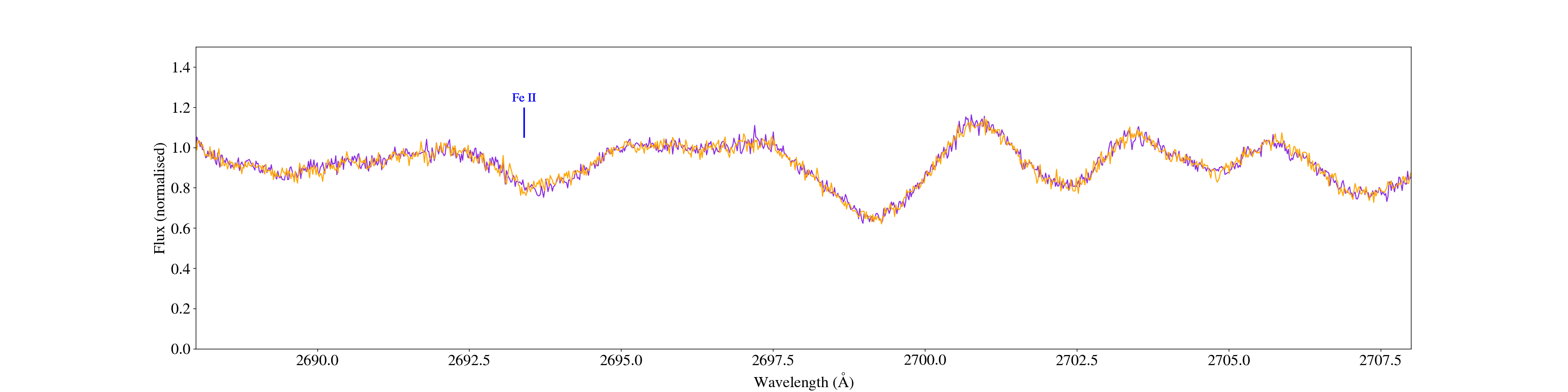}    
    \includegraphics[scale = 0.28,     trim = 80 0 80 40,clip]{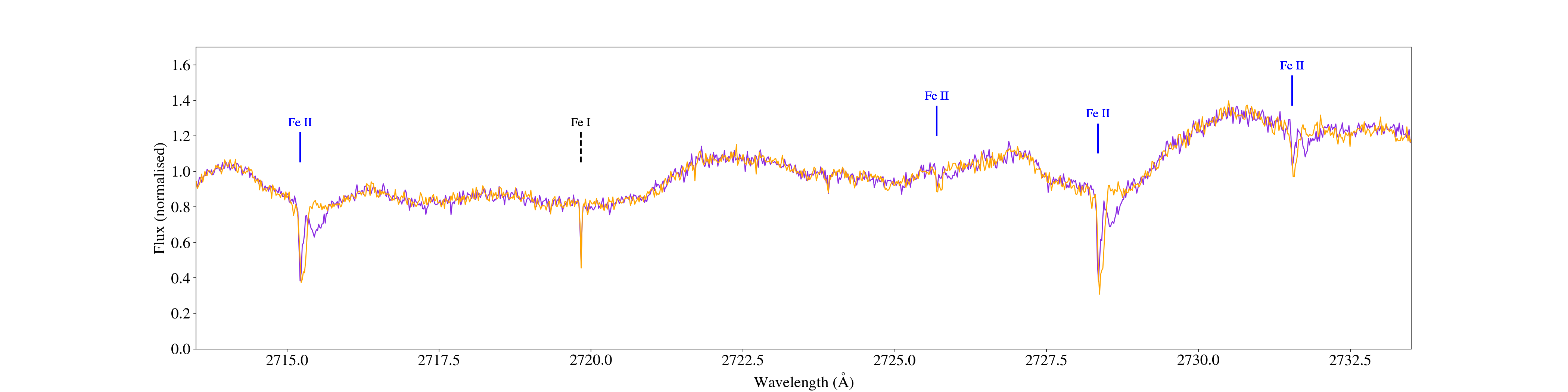}    
\end{figure*}
\small \textbf{Figure G.1}, continued.

\newpage

\begin{figure*}[h!]
\centering
    \includegraphics[scale = 0.28,     trim = 80 27 80 40,clip]{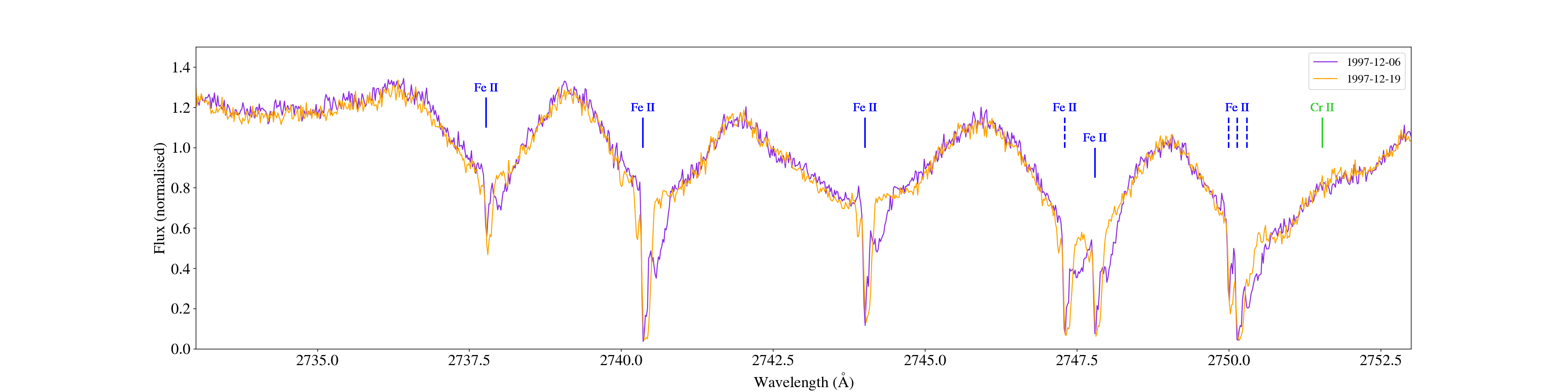}    
    \includegraphics[scale = 0.28,     trim = 80 27 80 40,clip]{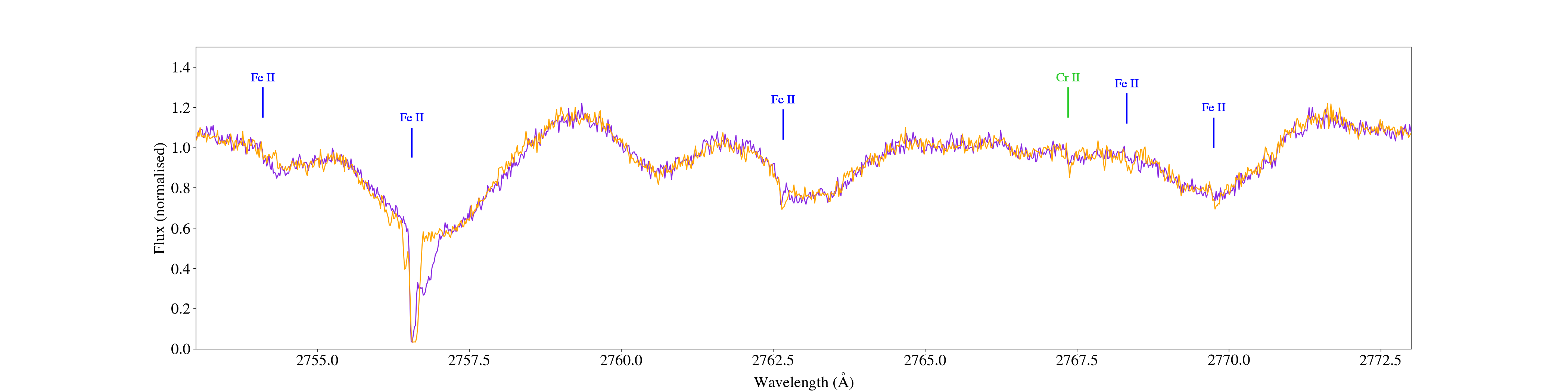}    
    \includegraphics[scale = 0.28,     trim = 80 27 80 40,clip]{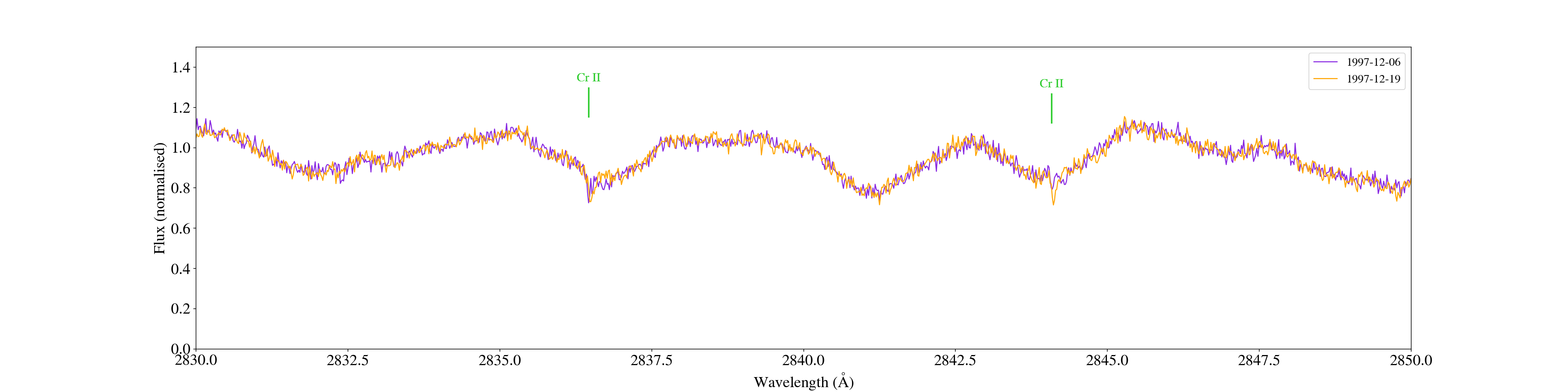}  
    \includegraphics[scale = 0.28,     trim = 80 0 80 40,clip]{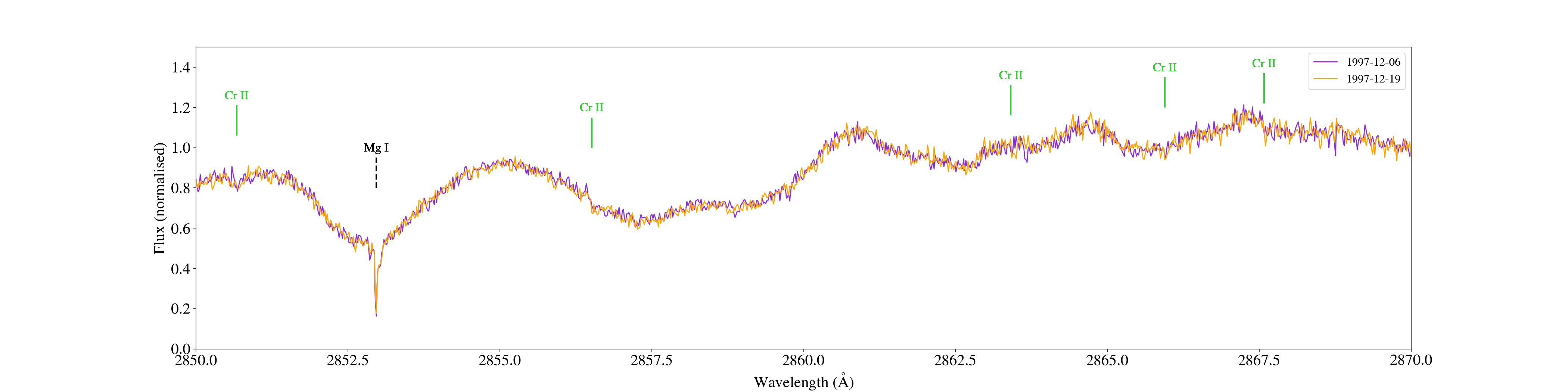}  
\end{figure*}
\small \textbf{Figure G.1}, continued.

\end{appendix}
\end{document}